\documentclass[12pt]{article}
\usepackage{makeidx}
\usepackage{amssymb}
\usepackage{amsfonts}
\usepackage{amsmath}
\usepackage{graphicx}
\usepackage{setspace}
\usepackage{authblk}
\usepackage{units}
\usepackage{appendix}
\usepackage[left=2cm,top=2.3cm,right=2cm,bottom=2.1cm]{geometry}
\usepackage{xcolor}
\usepackage[hyperfootnotes=false]{hyperref}
\hypersetup{colorlinks=true,citecolor=cyan,urlcolor=Red}
\renewcommand\familydefault\rmdefault
\newcommand{\lambdabar}{{\mkern0.75mu\mathchar '26\mkern -9.75mu\lambda}}

\begin{document}

\title{{\textbf{Vacuum instability in a constant inhomogeneous electric field. A new
example of exact nonperturbative calculations}}}
\author[1,2]{T. C. Adorno\thanks{adorno@hbu.edu.cn, tg.adorno@gmail.com}}
\author[2,3]{S. P. Gavrilov\thanks{gavrilovsergeyp@yahoo.com, gavrilovsp@herzen.spb.ru}}
\author[2,4,5]{D. M. Gitman\thanks{gitman@if.usp.br}}
\affil[1]{\textit{Department of Physics, College of Physical Sciences and Technology, Hebei University, Wusidong Road 180, 071002, Baoding, China;}}
\affil[2]{\textit{Department of Physics, Tomsk State University, Lenin Prospekt 36, 634050, Tomsk, Russia;}}
\affil[3]{\textit{Department of General and Experimental Physics, Herzen State Pedagogical
University of Russia, 191186, St. Petersburg, Russia;}}
\affil[4]{\textit{P. N. Lebedev Physical Institute, 53 Leninskiy prospekt, 119991, Moscow, Russia;}}
\affil[5]{\textit{Instituto de F\'{\i}sica, Universidade de S\~{a}o Paulo, Caixa Postal 66318, CEP 05508-090, S\~{a}o Paulo, S.P., Brazil.}}

\maketitle

\onehalfspacing

\begin{abstract}
Basic quantum processes (such as particle creation, reflection, and
transmission on the corresponding Klein steps) caused by inverse-square
electric fields are calculated. These results represent a new example of
exact nonperturbative calculations in the framework of QED. The
inverse-square electric field is time-independent, inhomogeneous in the $x$%
-direction, and is inversely proportional to $x$ squared. We find exact
solutions of the Dirac and Klein-Gordon equations with such a field and
construct corresponding \textrm{in}- and \textrm{out}-states. With the help
of these states and using the techniques developed in the framework of QED
with $x$-electric potential steps, we calculate characteristics of the
vacuum instability, such as differential and total mean numbers of particles
created from the vacuum and vacuum-to-vacuum transition probabilities. We
study the vacuum instability for two particular backgrounds: for fields
widely stretches over the $x$-axis (small-gradient configuration) and for
the fields sharply concentrates near the origin $x=0$ (sharp-gradient
configuration). We compare exact results with ones calculated numerically.
Finally, we consider the electric field configuration, composed by
inverse-square fields and by an $x$-independent electric field between them
to study the role of growing and decaying processes in the vacuum
instability.
\end{abstract}
{\textbf{Keywords}: Particle creation, Schwinger effect, Klein paradox,
inhomogeneous external\newline
electric field, Dirac and Klein-Gordon equations.}

\section{Introduction\label{Sec1}}

Particle creation from the vacuum by strong electromagnetic and
gravitational fields is a remarkable effect predicted by quantum field
theory (QFT). In the late 20s and the early 30s, Klein \cite{Klein27} and
Sauter \cite{Sauter31a} considered the effect in the framework of the
relativistic quantum mechanics. However, from the very beginning, it became
clear that all the questions could be answered only in the framework of QFT.
QFT with external backgrounds is, to a certain extent, an appropriate model
for such calculations. In the framework of such a model, the particle
creation is related to a violation of the vacuum stability with the time.
Backgrounds (external fields) that violate the vacuum stability are
electric-like fields that are able to produce nonzero work when interacting
with charged particles. Depending on the structure of such backgrounds,
different approaches for calculating the effect were proposed and realized.
From a quantum mechanical point of view, the most clear formulation of the
problem of particle production from the vacuum by external fields is
possible for time-dependent external electric fields that are switched on
and off at infinitely remote times $t\rightarrow \pm \infty $,
respectively.\ Such kind of external fields are called the $t$-electric
potential steps ($t$-step or $t$-steps). Scattering, particle creation from
the vacuum and particle annihilation by the $t$-steps were considered in the
framework of the relativistic quantum mechanics, see Refs. \cite%
{Nikis70a,Nikis79,GMR85,ruffini,GelTan16}, a more complete list of relevant
publications can be found in \cite{ruffini,GelTan16}. A general
nonperturbative with respect to the external background formulation of QED
with $t$-steps{\large \ }was developed in Ref. \cite{Gitman}.

In contrast to the $t$-electric potential steps, there are many physically
interesting situations where the external backgrounds are constant
(time-independent) but spatially inhomogeneous, for example, concentrated in
restricted space areas. The simplest type of such backgrounds is the
so-called $x$-electric potential steps ($x$-step or $x$-steps), in which the
field is inhomogeneous only in one space coordinate and represents a
spatial-like step for charged particles. The $x$-steps can also create
particles from the vacuum, the Klein paradox is closely related to this
process \cite{Klein27,Sauter31a,Sauter-pot}. Important calculations of the
particle creation by $x$-steps in the framework of the relativistic quantum
mechanics were presented by Nikishov in Refs. \cite{Nikis79,Nikis70b} and
later developed by Hansen and Ravndal in Refs. \cite{HansRavn81,Damour77}. A
general nonperturbative with respect to the external background formulation
of QED with $x$-steps{\large \ }was developed in Ref. \cite{x-case}. The
corresponding calculation is based on the existence of exact solutions of
Dirac or Klein-Gordon equation (wave equations, in what follows) with
corresponding external fields. When such solutions can be found and all the
calculations can be done, we refer these examples as exactly solvable cases.
Until now, there are known only few exactly solvable cases related to $t$%
-steps and to $x$-steps. In the case of the $t$-steps, these are particle
creation in the constant uniform electric field \cite{Sch51,Nikis70a}\emph{,}
in the adiabatic electric field $E\left( t\right) =E\cosh ^{-2}\left( t/T_{%
\mathrm{S}}\right) \,$\cite{NarNik70}, in the so-called $T$-constant
electric field \cite{BagGiS75,GavGit96}, in a periodic alternating in time
electric field \cite{NarNik74,MosFro74}, in an exponentially decaying
electric field \cite{AdoGavGit15}, in an exponentially growing and decaying
electric fields \cite{AdoGavGit16,AFGG17} (see Ref. \cite{AdoGavGit17} for
the review), in a composite electric field \cite{AFGG19}, and in an
inverse-square electric field (an electric field that is inversely
proportional to time squared \cite{AdoGavGit18}). In the case of $x$-steps
these are particle creation in the Sauter electric field \cite{x-case}, in
the so-called $L$-constant electric field \cite{GavGit16b}, and in the
inhomogeneous exponential peak field \cite{GavGitShi17}.

In this article, we present a new exactly solvable case in QED with $x$%
-steps where all nonperturbative characteristics of the vacuum instability
can be calculated and analyzed in detail. The electric field that
corresponds to this specific step is time-independent, it grows from zero in
the interval $x\in \left( -\infty ,0\right) $ inversely proportional to $x$
squared and decreases in the interval $x\in \left[ 0,+\infty \right) $ also
inversely proportional to $x$ squared, with $x$ being the coordinate $%
x=X^{1} $. For brevity, we hereinafter call such a field the inverse
potential step. An exact description of this field is given in Sec. \ref%
{Sec2}. There, we present exact solutions of the Dirac and Klein-Gordon
equations with such a step and construct corresponding \textrm{in}- and 
\textrm{out}-states. With the help of these states and using the techniques
developed in the work \cite{x-case}, we calculate pertinent quantities for
studying all the characteristics of the particle creation effect occurring
in the Klein zone, such as differential mean numbers of particles created
from the vacuum, total numbers and vacuum-to-vacuum transition
probabilities. These results are presented in Sec. \ref{Sec3}. Besides
processes related to the vacuum instability, in Sec. \ref{Sec4} we calculate
amplitudes and probabilities of basic processes occurring beyond the Klein
zone, namely reflection and transmission probabilities. Comparisons between
exact results (calculated numerically) and corresponding asymptotic
estimates are placed in Sec. \ref{Sec5}. In Sec. \ref{Sec7}, we discuss the
role of growing and decaying processes in the vacuum instability considering
various electric field configuration, composed by inverse-square fields and
by an $x$-independent electric field between them. The section \ref{Sec6} is
devoted to the concluding remarks. Useful formulas involving Whittaker
functions and some asymptotic representations of confluent hypergeometric
functions are placed in Appendix \ref{App1}. An unitary operator, connecting 
\textrm{in}- and \textrm{out}-vacua in Klein zone, is described in Appendix %
\ref{App2}.

\section{Solutions of wave equations with inverse potential steps\label{Sec2}%
}

\subsection{Inverse potential steps\label{Sec2.0}}

Here we consider wave equations with inverse potential steps and their
solutions. First of all, we describe more exactly the structure of the
electromagnetic field of inverse potential steps. Such a field is an
electric field in a $d=D+1$ dimensional Minkowski space-time. The latter
space-time is parameterized by coordinates $X=\left( X^{\mu }\,,\mu
=0,1,...,D\right) =\left( X^{0}=t,\mathbf{r}\right) $, $\mathbf{r}=\left(
X^{1}=x,\mathbf{r}_{\perp }\right) $, $\mathbf{r}_{\perp }=\left(
X^{2},...,X^{D}\right) $, the corresponding metric reads $\eta _{\mu \nu }=%
\mathrm{diag}\left( 1,-1,...,-1\right) $. The electric field is constant and
has only one component along the $x$-axis, $\mathbf{E}\left( X\right)
=\left( E^{1}\left( x\right) =E\left( x\right) ,0,...,0\right) $. The
corresponding electromagnetic potentials $A^{\mu }\left( X\right) $ are:%
\begin{equation}
A^{\mu }\left( X\right) =\left( A^{0}\left( x\right) \,,\ \ A^{k}=0\,,\ \
k=1,...,D\right) \,.  \notag
\end{equation}%
It is assumed that the electric field $E\left( x\right) =-\partial
_{x}A_{0}\left( x\right) >0$ is positive on the whole interval $x\in \mathbb{%
R=(}-\infty ,+\infty )$ and switches on and off at $x\rightarrow -\infty $
and $x\rightarrow -\infty $ respectively. At the same time, its potential $%
A^{0}\left( x\right) $ tends to certain, different in the general case,
constants values,%
\begin{equation}
\lim_{x\rightarrow \pm \infty }A^{0}\left( x\right) =A_{0}\left( \pm \infty
\right) ,\ A_{0}\left( -\infty \right) \neq A_{0}\left( +\infty \right) ,\
A_{0}\left( -\infty \right) >A_{0}\left( +\infty \right) \,.  \label{2.1}
\end{equation}

The field in question consists of two pieces, the first one is defined on
the interval $x\in \mathrm{I}=\left( -\infty ,0\right) $ while the second
one is defined on the interval $x\in \mathrm{II}=\left[ 0,+\infty \right) $,%
\begin{equation}
E\left( x\right) =E\left\{ 
\begin{array}{ll}
\left( 1-x/\xi _{1}\right) ^{-2}\,, & x\in \mathrm{I}\,, \\ 
\left( 1+x/\xi _{2}\right) ^{-2}\,, & x\in \mathrm{II}\,.%
\end{array}%
\right.  \label{2.2}
\end{equation}

\begin{figure}[th]
\begin{center}
\includegraphics[scale=0.3]{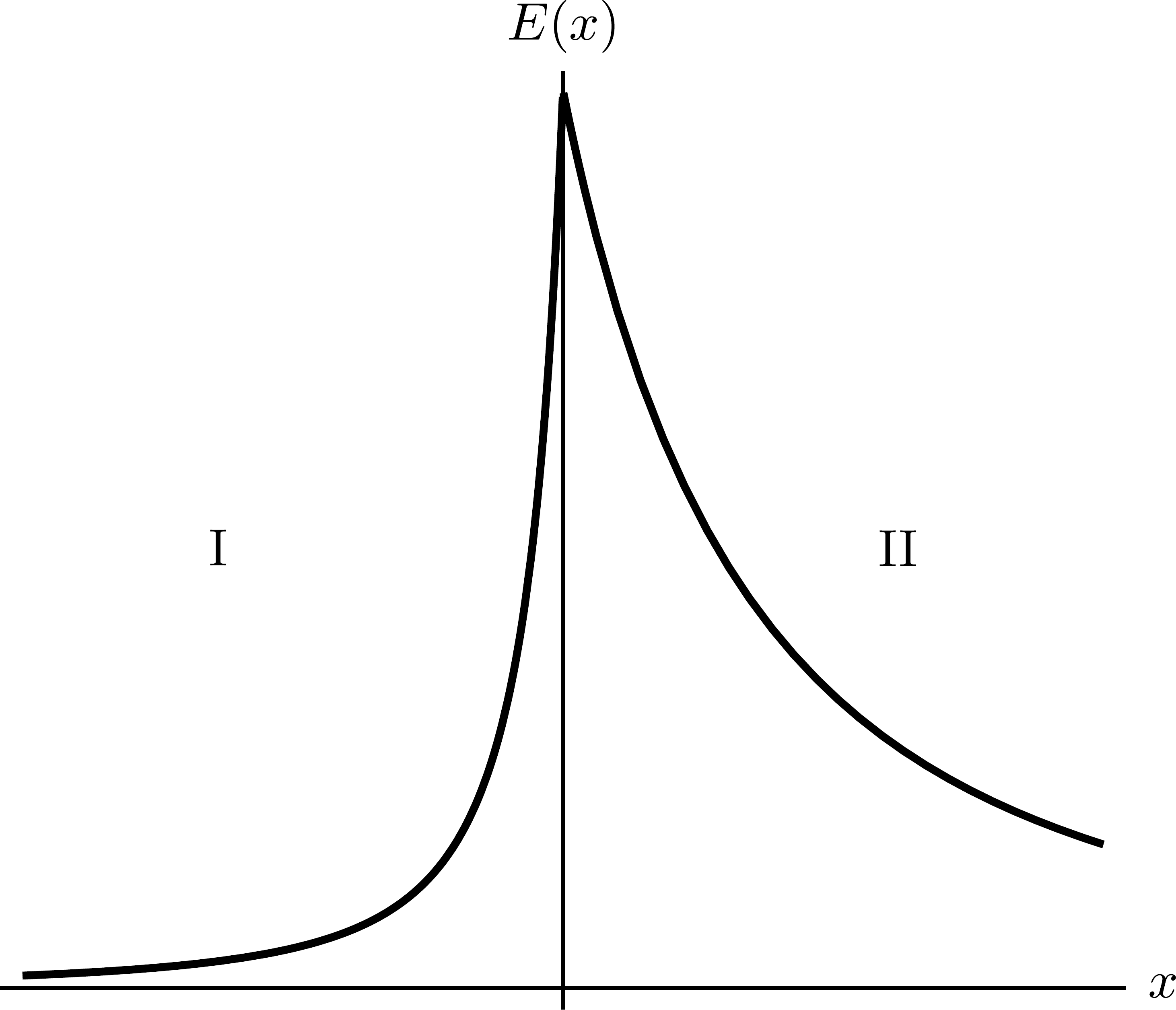}
\end{center}
\caption{Inverse-square electric field. In this picture, $\protect\xi _{2}>%
\protect\xi _{1}$.}
\label{Fig0}
\end{figure}

The corresponding potential reads:%
\begin{equation}
A_{0}\left( x\right) =E\left\{ 
\begin{array}{ll}
\xi _{1}\left[ 1-\left( 1-x/\xi _{1}\right) ^{-1}\right] \,, & x\in \mathrm{I%
}\,, \\ 
\xi _{2}\left[ \left( 1+x/\xi _{2}\right) ^{-1}-1\right] \,, & x\in \mathrm{%
II}\,.%
\end{array}%
\right.  \label{2.3}
\end{equation}%
The constants $\xi _{1,2}>0$ are length scales characterizing how
\textquotedblleft smooth\textquotedblright\ or \textquotedblleft
sharp\textquotedblright\ the electric field evolves from $x=-\infty $ to $%
x=0 $ and from $x=0$ to $x=+\infty $. At the same time, they characterize
the magnitude of the potential step. On Fig. \ref{Fig0}, we represent an
asymmetrical inverse-square electric field in which $\xi _{2}>\xi _{1}$.

The potential energy of an electron (with the charge $q=-e,$ $e>0$) in the
field of the step is $U\left( x\right) =-eA_{0}\left( x\right) $. It tends
to different in the general case constants values $U\left( -\infty \right) $
and $U\left( +\infty \right) $ as $x\rightarrow -\infty $ and $x\rightarrow
-\infty $ respectively,%
\begin{equation}
U\left( -\infty \right) \equiv U_{\mathrm{L}}=-eE\xi _{1}\,,\ \ U\left(
+\infty \right) \equiv U_{\mathrm{R}}=eE\xi _{2}\,.  \label{2.4}
\end{equation}%
\ 

The magnitude $\mathbb{U}$ of the potential step is given by the difference $%
U_{\mathrm{R}}-U_{\mathrm{L}}$:%
\begin{eqnarray}
&&\mathbb{U}=U_{\mathrm{R}}-U_{\mathrm{L}}=\Delta U_{1}+\Delta
U_{2}=eE\left( \xi _{1}+\xi _{2}\right) >0\,,  \notag \\
&&\Delta U_{1}=U\left( 0\right) -U\left( -\infty \right) =eE\xi _{1},\
\Delta U_{2}=U\left( +\infty \right) -U\left( 0\right) =eE\xi _{2}\ .
\label{2.4a}
\end{eqnarray}

Depending on the magnitude $\mathbb{U}$, the step is called noncritical or
critical one, see \cite{x-case},%
\begin{equation}
\mathbb{U}=\left\{ 
\begin{array}{l}
\mathbb{U}<\mathbb{U}_{c}=2m\,,\text{ \textrm{noncritical step}} \\ 
\mathbb{U}>\mathbb{U}_{c}\,,\text{ \textrm{critical step}}%
\end{array}%
\right. \,.  \label{2.5}
\end{equation}

\begin{figure}[th]
\begin{center}
\includegraphics[scale=0.6]{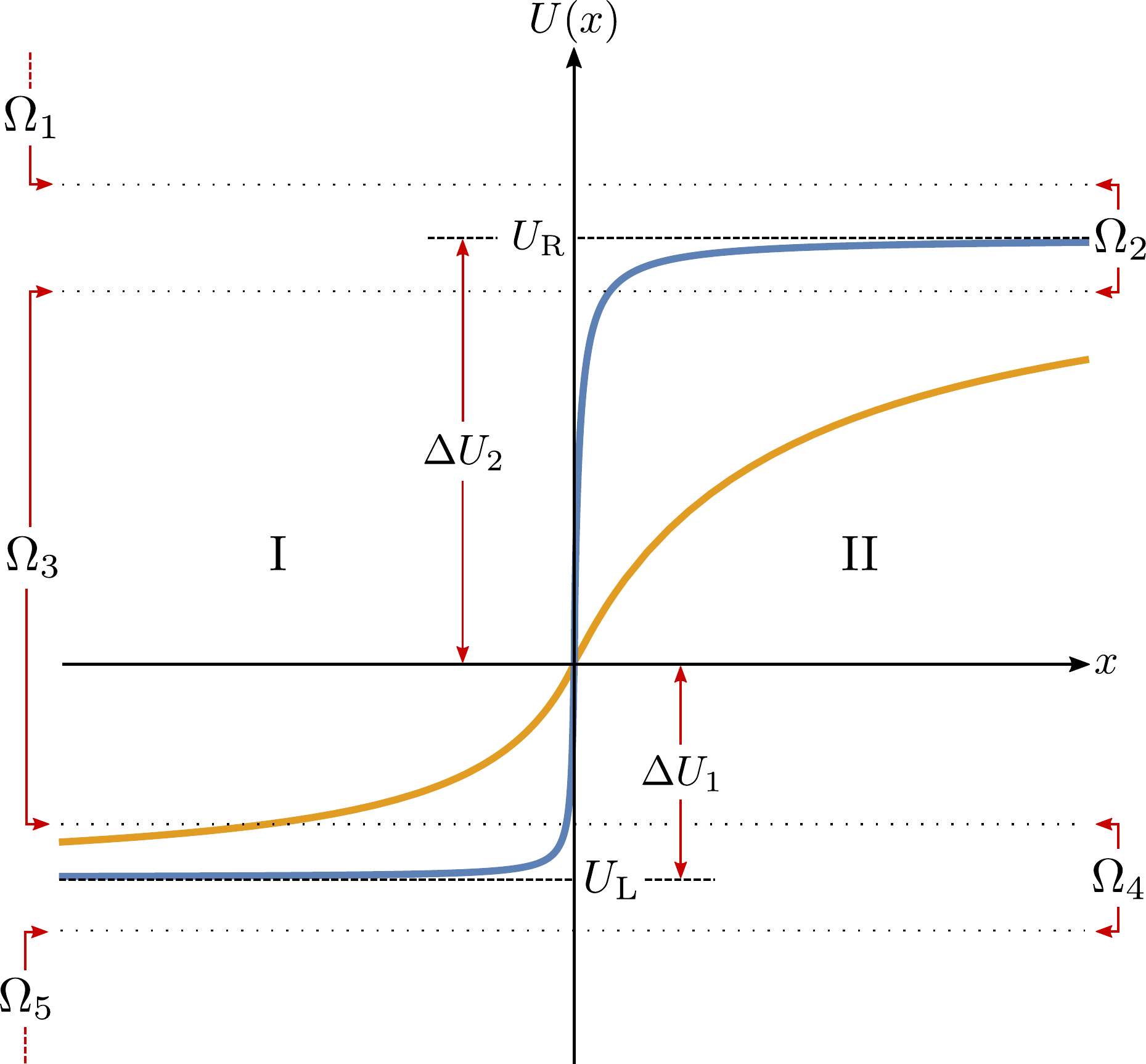}
\end{center}
\caption{Potential energies of an electron $U\left( x\right)$ in critical
inverse-square electric fields, corresponding to a \textquotedblleft
smooth\textquotedblright\ potential step (solid yellow line) and a
\textquotedblleft steep\textquotedblright\ potential step (solid blue line),
both with the same asymptotic values $U_{\mathrm{L/R}}$. The smaller
(larger) the value of the length scales $\protect\xi _{j}$, the steeper (the
smoother) the potential step. In both curves, $\protect\xi _{2}>\protect\xi %
_{1}$.}
\label{Fig1}
\end{figure}

As follows from Eqs. (\ref{2.4a}), this classification can be formulated in
terms of the sum ($\xi _{2}+\xi _{1}$) of the length scales $\xi _{1,2}$,%
\begin{eqnarray}
&&\xi _{2}+\xi _{1}<2\ell _{c}\,,\text{ \textrm{noncritical step,}}  \notag
\\
&&\xi _{2}+\xi _{1}>2\ell _{c}\,,\text{ \textrm{critical step,}}  \notag \\
&&\ell _{c}=\lambdabar_{c}E_{c}/E\,,\ \ E_{c}=m^{2}/e\,,\,\,\lambdabar_{c}=m^{-1}\ ,  \label{2.5a}
\end{eqnarray}%
where $E_{c}=m^{2}/e\approx 10^{16}\ \mathrm{V/cm}$ is the Schwinger
critical field and $\lambdabar_{c}$ is the Compton wave length of the electron. If the length scales $\xi
_{1,2}$ are large enough, the particle production from the vacuum could be
essential. On Fig. \ref{Fig1} we illustrate the potential energy $U\left(
x\right) $ for specific values of $\xi _{1,2}$ and electric field amplitude $%
E$.

In the Hamiltonian form, the Dirac equation with the inverse step reads:%
\begin{eqnarray}
&&i\partial _{0}\psi \left( X\right) =\hat{H}\psi \left( X\right) \,,  \notag
\\
&&\hat{H}=\gamma ^{0}\left( -i\gamma ^{j}\partial _{j}+m\right) +U\left(
x\right) \,,\ \ j=1,...,D\,,  \label{2.6}
\end{eqnarray}%
where the spinor field $\psi \left( X\right) $ has $2^{\left[ d/2\right] }$
components\footnote{$\left[ d/2\right] $ denotes the integer part of $d/2$.}
and $\gamma ^{\mu }$ are $2^{\left[ d/2\right] }\times 2^{\left[ d/2\right]
} $ Dirac matrices in $d=D+1$ dimensions,%
\begin{equation}
\left[ \gamma ^{\mu },\gamma ^{\nu }\right] _{+}=2\eta ^{\mu \nu }\,,\ \
\eta ^{\mu \nu }=\mathrm{diag}\underset{d}{\underbrace{\left(
+1,-1,...,-1\right) }}\,.  \label{2.7}
\end{equation}

Because electromagnetic potentials of the inverse steps have trivial
components $A_{0}=0$, $\mathbf{A}_{\perp }=0$, there exist solutions of the
Dirac equation in the form of stationary plane waves propagating along the
space-time directions $t$ and $\mathbf{r}_{\perp }$. In this case the Dirac
spinors can be represented as%
\begin{eqnarray}
\psi _{n}\left( X\right) &=&\exp \left( -ip_{0}t+i\mathbf{p}_{\perp }\mathbf{%
r}_{\perp }\right) \psi _{n}\left( x\right) \,,\ \ n=\left( p_{0},\mathbf{p}%
_{\perp },\sigma \right) \,,  \notag \\
\psi _{n}\left( x\right) &=&\left\{ \gamma ^{0}\left[ p_{0}-U\left( x\right) %
\right] +i\gamma ^{1}\partial _{x}-\boldsymbol{\gamma }_{\perp }\mathbf{p}%
_{\perp }+m\right\} \varphi _{n}\left( x\right) v_{\chi ,\sigma }\,,
\label{2.8}
\end{eqnarray}%
where the spinor $\psi _{n}\left( x\right) $ and the scalar function $%
\varphi _{n}\left( x\right) $ depend exclusively on $x$ while $v_{\chi
,\sigma }$ are eigenspinors of $\gamma ^{0}\gamma ^{1}$, satisfying $\gamma
^{0}\gamma ^{1}v_{\chi ,\sigma }=\chi v_{\chi ,\sigma }$, $\chi =\pm 1$.
Here, $\sigma =\left( \sigma _{s}=\pm 1\,,\ s=1,2,...,\left[ d/2\right]
-1\right) $ denotes a set of eigenvalues of spin operators compatible with $%
\gamma ^{0}\gamma ^{1}$, whose amount depends on the space-time
dimensionality\ $d$. For higher dimensions\footnote{%
The spinning degrees-of-freedom $\sigma $ are absent in $d=1+1$ or $d=2+1$
space-time dimensions.}, $d>3+1$, we may construct $J_{\left( d\right) }=2^{%
\left[ d/2\right] -1}$ additional spin operators and subject the constant
spinors $v_{\chi ,\sigma }$ to obey the following supplementary conditions:%
\begin{eqnarray}
&&i\gamma ^{2s}\gamma ^{2s+1}v_{\chi ,\sigma }=\sigma _{s}v_{\chi ,\sigma
}\,,\mathrm{\ \ for\ even\ }d\,,  \notag \\
&&i\gamma ^{2s+1}\gamma ^{2s+2}v_{\chi ,\sigma }=\sigma _{s}v_{\chi ,\sigma
}\,,\ \ \mathrm{for\ odd}\ d\,,  \notag \\
&&v_{\chi ^{\prime },\sigma ^{\prime }}^{\dagger }v_{\chi ,\sigma }=\delta
_{\chi ^{\prime }\chi }\delta _{\sigma ^{\prime }\sigma }\,.  \label{2.9b}
\end{eqnarray}%
Due to the compatibility of the spin operators with $\gamma ^{0}\gamma ^{1}$%
, the eigenvalues $\sigma _{s}$, in addition to $\chi $, parameterizes the
solutions. Plugging Eq. (\ref{2.8}) into Eq. (\ref{2.6}), one finds that
scalar functions $\varphi _{n}\left( x\right) $ obey the second-order
ordinary differential equation%
\begin{equation}
\left\{ \frac{d^{2}}{dx^{2}}+\left[ p_{0}-U\left( x\right) \right] ^{2}-\pi
_{\perp }^{2}+i\chi U^{\prime }\left( x\right) \right\} \varphi _{n}\left(
x\right) =0\,.  \label{2.10}
\end{equation}%
Here $\pi _{\perp }=\sqrt{\mathbf{p}_{\perp }^{2}+m^{2}}$ and by a prime a
differentiation with respect to $x$, $U^{\prime }\left( x\right) =dU\left(
x\right) /dx$, is denoted.

It should be noted that similar solutions of the Klein-Gordon equation can
be represented as:%
\begin{equation}
\psi _{n}\left( X\right) =\exp \left( -ip_{0}t+i\mathbf{p}_{\perp }\mathbf{r}%
_{\perp }\right) \varphi _{n}\left( x\right) \,,\ \ n=\left( p_{0},\mathbf{p}%
_{\perp }\right) \,,  \label{2.20}
\end{equation}%
where $\varphi _{n}\left( x\right) $ satisfy Eq. (\ref{2.10}) with $\chi =0$%
. Besides minor modifications in the normalization constants for scalar
particles (discussed in the next subsection) a formal transition to the
Klein-Gordon case can be done by setting $\chi =0$\ in all formulas above.

\subsection{Solutions with special left\ and right\ asymptotics\label{Sec2.1}%
}

Introducing new variables%
\begin{eqnarray}
z_{1}\left( x\right) &=&2i\left\vert p^{\mathrm{L}}\right\vert \xi
_{1}\left( 1-x/\xi _{1}\right) \,,\ \ x\in \mathrm{I}\,,  \notag \\
z_{2}\left( x\right) &=&2i\left\vert p^{\mathrm{R}}\right\vert \xi
_{2}\left( 1+x/\xi _{2}\right) \,,\ \ x\in \mathrm{II}\,,  \label{2.11}
\end{eqnarray}%
where $p^{\mathrm{L/R}}=\zeta \left\vert p^{\mathrm{L/R}}\right\vert $, $%
\zeta =\pm =\mathrm{sgn}\left( p^{\mathrm{L}}\right) =\mathrm{sgn}\left( p^{%
\mathrm{R}}\right) $, denotes real asymptotic momenta along the $x$-axis%
\footnote{%
Hereafter, the indexes \textquotedblleft $\mathrm{L}$\textquotedblright\ and
\textquotedblleft $\mathrm{R}$\textquotedblright\ (standing for
\textquotedblleft left\textquotedblright\ and \textquotedblleft
right\textquotedblright , respectively) are used to label quantities with
specific asymptotic properties at $x\rightarrow -\infty $ and $x\rightarrow
+\infty $, respectively.},%
\begin{equation}
\left\vert p^{\mathrm{L/R}}\right\vert =\sqrt{\pi _{0}\left( \mathrm{L/R}%
\right) ^{2}-\pi _{\perp }^{2}}\,,\ \ \pi _{0}\left( \mathrm{L/R}\right)
=p_{0}-U_{\mathrm{L/R}}\,,  \label{2.12}
\end{equation}%
differential equation (\ref{2.10}) reduces to a Whittaker differential
equation\footnote{%
The index \textquotedblleft $j$\textquotedblright\ distinguish quantities
associated with the interval $\mathrm{I}$ $\left( j=1\right) $ from ones
associated with the interval $\mathrm{II}$ $\left( j=2\right) $.} \cite%
{Erdelyi},%
\begin{equation}
\left( \frac{d^{2}}{dz_{j}^{2}}-\frac{1}{4}+\frac{\kappa _{j}}{z_{j}}+\frac{%
1/4-\mu _{j}^{2}}{z_{j}^{2}}\right) \varphi _{n}\left( z_{j}\right) =0\,,
\label{2.13}
\end{equation}%
whose parameters $\kappa _{j}$, $\mu _{j}$ are given by%
\begin{eqnarray}
&&\kappa _{1}=i\Delta U_{1}\xi _{1}\frac{\pi _{0}\left( \mathrm{L}\right) }{%
\left\vert p^{\mathrm{L}}\right\vert }\,,\ \ \kappa _{2}=-i\Delta U_{2}\xi
_{2}\frac{\pi _{0}\left( \mathrm{R}\right) }{\left\vert p^{\mathrm{R}%
}\right\vert }\,,  \notag \\
&&\mu _{j}=\left( -1\right) ^{j}\left( i\Delta U_{j}\xi _{j}-\chi /2\right)
\,.  \label{2.15}
\end{eqnarray}

General solutions of Eq. (\ref{2.13}) are chosen to be combinations of
Whittaker functions with regular asymptotics at infinity \cite{Erdelyi,NIST},%
\begin{equation}
W_{\kappa ,\mu }\left( z\right) =e^{-z/2}z^{\kappa }\left[ 1+O\left(
z^{-1}\right) \right] \,,\ \ z\rightarrow \infty \,,\ \ \left\vert \arg
z\right\vert \leq \frac{3\pi }{2}-0^{+}\,,  \label{2.13b}
\end{equation}%
such that $\varphi _{n}\left( z_{j}\right) =b_{j}^{1}W_{\kappa _{j},\mu
_{j}}\left( z_{j}\right) +b_{j}^{2}W_{-\kappa _{j},\mu _{j}}\left( e^{-i\pi
}z_{j}\right) $, with $b_{j}^{1,2}$ being arbitrary constants. The Whittaker
functions can be alternatively expressed in terms of confluent
hypergeometric functions (CHF) as follows:%
\begin{eqnarray}
&&W_{\kappa _{j},\mu _{j}}\left( z_{j}\right)
=e^{-z_{j}/2}z_{j}^{c_{j}/2}\Psi \left( a_{j},c_{j};z_{j}\right) \,,  \notag
\\
&&W_{-\kappa _{j},\mu _{j}}\left( e^{-i\pi }z_{j}\right) =e^{-i\pi
c_{j}/2}e^{z_{j}/2}z_{j}^{c_{j}/2}\Psi \left( c_{j}-a_{j},c_{j};e^{-i\pi
}z_{j}\right) \,,  \notag \\
&&a_{j}=\mu _{j}-\kappa _{j}+1/2,\ c_{j}=1+2\mu _{j}\,,  \label{2.14}
\end{eqnarray}%
and the Wronskian of the independent set $W_{\kappa _{j},\mu _{j}}\left(
z_{j}\right) $, $W_{-\kappa _{j},\mu _{j}}\left( e^{-i\pi }z_{j}\right) $ is
given by Eq. (13.14.30) in \cite{NIST}.

Due to local properties of equation (\ref{2.10}) at $x\rightarrow \mp \infty 
$ (where the electric field is zero), the scalar functions $\varphi
_{n}\left( x\right) $ have definite left \textquotedblleft \textrm{L}%
\textquotedblright\ and right \textquotedblleft \textrm{R}%
\textquotedblright\ asymptotics:%
\begin{eqnarray}
\ _{\zeta }\varphi _{n}\left( x\right) &=&\ _{\zeta }\mathcal{N}e^{i\zeta
\left\vert p^{\mathrm{L}}\right\vert x}\ \text{\textrm{as}}\ x\rightarrow
-\infty \,,  \notag \\
\ ^{\zeta }\varphi _{n}\left( x\right) &=&\ ^{\zeta }\mathcal{N}e^{i\zeta
\left\vert p^{\mathrm{R}}\right\vert x}\text{ \textrm{as}}\ \ x\rightarrow
+\infty \,.  \label{2.16}
\end{eqnarray}%
Here $p^{\mathrm{L}}$, $p^{\mathrm{R}}$ are asymptotic momenta along the $x$%
-axis, given by Eq. (\ref{2.12}), whereas $_{\zeta }\mathcal{N}$ and$\
^{\zeta }\mathcal{N}$ are some normalization constants. We label the scalar
functions by $\zeta $ related to the corresponding momenta.

For the Dirac spinors we have:%
\begin{eqnarray}
\hat{p}_{x}\ _{\zeta }\psi _{n}\left( X\right) &=&\zeta \left\vert p^{%
\mathrm{L}}\right\vert \ _{\zeta }\psi _{n}\left( X\right) \ \text{\textrm{as%
}}\ x\rightarrow -\infty \,,  \notag \\
\hat{p}_{x}\ ^{\zeta }\psi _{n}\left( X\right) &=&\zeta \left\vert p^{%
\mathrm{R}}\right\vert \ ^{\zeta }\psi _{n}\left( X\right) \ \text{\textrm{as%
}}\ x\rightarrow +\infty \,,  \label{2.17}
\end{eqnarray}%
and%
\begin{eqnarray}
&&\hat{H}^{\mathrm{kin}}\ _{\zeta }\psi _{n}\left( X\right) =\pi _{0}\left( 
\mathrm{L}\right) \ _{\zeta }\psi _{n}\left( X\right) \ \mathrm{as}\
x\rightarrow -\infty \,,  \notag \\
&&\hat{H}^{\mathrm{kin}}\ ^{\zeta }\psi _{n}\left( X\right) =\pi _{0}\left( 
\mathrm{R}\right) \ ^{\zeta }\psi _{n}\left( X\right) \ \mathrm{as}\
x\rightarrow +\infty \,,  \label{2.18}
\end{eqnarray}%
where $\hat{H}^{\mathrm{kin}}=\hat{H}-U\left( x\right) $ is the one-particle
quantum kinetic energy operator. Thus, nontrivial sets of Dirac spinors $%
\left\{ \ _{\zeta }\psi _{n}\left( X\right) \right\} $, $\left\{ \ ^{\zeta
}\psi _{n}\left( X\right) \right\} $ exist for quantum numbers $n$
satisfying the conditions%
\begin{equation}
\pi _{0}\left( \mathrm{L/R}\right) ^{2}>\pi _{\perp }^{2}\Leftrightarrow
\left\{ 
\begin{array}{l}
\pi _{0}\left( \mathrm{L/R}\right) >\pi _{\perp }\, \\ 
\pi _{0}\left( \mathrm{L/R}\right) <-\pi _{\perp }\,%
\end{array}%
\right. \,.  \label{2.19}
\end{equation}%
As a result of the above inequalities, the set of quantum numbers $n$ can be
divided in specific ranges $\Omega _{k}$, where the index $k$ labels
distinct ranges and the corresponding quantum numbers $n_{k}\in \Omega _{k}$%
. For critical steps, $\mathbb{U}>\mathbb{U}_{c}$, there are five ranges of
quantum numbers $\Omega _{k}$, $k=1,...,5$, composed by all spinning
degrees-of-freedom $\sigma $, unbounded perpendicular momenta $\mathbf{p}%
_{\perp }\in \left( -\infty ,+\infty \right) $, and by certain energies $%
p_{0}$, whose definitions and general properties are briefly listed below:

\begin{enumerate}
\item The ranges $\Omega _{1}$ and $\Omega _{5}$ are characterized by
energies bounded from below, \newline
$\Omega _{1}=\left\{ n:p_{0}\geq U_{\mathrm{R}}+\pi _{\perp }\right\} ,$ and
by energies bounded from above $\Omega _{5}=\left\{ n:p_{0}\leq U_{\mathrm{L}%
}-\pi _{\perp }\right\} $. In each one of these ranges, all relations from
Eq. (\ref{2.19}) are satisfied, which implies that nontrivial complete sets
of solutions $\left\{ \ _{\zeta }\psi _{n_{1}}\left( X\right) \,,\ _{\zeta
}\psi _{n_{5}}\left( X\right) \right\} $ and $\left\{ \ ^{\zeta }\psi
_{n_{1}}\left( X\right) \,,\ ^{\zeta }\psi _{n_{5}}\left( X\right) \right\} $
do exist.

\item The ranges $\Omega _{2}$ and $\Omega _{4}$ are characterized by
bounded energies, namely \newline
$\Omega _{2}=\left\{ n:U_{\mathrm{R}}-\pi _{\perp }<p_{0}<U_{\mathrm{R}}+\pi
_{\perp }\right\} $ and $\Omega _{4}=\left\{ n:U_{\mathrm{L}}-\pi _{\perp
}<p_{0}<U_{\mathrm{L}}+\pi _{\perp }\right\} $ if $\mathbb{U}\geq 2\pi
_{\perp }$ or $\Omega _{2}=\left\{ n:U_{\mathrm{L}}+\pi _{\perp }<p_{0}<U_{%
\mathrm{R}}+\pi _{\perp }\right\} $ and $\Omega _{4}=\left\{ n:U_{\mathrm{L}%
}-\pi _{\perp }<p_{0}<U_{\mathrm{R}}-\pi _{\perp }\right\} $ if $\mathbb{U}%
<2\pi _{\perp }$. The relation $\pi _{0}\left( \mathrm{L}\right) >\pi
_{\perp }$ is satisfied only for quantum numbers from $\Omega _{2}$ while
the relation $\pi _{0}\left( \mathrm{R}\right) <-\pi _{\perp }$ is satisfied
only for quantum numbers from $\Omega _{4}$, which means that in $\Omega
_{2} $ there exist solutions only with left asymptotics $\left\{ \ _{\zeta
}\psi _{n_{2}}\left( X\right) \right\} $ while in $\Omega _{4}$ there exist
solutions only with right asymptotics $\left\{ \ ^{\zeta }\psi
_{n_{4}}\left( X\right) \right\} $.

\item The range $\Omega _{3}$ is nontrivial only for critical steps and
perpendicular momenta $\mathbf{p}_{\perp }$ restricted by the inequality $%
2\pi _{\perp }\leq \mathbb{U}$. This range is characterized by bounded
energies, \newline
$\Omega _{3}=\left\{ n:U_{\mathrm{L}}+\pi _{\perp }\leq p_{0}\leq U_{\mathrm{%
R}}-\pi _{\perp }\right\} $. In this range, the relations $\pi _{0}\left( 
\mathrm{L}\right) \geq \pi _{\perp }$ and $\pi _{0}\left( \mathrm{R}\right)
\leq -\pi _{\perp }$ hold true which means that both sets of solutions $%
\left\{ \ _{\zeta }\psi _{n_{3}}\left( X\right) \right\} $ and $\left\{ \
^{\zeta }\psi _{n_{3}}\left( X\right) \right\} $ do exist.
\end{enumerate}

The assumption about the completeness of solutions in some ranges refers to
their asymptotic properties at infinitely remote distances. Because of the
properties of the Whittaker functions with large arguments (\ref{2.13b}),
sets of solutions in the ranges $\Omega _{1}$, $\Omega _{3}$, and $\Omega
_{5}$ are complete asymptotically. Moreover, because of the triviality of
right solutions in $\Omega _{2}$ and left solutions in $\Omega _{4}$,
certain restrictions on the form of solutions apply in these ranges. The
manifold of all the quantum numbers $n$ is denoted by $\Omega =\Omega
_{1}\cup \cdots \cup \Omega _{5}$. For noncritical steps $\mathbb{U}<\mathbb{%
U}_{c}$, the range $\Omega _{3}$ is absent. For the correct interpretation
of the states $\left\{ \ _{\zeta }\psi _{n}\left( X\right) \right\} $ and $%
\left\{ \ ^{\zeta }\psi _{n}\left( X\right) \right\} $ as wave functions
describing electrons and positrons as well as for a complete discussion
about the ranges and further properties, see Ref. \cite{x-case}.

In view of the asymptotic behavior of the Whittaker functions with large
argument (\ref{2.13b}) and the properties discussed above, it is possible to
classify solutions in the first $\mathrm{I}$ and in the second $\mathrm{II}$
intervals according to the sign $\zeta =\pm $ of the asymptotic momenta $p^{%
\mathrm{L/R}}$. Denoting scalar functions in $\mathrm{I}$, $\mathrm{II}$ as$%
\ _{\zeta }\varphi _{n}\left( x\right) $ and$\ ^{\zeta }\varphi _{n}\left(
x\right) $, respectively, we have:%
\begin{eqnarray}
\ _{+}\varphi _{n}\left( x\right) &=&\ _{+}\mathcal{N}W_{\kappa _{1},\mu
_{1}}\left( z_{1}\right) \,,\ _{-}\varphi _{n}\left( x\right) =\ _{-}%
\mathcal{N}W_{-\kappa _{1},\mu _{1}}\left( e^{-i\pi }z_{1}\right) \,,\ \
x\in \mathrm{I}\,,  \notag \\
\ ^{+}\varphi _{n}\left( x\right) &=&\ ^{+}\mathcal{N}W_{-\kappa _{2},\mu
_{2}}\left( e^{-i\pi }z_{2}\right) \,,\ ^{-}\varphi _{n}\left( x\right) =\
^{-}\mathcal{N}W_{\kappa _{2},\mu _{2}}\left( z_{2}\right) \,,\ \ x\in 
\mathrm{II}\,.  \label{2.21}
\end{eqnarray}

Once the electric field is homogeneous in time and in the coordinates
perpendicular to the field $\mathbf{r}_{\perp }$, the normalization
constants $_{\zeta }\mathcal{N}$ and$\ ^{\zeta }\mathcal{N}$ are calculated
with respect to the inner product on the $x$-constant hyperplane%
\begin{equation}
\left( \psi ,\psi ^{\prime }\right) _{x}=\int \psi ^{\dagger }\left(
X\right) \gamma ^{0}\gamma ^{1}\psi ^{\prime }\left( X\right) dtd\mathbf{r}%
_{\perp }\,.  \label{2.22}
\end{equation}

To calculate the inner product, we consider our system in a large space-time
box of the volume $V_{\perp }=\prod_{j=2}^{D}K_{j}$ and over time $T$, where
all length scales $K_{j}$, $T$ are macroscopically large. Moreover, we
impose periodic boundary conditions on the Dirac spinors $\psi \left(
X\right) $ in the variables $t$ and $X^{j}$, $j=2,...,D$, so that all
solutions are periodic under transitions from one box to another. Then, the
integrations over the transverse coordinates are performed from $-K_{j}/2$
to $+K_{j}/2$ and from $-T/2$ to $+T/2$, where the limits $K_{j}\rightarrow
\infty $, $T\rightarrow \infty $ are assumed in final expressions; see Ref. 
\cite{x-case} for details. Under these conditions, inner product (\ref{2.22}%
) is $x$-independent and can be expressed in terms of the scalar functions as%
\footnote{%
For $\psi ^{\prime }=\psi $, the inner product (\ref{2.22}) divided by $T$
coincides with the definition of the current density across \ the hyperplane 
$x=\mathrm{const.}$} follows:%
\begin{eqnarray}
&&\left( \psi _{n},\psi _{n^{\prime }}^{\prime }\right) _{x}=V_{\perp
}T\delta _{nn^{\prime }}\mathcal{I}_{n}\,,  \notag \\
&&\mathcal{I}_{n}=\varphi _{n}^{\ast }\left( x\right) \left( i\overleftarrow{%
\partial }_{x}-i\overrightarrow{\partial }_{x}\right) \left[ p_{0}-U\left(
x\right) +i\chi \partial _{x}\right] \varphi _{n}^{\prime }\left( x\right)
\,.  \label{2.23}
\end{eqnarray}%
According to general properties of the left and right asymptotics outlined
in the previous section, the solutions $\left\{ \ _{\zeta }\psi _{n}\left(
X\right) \right\} $ and $\left\{ \ ^{\zeta }\psi _{n}\left( X\right)
\right\} $ can be subjected to the orthonormalization conditions%
\begin{eqnarray}
\left( \ _{\zeta }\psi _{n},\ _{\zeta }\psi _{n^{\prime }}\right) _{x}
&=&\zeta \eta _{\mathrm{L}}\delta _{\zeta \zeta ^{\prime }}\delta
_{nn^{\prime }}\,,\ \ n\in \Omega _{1}\cup \Omega _{2}\cup \Omega _{3}\cup
\Omega _{5}\,,  \notag \\
\left( \ ^{\zeta }\psi _{n},\ ^{\zeta }\psi _{n^{\prime }}\right) _{x}
&=&\zeta \eta _{\mathrm{R}}\delta _{\zeta \zeta ^{\prime }}\delta
_{nn^{\prime }}\,,\ \ n\in \Omega _{1}\cup \Omega _{3}\cup \Omega _{4}\cup
\Omega _{5}\,,  \label{2.24}
\end{eqnarray}%
where $\eta _{\mathrm{L}}=\mathrm{sgn}\pi _{0}\left( \mathrm{L}\right) $ and 
$\eta _{\mathrm{R}}=\mathrm{sgn}\pi _{0}\left( \mathrm{R}\right) $. Using
asymptotic properties of the Whittaker functions (\ref{2.13b}) and the above
conditions, the normalization constants $_{\zeta }\mathcal{N}$ and $\
^{\zeta }\mathcal{N}$ are%
\begin{eqnarray}
&&\ _{\zeta }\mathcal{N}=\ _{\zeta }CY\,,\ \ ^{\zeta }\mathcal{N}=\ ^{\zeta
}CY\,,\ \ Y=\left( V_{\perp }T\right) ^{-1/2}\,,  \notag \\
&&\ _{\zeta }C=\frac{\exp \left( -i\pi \kappa _{1}/2\right) }{\sqrt{%
2\left\vert p^{\mathrm{L}}\right\vert \left\vert \pi _{0}\left( \mathrm{L}%
\right) -\zeta \chi \left\vert p^{\mathrm{L}}\right\vert \right\vert }}\,,\
\ ^{\zeta }C=\frac{\exp \left( -i\pi \kappa _{2}/2\right) }{\sqrt{%
2\left\vert p^{\mathrm{R}}\right\vert \left\vert \pi _{0}\left( \mathrm{R}%
\right) -\zeta \chi \left\vert p^{\mathrm{R}}\right\vert \right\vert }}\,.
\label{2.24b}
\end{eqnarray}

Because spinors $\ _{\zeta }\psi _{n}\left( X\right) $ and $\ ^{\zeta }\psi
_{n}\left( X\right) $ with quantum numbers $n\in \Omega _{1}\cup \Omega
_{3}\cup \Omega _{5}$ are complete, we can decompose solutions from one set
onto another as%
\begin{eqnarray}
\eta _{\mathrm{L}}\ ^{\zeta }\psi _{n}\left( X\right) &=&\ _{+}\psi
_{n}\left( X\right) g\left( _{+}|^{\zeta }\right) -\ _{-}\psi _{n}\left(
X\right) g\left( _{-}|^{\zeta }\right) \,,  \notag \\
\eta _{\mathrm{R}}\ _{\zeta }\psi _{n}\left( X\right) &=&\ ^{+}\psi
_{n}\left( X\right) g\left( ^{+}|_{\zeta }\right) -\ ^{-}\psi _{n}\left(
X\right) g\left( ^{-}|_{\zeta }\right) \,,  \label{2.25}
\end{eqnarray}%
where the decomposition coefficients $g$ are given by%
\begin{equation}
g\left( ^{\zeta ^{\prime }}|_{\zeta }\right) ^{\ast }=g\left( _{\zeta
}|^{\zeta ^{\prime }}\right) =\left( \ _{\zeta }\psi _{n},\ ^{\zeta ^{\prime
}}\psi _{n}\right) _{x}\,,\ \ n\in \Omega _{1}\cup \Omega _{3}\cup \Omega
_{5}\,.  \label{2.26}
\end{equation}%
Substituting decompositions (\ref{2.25}) in normalization conditions (\ref%
{2.24}) we find%
\begin{eqnarray}
&&g\left( ^{\zeta ^{\prime }}|_{+}\right) g\left( _{+}|^{\zeta }\right)
-g\left( ^{\zeta ^{\prime }}|_{-}\right) g\left( _{-}|^{\zeta }\right)
=\zeta \eta _{\mathrm{L}}\eta _{\mathrm{R}}\delta _{\zeta ,\zeta ^{\prime
}}\,,  \notag \\
&&g\left( _{\zeta ^{\prime }}|^{+}\right) g\left( ^{+}|_{\zeta }\right)
-g\left( _{\zeta ^{\prime }}|^{-}\right) g\left( ^{-}|_{\zeta }\right)
=\zeta \eta _{\mathrm{L}}\eta _{\mathrm{R}}\delta _{\zeta ,\zeta ^{\prime
}}\,.  \label{2.26b}
\end{eqnarray}%
The latter relations imply a number of equations on $g$-coefficients, in
particular,%
\begin{eqnarray}
&&\left\vert g\left( _{-}|^{+}\right) \right\vert ^{2}=\left\vert g\left(
_{+}|^{-}\right) \right\vert ^{2}\,,\ \ \left\vert g\left( _{+}|^{+}\right)
\right\vert ^{2}=\left\vert g\left( _{-}|^{-}\right) \right\vert ^{2}\,, 
\notag \\
&&\left\vert g\left( _{+}|^{-}\right) \right\vert ^{2}-\left\vert g\left(
_{+}|^{+}\right) \right\vert ^{2}=-\eta _{\mathrm{L}}\eta _{\mathrm{R}}\,.
\label{2.27}
\end{eqnarray}

From Eqs. (\ref{2.8}) and (\ref{2.25}), one finds similar decompositions
between the left and right scalar functions,%
\begin{equation}
\ ^{-}\varphi _{n}\left( x\right) =\left\{ 
\begin{array}{ll}
\eta _{\mathrm{L}}\left[ \ _{+}\varphi _{n}\left( x\right) g\left(
_{+}|^{-}\right) -\ _{-}\varphi _{n}\left( x\right) g\left( _{-}|^{-}\right) %
\right] \,, & x\in \mathrm{I} \\ 
\ ^{-}\mathcal{N}W_{\kappa _{2},\mu _{2}}\left( z_{2}\right) \,, & x\in 
\mathrm{II}%
\end{array}%
\right. ,  \label{2.28}
\end{equation}%
and%
\begin{equation}
\ _{+}\varphi _{n}\left( x\right) =\left\{ 
\begin{array}{ll}
\ _{+}\mathcal{N}W_{\kappa _{1},\mu _{1}}\left( z_{1}\right) \,, & x\in 
\mathrm{I} \\ 
\eta _{\mathrm{R}}\left[ \ ^{+}\varphi _{n}\left( x\right) g\left(
^{+}|_{+}\right) -\ ^{-}\varphi _{n}\left( x\right) g\left( ^{-}|_{+}\right) %
\right] \,, & x\in \mathrm{II}%
\end{array}%
\right.  \label{2.29}
\end{equation}%
The $g$-coefficient can be obtained imposing continuity conditions of
functions and their derivatives at $x=0$, namely $\ \left. \ _{+}^{-}\varphi
_{n}\left( x\right) \right\vert _{x-0}=\left. \ _{+}^{-}\varphi _{n}\left(
x\right) \right\vert _{x+0}$ and $\left. \partial _{x}\ _{+}^{-}\varphi
_{n}\left( x\right) \right\vert _{x-0}=\left. \partial _{x}\ _{+}^{-}\varphi
_{n}\left( x\right) \right\vert _{x+0}$. Thus, we obtain:%
\begin{eqnarray}
&&g\left( _{+}|^{-}\right) =2\eta _{\mathrm{L}}e^{i\theta _{+}}e^{-i\pi \chi
/2}\sqrt{\frac{\xi _{1}\left\vert \pi _{0}\left( \mathrm{L}\right) -\chi
\left\vert p^{\mathrm{L}}\right\vert \right\vert \xi _{2}}{\left\vert \pi
_{0}\left( \mathrm{R}\right) +\chi \left\vert p^{\mathrm{R}}\right\vert
\right\vert }}\left( \frac{\left\vert p^{\mathrm{L}}\right\vert \xi _{1}}{%
\left\vert p^{\mathrm{R}}\right\vert \xi _{2}}\right) ^{\chi /2}  \notag \\
&&\times \exp \left[ -\frac{\pi }{2}\left( \nu _{1}^{-}+\nu _{2}^{+}\right) %
\right] \Delta \left( _{+}|^{-}\right) \left( 0\right) \,,  \notag \\
&&\Delta \left( _{+}|^{-}\right) \left( x\right) =\Psi \left(
a_{2},c_{2};z_{2}\right) f_{\mathrm{L}}^{+}\left( x\right) +\Psi \left(
c_{1}-a_{1},c_{1};e^{-i\pi }z_{1}\right) f_{\mathrm{R}}^{-}\left( x\right)
\,,  \label{2.30}
\end{eqnarray}%
and%
\begin{eqnarray}
&&g\left( ^{-}|_{+}\right) =-2\eta _{\mathrm{R}}e^{i\theta _{-}}e^{i\pi \chi
/2}\sqrt{\frac{\xi _{1}\left\vert \pi _{0}\left( \mathrm{R}\right) +\chi
\left\vert p^{\mathrm{R}}\right\vert \right\vert \xi _{2}}{\left\vert \pi
_{0}\left( \mathrm{L}\right) -\chi \left\vert p^{\mathrm{L}}\right\vert
\right\vert }}\left( \frac{\left\vert p^{\mathrm{L}}\right\vert \xi _{1}}{%
\left\vert p^{\mathrm{R}}\right\vert \xi _{2}}\right) ^{\chi /2}  \notag \\
&&\times \exp \left[ \frac{\pi }{2}\left( \nu _{1}^{+}+\nu _{2}^{-}\right) %
\right] \Delta \left( ^{-}|_{+}\right) \left( 0\right) \,,  \notag \\
&&\Delta \left( ^{-}|_{+}\right) \left( x\right) =\Psi \left(
a_{1},c_{1};z_{1}\right) f_{\mathrm{R}}^{+}\left( x\right) +\Psi \left(
c_{2}-a_{2},c_{2};e^{-i\pi }z_{2}\right) f_{\mathrm{L}}^{-}\left( x\right)
\,,  \label{2.31}
\end{eqnarray}%
where%
\begin{eqnarray*}
&&\theta _{\pm }=\pm \left( \left\vert p^{\mathrm{L}}\right\vert \xi
_{1}-\left\vert p^{\mathrm{R}}\right\vert \xi _{2}\right) -eE\xi _{1}^{2}\ln
\left( 2\left\vert p^{\mathrm{L}}\right\vert \xi _{1}\right) +eE\xi
_{2}^{2}\ln \left( 2\left\vert p^{\mathrm{R}}\right\vert \xi _{2}\right) , \\
&&\nu _{1/2}^{\pm }=eE\xi _{1/2}^{2}\left( 1\pm \pi _{0}\left( \mathrm{L/R}%
\right) /\left\vert p^{\mathrm{L/R}}\right\vert \right) ,
\end{eqnarray*}%
such that%
\begin{eqnarray}
f_{\mathrm{L/R}}^{+}\left( x\right) &=&\left\vert p^{\mathrm{L/R}%
}\right\vert \left[ \frac{1}{2}\left( 1+\frac{c_{1/2}}{z_{1/2}}\right) +%
\frac{d}{dz_{1/2}}\right] \Psi \left( c_{1/2}-a_{1/2},c_{1/2};e^{-i\pi
}z_{1/2}\right) \,,  \notag \\
f_{\mathrm{L/R}}^{-}\left( x\right) &=&\left\vert p^{\mathrm{L/R}%
}\right\vert \left[ \frac{1}{2}\left( -1+\frac{c_{1/2}}{z_{1/2}}\right) +%
\frac{d}{dz_{1/2}}\right] \Psi \left( a_{1/2},c_{1/2};z_{1/2}\right) \,.
\label{2.32}
\end{eqnarray}%
One can map $g\left( _{+}|^{-}\right) $ onto its complex conjugate $g\left(
^{-}|_{+}\right) $ exchanging $p_{0}\rightleftarrows -p_{0}$ and $\xi
_{1}\rightleftarrows \xi _{2}$, simultaneously, to realize that $\left\vert
g\left( _{+}|^{-}\right) \right\vert ^{2}$ is an invariant. For example,
employing Kummer transformations $\Psi \left( a,c;z\right) =z^{1-c}\Psi
\left( a-c+1,2-c,z\right) $\ to transformed CHF $\Psi \left(
c_{1}-a_{1},c_{1};e^{-i\pi }z_{1}\right) \rightleftarrows \Psi \left(
1-a_{2},2-c_{2};e^{-i\pi }z_{2}\right) $ and $\Psi \left(
a_{2},c_{2};z_{2}\right) \rightleftarrows \Psi \left(
a_{1}-c_{1}+1,2-c_{1};z_{1}\right) $, one finds that $\Delta \left(
_{+}|^{-}\right) \left( x\right) \rightleftarrows e^{i\pi \left(
1-c_{2}\right) }z_{1}^{c_{1}-1}z_{2}^{c_{2}-1}\Delta \left( ^{-}|_{+}\right)
\left( x\right) $ to conclude that $g\left( _{+}|^{-}\right)
\rightleftarrows g\left( ^{-}|_{+}\right) $ for Fermions. This property
simplify the calculation of differential quantities since one can select a
particular sign of $p_{0}$ to study $g\left( _{+}|^{-}\right) $ and
generalize results to the opposite sign this symmetry, as shall be discussed
in Sec. \ref{Sec3}.

To study arbitrary quantum processes, sometimes it is useful to consider
other $g$-coefficients in addition to the coefficients calculated above. For
example, to study amplitudes of particle scattering, one may find convenient
to use the expression for $g\left( _{+}|^{+}\right) $ rather than the
relation (\ref{2.27}) once $g\left( _{+}|^{-}\right) $ has been calculated.
For such cases, the coefficient $g\left( _{+}|^{+}\right) $ has the form%
\begin{eqnarray}
&&g\left( _{+}|^{+}\right) =-2ie^{i\tilde{\theta}}\eta _{\mathrm{L}}\sqrt{%
\frac{\xi _{1}\left\vert \pi _{0}\left( \mathrm{L}\right) -\chi \left\vert
p^{\mathrm{L}}\right\vert \right\vert \xi _{2}}{\left\vert \pi _{0}\left( 
\mathrm{R}\right) -\chi \left\vert p^{\mathrm{R}}\right\vert \right\vert }}%
\left( \frac{\left\vert p^{\mathrm{L}}\right\vert \xi _{1}}{\left\vert p^{%
\mathrm{R}}\right\vert \xi _{2}}\right) ^{\chi /2}  \notag \\
&&\times \exp \left[ -\frac{\pi }{2}\left( \nu _{1}^{-}-\nu _{2}^{-}\right) %
\right] \Delta \left( _{+}|^{+}\right) \left( 0\right) \,.  \notag \\
&&\Delta \left( _{+}|^{+}\right) \left( x\right) =\Psi \left(
c_{2}-a_{2},c_{2};e^{-i\pi }z_{2}\right) f_{\mathrm{L}}^{+}\left( x\right)
+\Psi \left( c_{1}-a_{1},c_{1};e^{-i\pi }z_{1}\right) f_{\mathrm{R}%
}^{+}\left( x\right) \,,  \label{2.32a}
\end{eqnarray}%
where $\tilde{\theta}=\left\vert p^{\mathrm{L}}\right\vert \xi
_{1}+\left\vert p^{\mathrm{R}}\right\vert \xi _{2}-eE\xi _{1}^{2}\ln \left(
2\left\vert p^{\mathrm{L}}\right\vert \xi _{1}\right) +eE\xi _{2}^{2}\ln
\left( 2\left\vert p^{\mathrm{R}}\right\vert \xi _{2}\right) $. It can be
obtained through the same continuity conditions considered above but applied
to appropriate decompositions between the left and right solutions, similar
those given by Eqs. (\ref{2.30}) and (\ref{2.31}).

With minor modifications, one may extract results from Eqs. (\ref{2.30}) and
(\ref{2.31}) to obtain corresponding expressions for scalar particles. For
example, on account of the inner product of the solutions of the
Klein-Gordon equation on the hyperplane $x$-constant \cite{x-case}, the
orthonormalization conditions are identical to the ones in Eqs. (\ref{2.24})
but with $\eta _{\mathrm{L}}=\eta _{\mathrm{R}}=1$. As a result, relations
between the $g$-coefficients for the scalar case can be extracted from Eqs. (%
\ref{2.26b}) and (\ref{2.27}) setting $\eta _{\mathrm{L}}=\eta _{\mathrm{R}%
}=1$. Moreover, the normalization constants $_{\zeta }\mathcal{N}=\ _{\zeta
}CY$ and $^{\zeta }\mathcal{N}=\ ^{\zeta }CY$ are simpler in this case%
\begin{equation}
\ _{\zeta }C=\frac{\exp \left( -i\pi \kappa _{1}/2\right) }{\sqrt{%
2\left\vert p^{\mathrm{L}}\right\vert }}\,,\ \ ^{\zeta }C=\frac{\exp \left(
-i\pi \kappa _{2}/2\right) }{\sqrt{2\left\vert p^{\mathrm{R}}\right\vert }}%
\,,  \label{2.32d}
\end{equation}%
such that coefficients (\ref{2.30}) and (\ref{2.31}) have the form:%
\begin{eqnarray}
g\left( _{+}|^{-}\right) &=&2\sqrt{\xi _{1}\xi _{2}}\exp \left[ -\frac{\pi }{%
2}\left( \nu _{1}^{-}+\nu _{2}^{+}\right) \right] e^{i\theta _{+}}\left.
\Delta \left( _{+}|^{-}\right) \left( 0\right) \right\vert _{\chi =0}\,, 
\notag \\
g\left( ^{-}|_{+}\right) &=&-2\sqrt{\xi _{1}\xi _{2}}\exp \left[ \frac{\pi }{%
2}\left( \nu _{1}^{+}+\nu _{2}^{-}\right) \right] e^{i\theta _{-}}\left.
\Delta \left( ^{-}|_{+}\right) \left( 0\right) \right\vert _{\chi =0}\,.
\label{2.32c}
\end{eqnarray}%
In contrast to Fermions, one can easily show that $g\left( _{+}|^{-}\right)
\rightleftarrows -g\left( ^{-}|_{+}\right) $ for Bosons, under the exchanges 
$p_{0}\rightleftarrows -p_{0}$ and $\xi _{1}\rightleftarrows \xi _{2}$.
Hence, the absolute square value $\left\vert g\left( _{+}|^{-}\right)
\right\vert ^{2}$ is also invariant for Bosons. Due to the opposite signs
between $g\left( _{+}|^{-}\right) $ and $g\left( ^{-}|_{+}\right) $ under
these exchanges in the Dirac and Klein-Gordon cases, we conveniently
introduce a constant\footnote{%
This constant should not be confused with the parameters $\kappa _{1}$ and $%
\kappa _{2}$ defined in Eqs. (\ref{2.15}).} $\kappa $ to represent the
transformations as follows\footnote{%
By \textquotedblleft Fermi\textquotedblright\ and \textquotedblleft
Bose\textquotedblright\ we mean \textquotedblleft Dirac
particles\textquotedblright\ and \textquotedblleft Klein-Gordon
particles\textquotedblright , respectively. For the sake of convenience, we
employ this terminology everywhere in the text.}:%
\begin{equation}
g\left( _{+}|^{-}\right) \rightleftarrows \kappa g\left( ^{-}|_{+}\right)
\,,\ \ \kappa =\left\{ 
\begin{array}{l}
+1\ \mathrm{Fermi} \\ 
-1\text{\ \textrm{Bose}}%
\end{array}%
\right. \,.  \label{2.32b}
\end{equation}%
Thus, besides the constant $\chi $, the above constant is frequently used to
map results from the Dirac to the Klein-Gordon cases, as we will see below.
The coefficient $g\left( _{+}|^{+}\right) $ for Bosons can be extracted from
Eq. (\ref{2.32a}) setting $\chi =0$, $\eta _{\mathrm{L}}=\eta _{\mathrm{R}%
}=0 $ and, besides, using the normalization constants (\ref{2.32d}) instead
Eq. (\ref{2.24b}). Its representation in terms of Whittaker functions can be
found in Appendix \ref{App1}; cf. Eq. (\ref{app1.0b}).

\subsection{In\ and out-states\label{Sec2.2}}

In contrast to time-dependent electric backgrounds\footnote{%
As it is well known, for time-dependent electric backgrounds (more
precisely, $t$-electric potential steps) the quantization of
Dirac/Klein-Gordon fields is performed using exact solutions describing
particle and antiparticle states at asymptotic times; see e.g., Refs. \cite%
{GMR85,Gitman,AdoGavGit17}.}, a quantization of Dirac and Klein-Gordon
fields with $x$-electric potential steps is performed with the help of
solutions describing particles moving to the steps from infinitely remote
distances or leaving the step to infinitely remote distances. In-solutions
are defined as incoming waves (that is, waves going toward the step) while
out-solutions are classified as outgoing waves (that is, waves going
outwards the step). Since there are five distinct ranges of quantum numbers $%
\Omega _{k}$, definitions of in- or out-sets are different. For some of the
ranges, these definitions are similar to the one-particle relativistic
quantum theory. In the case under consideration, the classification is the
following\footnote{%
Similar classification holds for the scalar case, but in $\Omega _{3}$, $%
_{+}\psi _{n_{3}}\,,\ ^{+}\psi _{n_{3}}$ are in-solutions whereas $_{-}\psi
_{n_{3}}\,,\ ^{-}\psi _{n_{3}}$ out-solutions.}:%
\begin{equation}
\begin{array}{ll}
\text{\textrm{in-solutions:}} & \ _{+}\psi _{n_{1}}\,,\ ^{-}\psi
_{n_{1}}\,;\ _{-}\psi _{n_{5}}\,,\ ^{+}\psi _{n_{5}}\,;\ _{-}\psi
_{n_{3}}\,,\ ^{-}\psi _{n_{3}}\,, \\ 
\text{\textrm{out-solutions:}} & \ _{-}\psi _{n_{1}}\,,\ ^{+}\psi
_{n_{1}}\,;\ _{+}\psi _{n_{5}}\,,\ ^{-}\psi _{n_{5}}\,;\ _{+}\psi
_{n_{3}}\,,\ ^{+}\psi _{n_{3}}\,.%
\end{array}
\label{2.33}
\end{equation}%
The sets $\left\{ \ _{+}\psi _{n_{1}}\,,\ ^{-}\psi _{n_{1}}\right\} $ and $%
\left\{ \ _{-}\psi _{n_{1}}\,,\ ^{+}\psi _{n_{1}}\right\} $ describe
incoming and outgoing electron states respectively, while $\left\{ \
_{-}\psi _{n_{5}}\,,\ ^{+}\psi _{n_{5}}\right\} $ and $\left\{ \ _{-}\psi
_{n_{5}}\,,\ ^{+}\psi _{n_{5}}\right\} $ describes incoming and outgoing
positron states respectively.

One can demonstrate that the sets of solutions are complete and orthogonal
with respect to the inner product on the $t$-constant hyperplane%
\begin{equation}
\left( \psi _{n},\psi _{n^{\prime }}^{\prime }\right) =\int_{V_{\perp }}d%
\mathbf{r}_{\perp }\int_{-K^{\left( \mathrm{L}\right) }}^{K^{\left( \mathrm{R%
}\right) }}dx\psi _{n}^{\dagger }\left( X\right) \psi _{n^{\prime }}^{\prime
}\left( X\right) \,,\ \ V_{\perp }=\prod_{j=2}^{D}K_{j}\,,  \label{2.34}
\end{equation}%
where the lower/upper cutoffs $K^{\left( \mathrm{L/R}\right) }$ are supposed
to admit the limits $K^{\left( \mathrm{L}\right) }\sim T$ and $K^{\left( 
\mathrm{R}\right) }\sim T$ (and $T\rightarrow \infty $) in final
expressions; see Ref. \cite{GavGit19} for details. In particular,%
\begin{eqnarray}
&&\left( \ _{\zeta }\psi _{n},\ _{\zeta }\psi _{n^{\prime }}\right) =\left(
\ ^{\zeta }\psi _{n},\ ^{\zeta }\psi _{n^{\prime }}\right) =\delta
_{n,n^{\prime }}\mathcal{M}_{n}\,,\ \ n,n^{\prime }\in \Omega _{1}\cup
\Omega _{3}\cup \Omega _{5}\,,  \notag \\
&&\left( \ \psi _{n},\ \psi _{n^{\prime }}\right) =\delta _{n,n^{\prime
}}\,,\ \ n,n^{\prime }\in \Omega _{2}\cup \Omega _{4}\,,  \notag \\
&&\left( \ _{\zeta }\psi _{n},\ ^{-\zeta }\psi _{n}\right) =0\,,\ \ n\in
\Omega _{1}\cup \Omega _{5}\,,\ \ \left( \ _{\zeta }\psi _{n},\ ^{\zeta
}\psi _{n}\right) =0\,,\ \ n\in \Omega _{3}\,,  \notag \\
&&\mathcal{M}_{n}=\left\vert g\left( _{+}|^{+}\right) \right\vert ^{2},\ \
n\in \Omega _{1}\cup \Omega _{5}\,,\ \ \mathcal{M}_{n}=\left\vert g\left(
_{+}|^{-}\right) \right\vert ^{2},\ \ n\in \Omega _{3}\,,  \label{2.34b}
\end{eqnarray}%
where we have $\delta _{n,n^{\prime }}=\delta _{\sigma ,\sigma ^{\prime
}}\delta \left( p_{0}-p_{0}^{\prime }\right) \delta \left( \mathbf{p}_{\perp
}-\mathbf{p}_{\perp }^{\prime }\right) $ in the limit $K^{\left( \mathrm{L/R}%
\right) }\rightarrow \infty $. For each set of quantum numbers $n$ there
exist one or two complete sets of solutions:

\begin{enumerate}
\item[(a)] For $\forall \,n\in \Omega _{1}\cup \Omega _{5}$, there are two $%
\left( \zeta =\pm \right) $ independent sets of solutions: $\left\{ \
_{\zeta }\psi _{n}\left( X\right) \,,\ ^{-\zeta }\psi _{n}\left( X\right)
\right\} $;

\item[(b)] For $\forall \,n\in \Omega _{3}$, there are two $\left( \zeta
=\pm \right) $ independent sets of solutions: $\left\{ \ _{\zeta }\psi
_{n}\left( X\right) \,,\ ^{\zeta }\psi _{n}\left( X\right) \right\} $;

\item[(c)] For $\forall \,n\in \Omega _{2}\cup \Omega _{4}$, there is one
set of solutions $\left\{ \psi _{n}\left( X\right) \right\} $.
\end{enumerate}

The classification of solutions (\ref{2.33}), together with the above
properties, allows us to quantize the Dirac and Klein-Gordon fields in terms
of particles and antiparticles. To quantize the Dirac field operator $\hat{%
\Psi}\left( X\right) $, we decompose it using the sets of solutions
discussed above on the hyperplane $t=\mathrm{const.}$, in which the $x$%
-independent decomposition coefficients are creation and annihilation
operators of particles or antiparticles. Because there are two independent
sets of solution for states within $\Omega _{1}\cup \Omega _{3}\cup \Omega
_{5}$, two possible quantizations exist, one formed exclusively with in\
operators and another formed exclusively with out operators, namely%
\begin{equation}
\begin{array}{ll}
\text{\textrm{in-set:}} & \ _{+}a_{n_{1}}\left( \mathrm{in}\right) \,,\
^{-}a_{n_{1}}\left( \mathrm{in}\right) \,;\ _{-}b_{n_{5}}\left( \mathrm{in}%
\right) \,,\ ^{+}b_{n_{5}}\left( \mathrm{in}\right) \,;\ _{-}b_{n_{3}}\left( 
\mathrm{in}\right) \,,\ ^{-}a_{n_{3}}\left( \mathrm{in}\right) \,, \\ 
\text{\textrm{out-set:}} & \ _{-}a_{n_{1}}\left( \mathrm{out}\right) \,,\
^{+}a_{n_{1}}\left( \mathrm{out}\right) \,;\ _{+}b_{n_{5}}\left( \mathrm{out}%
\right) \,,\ ^{-}b_{n_{5}}\left( \mathrm{out}\right) \,;\
_{+}b_{n_{3}}\left( \mathrm{out}\right) \,,\ ^{+}a_{n_{3}}\left( \mathrm{out}%
\right) \,.%
\end{array}
\label{2.35}
\end{equation}%
For states in $\Omega _{2}$, we have only pairs of creation/annihilation
operators of particles $\left\{ a_{n_{2}}^{\dagger },a_{n_{2}}\right\} $\
whereas for states in $\Omega _{4}$ we have only pairs of
creation/annihilation operators of antiparticles $\left\{ b_{n_{4}}^{\dagger
},b_{n_{4}}\right\} $. All $a$'s and $b$'s are interpreted as annihilation
operators of particles and antiparticles, respectively; their adjoints, $%
a^{\dagger }$'s and $b^{\dagger }$'s, are interpreted as creation operators
of particles and antiparticles, respectively. Operators labeled by the
argument \textrm{in} are in-operators while the ones labeled by the argument 
\textrm{out} are out-operators. All creation and annihilation operators with
different quantum numbers or from different ranges $\Omega _{i}$ anticommute
between themselves. For example, the only nontrivial anticommutation
relations for in- operators are:%
\begin{eqnarray}
&&\left[ \ _{+}a_{n_{1}^{\prime }}\left( \mathrm{in}\right) ,\
_{+}a_{n_{1}}^{\dagger }\left( \mathrm{in}\right) \right] _{+}=\left[ \
^{-}a_{n_{1}^{\prime }}\left( \mathrm{in}\right) ,\ ^{-}a_{n_{1}}^{\dagger
}\left( \mathrm{in}\right) \right] _{+}=\delta _{n_{1}^{\prime }n_{1}}\,, 
\notag \\
&&\left[ \ ^{-}a_{n_{3}^{\prime }}\left( \mathrm{in}\right) ,\
^{-}a_{n_{3}}^{\dagger }\left( \mathrm{in}\right) \right] _{+}=\left[ \
_{-}b_{n_{3}^{\prime }}\left( \mathrm{in}\right) ,\ _{-}b_{n_{3}}^{\dagger
}\left( \mathrm{in}\right) \right] _{+}=\delta _{n_{3}^{\prime }n_{3}}\,, 
\notag \\
&&\left[ \ _{-}b_{n_{5}^{\prime }}\left( \mathrm{in}\right) ,\
_{-}b_{n_{5}}^{\dagger }\left( \mathrm{in}\right) \right] _{+}=\left[ \
^{+}b_{n_{5}^{\prime }}\left( \mathrm{in}\right) ,\ ^{+}b_{n_{5}}^{\dagger
}\left( \mathrm{in}\right) \right] _{+}=\delta _{n_{5}^{\prime }n_{5}}\,, 
\notag \\
&&\left[ a_{n_{2}^{\prime }},a_{n_{2}}^{\dagger }\right] _{+}=\delta
_{n_{2}^{\prime }n_{2}}\,,\ \ \left[ b_{n_{4}^{\prime }},b_{n_{4}}^{\dagger }%
\right] _{+}=\delta _{n_{4}^{\prime }n_{4}}\,.  \label{2.36}
\end{eqnarray}%
Furthermore, the in-vacuum state $\left\vert 0,\mathrm{in}\right\rangle $,%
\begin{equation}
\left\vert 0,\mathrm{in}\right\rangle =\prod_{i=1,3,5}\otimes \left\vert 0,%
\mathrm{in}\right\rangle ^{\left( i\right) }\otimes \left\vert
0\right\rangle ^{\left( 2\right) }\otimes \left\vert 0\right\rangle ^{\left(
4\right) }\,,  \label{2.37}
\end{equation}%
is defined as the direct product of partial in-vacuum states $\left\vert 0,%
\mathrm{in}\right\rangle ^{\left( i\right) }\ $states; all vacua annihilated
by corresponding annihilation operators%
\begin{eqnarray}
&&\ _{+}^{-}a_{n_{1}}\left( \mathrm{in}\right) \left\vert 0,\mathrm{in}%
\right\rangle ^{\left( 1\right) }=\ _{+}^{-}a_{n_{1}}\left( \mathrm{in}%
\right) \left\vert 0,\mathrm{in}\right\rangle =0\,,  \notag \\
&&\ ^{-}a_{n_{3}}\left( \mathrm{in}\right) \left\vert 0,\mathrm{in}%
\right\rangle ^{\left( 3\right) }=\ ^{-}a_{n_{3}}\left( \mathrm{in}\right)
\left\vert 0,\mathrm{in}\right\rangle =0\,,  \notag \\
&&\ _{-}b_{n_{3}}\left( \mathrm{in}\right) \left\vert 0,\mathrm{in}%
\right\rangle ^{\left( 3\right) }=\ _{-}b_{n_{3}}\left( \mathrm{in}\right)
\left\vert 0,\mathrm{in}\right\rangle =0\,,  \notag \\
&&\ _{-}^{+}b_{n_{5}}\left( \mathrm{in}\right) \left\vert 0,\mathrm{in}%
\right\rangle ^{\left( 5\right) }=\ _{-}^{+}b_{n_{5}}\left( \mathrm{in}%
\right) \left\vert 0,\mathrm{in}\right\rangle =0\,,  \notag \\
&&\ a_{n_{2}}\left\vert 0\right\rangle ^{\left( 2\right)
}=a_{n_{2}}\left\vert 0,\mathrm{in}\right\rangle =0\,,\ \
b_{n_{4}}\left\vert 0\right\rangle ^{\left( 4\right) }=b_{n_{4}}\left\vert 0,%
\mathrm{in}\right\rangle =0\,.  \label{2.38}
\end{eqnarray}%
Anticommutation relations for out-operators and out-vacuum states $%
\left\vert 0,\mathrm{out}\right\rangle ^{\left( i\right) }$ can be
introduced following the same considerations above.

Due to the quantization of the Dirac/Klein-Gordon fields and the canonical
transformations between in and out sets of creation and annihilation
operators in the ranges $\Omega _{1}$, $\Omega _{2}$, $\Omega _{4}$ and $%
\Omega _{5}$ \cite{x-case}, each partial in-vacuum $\left\vert 0,\mathrm{in}%
\right\rangle ^{\left( i\right) },\ i=1,2,4,5$ differs from its
corresponding out-vacuum $\left\vert 0,\mathrm{out}\right\rangle ^{\left(
i\right) },\ i=1,2,4,5$ by a complex phase which, without loss of
generality, can be selected to match one another, namely $\left\vert 0,%
\mathrm{in}\right\rangle ^{\left( i\right) }=\left\vert 0,\mathrm{out}%
\right\rangle ^{\left( i\right) }\,,\ \ i=1,2,4,5$. Therefore, we
conveniently represent the direct product of all partial vacua by%
\begin{equation*}
\left\vert 0\right\rangle =\prod_{i=1,2,4,5}\otimes \left\vert 0,\mathrm{in}%
\right\rangle ^{\left( i\right) }=\prod_{i=1,2,4,5}\otimes \left\vert 0,%
\mathrm{out}\right\rangle ^{\left( i\right) }\,.
\end{equation*}

This is not the case of the partial vacua $\left\vert 0,\mathrm{in}%
\right\rangle ^{\left( 3\right) }$ and $\left\vert 0,\mathrm{out}%
\right\rangle ^{\left( 3\right) }$ in the range $\Omega _{3}$ (Klein zone),
which are different by reasons that shall be briefly discussed below. That
is why the total vacuum-vacuum transition amplitude $c_{v}$%
\begin{equation}
c_{v}=\left\langle 0,\mathrm{out}|0,\mathrm{in}\right\rangle =c_{v}^{\left(
3\right) }=\ ^{\left( 3\right) }\left\langle 0,\mathrm{out}|0,\mathrm{in}%
\right\rangle ^{\left( 3\right) }\,,  \label{2.39}
\end{equation}%
coincides with the vacuum-vacuum transition amplitude in the Klein zone $%
c_{v}^{\left( 3\right) }$.

Notice that in the scalar case, all anticommutation relations must be
replaced by commutation relations and definitions of in- or
out-creation/annihilation operators in $\Omega _{3}$ are different, namely$\
^{+}a_{n_{3}}\left( \mathrm{in}\right) $ and $_{+}b_{n_{3}}\left( \mathrm{in}%
\right) $, are in-operators, while$\ ^{-}a_{n_{3}}\left( \mathrm{out}\right) 
$ and$\ _{-}b_{n_{3}}\left( \mathrm{out}\right) $ out-operators.

\section{Processes in the Klein zone\label{Sec3}}

Particle creation from the vacuum occurs exclusively in the Klein zone (the
range $\Omega _{3}$), defined by bounded energies $p_{0}$%
\begin{equation}
U_{\mathrm{L}}+\pi _{\perp }\leq p_{0}\leq U_{\mathrm{R}}-\pi _{\perp }\,,
\label{3.0}
\end{equation}%
and quantum numbers $\mathbf{p}_{\perp },\sigma $ that may be arbitrary. We
recapitulate below the calculation of main quantities characterizing the
vacuum instability as well as elementary processes occurring in the Klein
zone.

According to the partial decompositions of the quantized Dirac field $\hat{%
\Psi}\left( X\right) $ in the range $\Omega _{3}$ \cite{x-case},%
\begin{eqnarray}
\hat{\Psi}_{3}\left( X\right) &=&\sum_{n\in \Omega _{3}}\mathcal{M}%
_{n}^{-1/2}\left[ \ ^{-}a_{n}\left( \mathrm{in}\right) \ ^{-}\psi _{n}\left(
X\right) +\ _{-}b_{n}^{\dagger }\left( \mathrm{in}\right) \ _{-}\psi
_{n}\left( X\right) \right] \,,  \notag \\
&=&\sum_{n\in \Omega _{3}}\mathcal{M}_{n}^{-1/2}\left[ \ ^{+}a_{n}\left( 
\mathrm{out}\right) \ ^{+}\psi _{n}\left( X\right) +\ _{+}b_{n}^{\dagger
}\left( \mathrm{out}\right) \ _{+}\psi _{n}\left( X\right) \right] \,,
\label{3.1}
\end{eqnarray}%
and the relations (\ref{2.25}), specialized to $\Omega _{3}$ (where $\eta _{%
\mathrm{L}}=1=-\eta _{\mathrm{R}}$), one may express the independent set $%
\left\{ \ ^{+}\psi _{n_{3}}\left( X\right) ,\ _{+}\psi _{n_{3}}\left(
X\right) \right\} $ in terms of the independent set $\left\{ \ ^{-}\psi
_{n_{3}}\left( X\right) ,\ _{-}\psi _{n_{3}}\left( X\right) \right\} $ (or 
\textit{vice-versa}) and use inner products on $t$-constant hyperplane (\ref%
{2.34b}) to establish linear canonical transformations between in- and
out-operators in $\Omega _{3}$. These transformations are specified by Eqs.
(7.4) in \cite{x-case}. With the aid of these transformations we may
introduce, for instance, the differential mean number of in-particles $%
N_{n_{3}}^{a}\left( \mathrm{in}\right) $ created from the out-vacuum 
\begin{equation}
N_{n}^{a}\left( \mathrm{in}\right) =\left\langle 0,\mathrm{out}\left\vert \
^{-}a_{n}^{\dagger }\left( \mathrm{in}\right) \ ^{-}a_{n}\left( \mathrm{in}%
\right) \right\vert 0,\mathrm{out}\right\rangle =\left\vert g\left(
_{+}|^{-}\right) \right\vert ^{-2}\,,\ \ n\in \Omega _{3}\,,  \label{3.3}
\end{equation}%
and use the identity $\left\vert g\left( _{-}|^{+}\right) \right\vert
^{2}=\left\vert g\left( _{+}|^{-}\right) \right\vert ^{2}$ to realize that
Eq. (\ref{3.3}) is identically equal to the differential mean number of
in-antiparticles created from the out-vacuum $N_{n_{3}}^{b}\left( \mathrm{in}%
\right) $, the differential mean number of out-particles $%
N_{n_{3}}^{a}\left( \mathrm{out}\right) $ and out-antiparticles $%
N_{n_{3}}^{b}\left( \mathrm{out}\right) $ created from the in-vacuum. That
is why we denote all these quantities by $N_{n}^{\mathrm{cr}}\equiv
N_{n}^{a}\left( \mathrm{in}\right) =N_{n}^{b}\left( \mathrm{in}\right)
=N_{n}^{a}\left( \mathrm{out}\right) =N_{n}^{b}\left( \mathrm{out}\right) $.
Thus, the total number of particles created $N^{\mathrm{cr}}$ correspond to
the summation of the mean numbers $N_{n}^{\mathrm{cr}}$ over the quantum
numbers within $\Omega _{3}$%
\begin{equation}
N^{\mathrm{cr}}=\sum_{n\in \Omega _{3}}N_{n}^{\mathrm{cr}}=\frac{J_{\left(
d\right) }V_{\perp }T}{\left( 2\pi \right) ^{d-1}}\int_{\Omega _{3}}dp_{0}d%
\mathbf{p}_{\perp }N_{n}^{\mathrm{cr}}\,,  \label{3.4}
\end{equation}%
where $V_{\perp }$ is the space volume perpendicular to the direction of the
electric field, $T$ its time duration and $J_{\left( d\right) }=2^{\left[ d/2%
\right] -1}$ the total number of spinning degrees of freedom ($J_{\left(
d\right) }=1$ for scalar particles), that factorizes out as a multiplicative
constant since the field does not mix different spin polarizations. In the
rightmost equality, the summations were converted into multiple integrals in
the standard way, viz. $\left( V_{\perp }T\right) ^{-1}\sum {}_{p_{0},%
\mathbf{p}_{\perp }\in \Omega _{3}}\leftrightarrow \left( 2\pi \right)
^{1-d}\int dp_{0}d\mathbf{p}_{\perp }$, in which $V_{\perp }$, $T$ are
macroscopically large.

Besides particle creation, there are other elementary processes in the Klein
zone worth of consideration, such as particle scattering, creation of a
particle-antiparticle pair and annihilation of a particle-antiparticle pair.
The relative (with respect to the in-vacuum $\left\vert 0,\mathrm{in}%
\right\rangle $ and the out-vacuum $\left\vert 0,\mathrm{out}\right\rangle $%
) amplitudes of particle scattering $w_{n}\left( +|+\right) $, antiparticle
scattering $w_{n}\left( -|-\right) $, particle-antiparticle creation $%
w_{n}\left( +-|0\right) $ and particle-antiparticle annihilation $%
w_{n}\left( 0|-+\right) $ are defined by Eqs. (7.17) and (A9) in \cite%
{x-case}. In terms of $g$-coefficients, the corresponding probabilities can
be expressed as follows:%
\begin{eqnarray}
&&\left\vert w_{n}\left( +|+\right) \right\vert ^{2}=\left\vert w_{n}\left(
-|-\right) \right\vert ^{2}=\left\vert g\left( _{+}|^{-}\right) \right\vert
^{2}\left\vert g\left( _{+}|^{+}\right) \right\vert ^{-2}=\frac{1}{1-\kappa
N_{n}^{\mathrm{cr}}}\,,  \notag \\
&&\left\vert w_{n}\left( +-|0\right) \right\vert ^{2}=\left\vert w_{n}\left(
0|-+\right) \right\vert ^{2}=\left\vert g\left( _{+}|^{+}\right) \right\vert
^{-2}=\frac{N_{n}^{\mathrm{cr}}}{1-\kappa N_{n}^{\mathrm{cr}}}\,.
\label{3.4b}
\end{eqnarray}%
It is noteworthy that total reflection, which is a direct consequence of the
quantization of the Dirac and Klein-Gordon fields in $\Omega _{3}$ (\ref{3.1}%
), \cite{x-case}, is the only possible form of particle scattering in $%
\Omega _{3}$. Particle reflection and particle transmission are allowed
beyond the Klein zone, as shall be discussed in Sec. \ref{Sec4}.

The most important quantity characterizing the vacuum instability is the
vacuum-vacuum transition probability $P_{v}$%
\begin{equation}
P_{v}=\left\vert c_{v}\right\vert ^{2}=\left\vert \left\langle 0,\mathrm{out}%
\left\vert V_{\Omega _{3}}\right\vert 0,\mathrm{out}\right\rangle
\right\vert ^{2}=\left\vert \left\langle 0,\mathrm{in}\left\vert V_{\Omega
_{3}}\right\vert 0,\mathrm{in}\right\rangle \right\vert ^{2}\,,  \label{3.5}
\end{equation}%
because $P_{v}\neq 1$ $\left( P_{v}<1\right) $\ indicates that pairs were
created from the vacuum by the external field. Here, $c_{v}$ denotes the
vacuum-vacuum transition amplitude and $V_{\Omega _{3}}$ an unitary operator
connecting the in and out vacua $\left\vert 0,\mathrm{out}\right\rangle
=V_{\Omega _{3}}^{\dagger }\left\vert 0,\mathrm{in}\right\rangle $%
\begin{eqnarray}
V_{\Omega _{3}} &=&\exp \left[ -\ ^{-}a_{n}^{\dagger }\left( \mathrm{in}%
\right) w_{n}\left( +-|0\right) \ _{-}b_{n}^{\dagger }\left( \mathrm{in}%
\right) \right] \exp \left[ -\ ^{-}a_{n}\left( \mathrm{in}\right) \ln
w_{n}\left( +|+\right) \ ^{-}a_{n}^{\dagger }\left( \mathrm{in}\right) %
\right]  \notag \\
&\times &\exp \left[ \ _{-}b_{n}^{\dagger }\left( \mathrm{in}\right) \ln
w_{n}\left( -|-\right) \ _{-}b_{n}\left( \mathrm{in}\right) \right] \exp %
\left[ -\ _{-}b_{n}\left( \mathrm{in}\right) w_{n}\left( 0|-+\right) \
^{-}a_{n}\left( \mathrm{in}\right) \right] \,,  \label{3.6}
\end{eqnarray}%
which, in turn, defines a transformation between in and out operators%
\begin{equation}
\left\{ \ ^{+}a_{n}^{\dagger }\left( \mathrm{out}\right) ,\ ^{+}a_{n}\left( 
\mathrm{out}\right) ,\ _{+}b_{n}^{\dagger }\left( \mathrm{out}\right) ,\
_{+}b_{n}\left( \mathrm{out}\right) \right\} =V_{\Omega _{3}}^{\dagger
}\left\{ \ ^{-}a_{n}^{\dagger }\left( \mathrm{in}\right) ,\ ^{-}a_{n}\left( 
\mathrm{in}\right) ,\ _{-}b_{n}^{\dagger }\left( \mathrm{in}\right) ,\
_{-}b_{n}\left( \mathrm{in}\right) \right\} V_{\Omega _{3}}\,.  \label{3.7}
\end{equation}%
The representation of $V_{\Omega _{3}}$ in terms of in-operators (\ref{3.6})
is complementary to the one given in terms of out-operators; cf. Eq. (7.20)
in Ref. \cite{x-case}. Details on the calculation of $V_{\Omega _{3}}$ for
Bosons can be found in Appendix \ref{App2}.

Using the explicit representations (\ref{3.6}) and (\ref{app2.17}), the
probability of a vacuum remains vacuum $P_{v}$ reads%
\begin{equation}
P_{v}=\exp \left( \sum_{n\in \Omega _{3}}\ln p_{v}^{n}\right) \,,\ \
p_{v}^{n}=\left( 1-\kappa N_{n}^{\mathrm{cr}}\right) ^{\kappa }\,.
\label{3.8}
\end{equation}

One of the consequences of particle creation occurring exclusively within
the Klein zone is the diminishing of $N_{n}^{\mathrm{cr}}$ near the
boundaries of $\Omega _{3}$. This is a local property resulting from the
behavior of the asymptotic momenta $\left\vert p^{\mathrm{L}}\right\vert $, $%
\left\vert p^{\mathrm{R}}\right\vert $ at the boundaries of the Klein zone,
namely $\left\vert p^{\mathrm{R}}\right\vert \rightarrow 0$ at the vicinity
of $\Omega _{2}$ while $\left\vert p^{\mathrm{L}}\right\vert \rightarrow 0$
at the vicinity of $\Omega _{4}$. Accordingly, the coefficient $g\left(
_{+}|^{-}\right) $ (or $g\left( _{-}|^{+}\right) $) diverges, which means
that $N_{n}^{\mathrm{cr}}\rightarrow 0$ near both boundaries. This is more
clearly seen if one expresses $g\left( _{+}|^{-}\right) $ in terms of
Whittaker functions, whose expressions are given by Eqs. (\ref{app1.0a}) and
(\ref{app1.0b}) in Appendix \ref{App1}, because $\left\vert g\left(
_{+}|^{-}\right) \right\vert ^{-2}=O\left( \left\vert p^{\mathrm{R}%
}\right\vert \left\vert p^{\mathrm{L}}\right\vert \right) \rightarrow 0$
near the boundaries. This is also true for the Klein-Gordon case. In the
next subsection \ref{Sec3.1}, we study this property in the most favorable
configuration for particle creation, wherein the differential mean numbers $%
N_{n}^{\mathrm{cr}}$ are not necessarily small over a sufficiently wide
range of quantum numbers in the Klein zone.

\subsection{Small-gradient configuration\label{Sec3.1}}

Here, we study the case when a strong electric field is concentrated in a
wide region on the $x$-direction with a sufficiently strong amplitude $E$
and sufficiently large length scales $\xi _{j}$ so that the parameters $%
\left\vert U_{\mathrm{L}}\right\vert \xi _{1}$ and $U_{\mathrm{R}}\xi _{2}$
are both large, satisfying the conditions%
\begin{equation}
\min \left( \left\vert U_{\mathrm{L}}\right\vert \xi _{1},U_{\mathrm{R}}\xi
_{2}\right) \gg \max \left( 1,\frac{m^{2}}{eE}\right) \,,\ \ \frac{\xi _{2}}{%
\xi _{1}}=\text{\textrm{fixed}}\,.  \label{31.1}
\end{equation}%
This configuration represents an almost symmetrical electric field that is
growing \textquotedblleft smoothly\textquotedblright\ from $x=-\infty $ to $%
x=0$ and then is decreasing \textquotedblleft smoothly\textquotedblright\ to 
$x=+\infty $. The configuration may be considered as a two-parameter
regularization of a uniform electric field\footnote{%
Because the lenght scales $\xi _{j}$ are large enough, variations of the
derivative $E^{\prime }\left( x\right) $ in a neighborhood of any point $x$
is small enough to treat the field as constant, for instance $\delta
E^{\prime }\left( x\right) =E^{\prime }\left( x+\delta x\right) -E^{\prime
}\left( x\right) =\left( 2E/\xi _{2}\right) \left[ 3\left( 1+x/\xi
_{2}\right) ^{-4}\left( \delta x/\xi _{2}\right) +O\left( \left( \delta
x/\xi _{2}\right) ^{2}\right) \right] \ll 1$ since $\left( \delta x/\xi
_{2}\right) \rightarrow 0$. The combination of small-gradient behavior with
strong amplitudes $E$ is commonly referred in literature by locally constant
field approximation (LCFA).}. The resulting potential energy of the electron
in this field is illustrated by the yellow solid line in Fig. \ref{Fig1}.

To study quantities characterizing the vacuum instability, one has to
compare the above parameters with parameters involving other quantum
numbers. Since particle creation is directly related to the extent of the
Klein zone, which is parameterized by the asymptotic potential energies $U_{%
\mathrm{L,R}}$ and the perpendicular energies $\pi _{\perp }$, we set an
upper bound to the perpendicular momenta $\mathbf{p}_{\perp }$ in order to
consider a sufficiently wide Klein zone, say $\pi _{\perp }\leq K_{\perp }$,
where $K_{\perp }$ is a number satisfying the inequality $\min \left( eE\xi
_{1}^{2},eE\xi _{2}^{2}\right) \gg K_{\perp }>\max \left( 1,m^{2}/eE\right) $%
. As for the parameter $p_{0}/\sqrt{eE}$, we restrict the consideration to
positive energies $0\leq p_{0}\leq U_{\mathrm{R}}-\pi _{\perp }$ and
generalize results to negative energies using the properties of the
coefficient $g\left( _{+}|^{-}\right) $ (and, therefore, its absolute square
value (\ref{3.3})) discussed at the end of Sec. \ref{Sec2.1}. In this case,
the left kinetic energy is always large and positive $\left\vert U_{\mathrm{L%
}}\right\vert \leq \pi _{0}\left( \mathrm{L}\right) \leq \mathbb{U}-\pi
_{\perp }$ , while the right kinetic energy is always negative, $\pi _{\perp
}\leq \left\vert \pi _{0}\left( \mathrm{R}\right) \right\vert \leq U_{%
\mathrm{R}}$. Within this range, the differential mean numbers $N_{n}^{%
\mathrm{cr}}$\ are significant only in a subrange $\beta \sqrt{\lambda }%
<\left\vert \pi _{0}\left( \mathrm{R}\right) \right\vert /\sqrt{eE}\leq U_{%
\mathrm{R}}/\sqrt{eE}$, which can be divided as follows:%
\begin{eqnarray}
&&\left( \mathrm{a}\right) \ \frac{U_{\mathrm{R}}}{\sqrt{eE}}-\frac{\delta }{%
\sqrt{2}}\leq \frac{\left\vert \pi _{0}\left( \mathrm{R}\right) \right\vert 
}{\sqrt{eE}}\leq \frac{U_{\mathrm{R}}}{\sqrt{eE}}\,,  \notag \\
&&\left( \mathrm{b}\right) \ \sqrt{\lambda }+\frac{U_{\mathrm{R}}}{\sqrt{eE}}%
\left( 1-\Upsilon \right) <\frac{\left\vert \pi _{0}\left( \mathrm{R}\right)
\right\vert }{\sqrt{eE}}<\frac{U_{\mathrm{R}}}{\sqrt{eE}}-\frac{\delta }{%
\sqrt{2}}\,,  \notag \\
&&\left( \mathrm{c}\right) \ \beta \sqrt{\lambda }<\frac{\left\vert \pi
_{0}\left( \mathrm{R}\right) \right\vert }{\sqrt{eE}}\leq \sqrt{\lambda }+%
\frac{U_{\mathrm{R}}}{\sqrt{eE}}\left( 1-\Upsilon \right) \,,  \label{31.2}
\end{eqnarray}%
where $\delta $ is a sufficiently small number $0<\delta \ll 1$, $\Upsilon $
is a fixed number $\pi _{\perp }/U_{\mathrm{R}}<\Upsilon <1,$ and $\beta $
is slightly larger than the unity, $1<\beta \ll 1+\left( 1-\Upsilon \right)
U_{\mathrm{R}}/\pi _{\perp }$. To study local properties of the mean numbers 
$N_{n}^{\mathrm{cr}}$, we introduce two new sets of variables%
\begin{equation}
\eta _{1}=\frac{e^{-i\pi }z_{1}\left( 0\right) }{c_{1}}\,,\ \ \eta _{2}=%
\frac{z_{2}\left( 0\right) }{c_{2}}\,,\ \ \mathcal{Z}_{j}=\left( \eta
_{j}-1\right) \mathcal{W}_{j}\sqrt{c_{j}}\,,  \label{31.3}
\end{equation}%
where $\mathcal{W}_{j}=\left\vert \eta _{j}-1\right\vert ^{-1}\sqrt{2\left(
\eta _{j}-1-\ln \eta _{j}\right) }$, \textrm{sgn}$\mathcal{Z}_{j}=\mathrm{sgn%
}\left( \eta _{j}-1\right) $, and take into account that $\pi _{0}\left( 
\mathrm{L}\right) /\sqrt{eE}$ is large and positive for positive energies,
which means that $c_{1}-a_{1}$ is fixed while $z_{1}$ and $c_{1}$ are
simultaneously large throughout the subranges $\left( \mathrm{a}\right)
-\left( \mathrm{c}\right) $.

The subrange $\left( \mathrm{a}\right) $ is characterized by sufficiently
small energies ($p_{0}/\sqrt{eE}$ small) and $\eta _{1}$ and $\eta _{2}$ are
sufficiently close to the unity%
\begin{equation}
\left( \mathrm{a}\right) \ 1>\eta _{2}\geq 1-\frac{\delta }{\sqrt{2eE}\xi
_{2}}\,,\ \ 1<\eta _{1}\leq 1+\frac{\delta }{\sqrt{2eE}\xi _{1}}\,,
\label{31.4}
\end{equation}%
such that $\mathcal{Z}_{1}$ and $\mathcal{Z}_{2}$ are small in this range, $%
\left\vert \mathcal{Z}_{1}\right\vert \gtrsim \left\vert \mathcal{Z}%
_{2}\right\vert $, $\left\vert \mathcal{Z}_{2}\right\vert \approx \delta $.
Thus, one can use an asymptotic approximation given by Eq. (66) in \cite%
{AdoGavGit18} and the approximations $\nu _{1}^{-}=-\left( \lambda /2\right) %
\left[ 1+O\left( \sqrt{eE}\xi _{1}\right) ^{-1}\right] $ and $\nu
_{2}^{+}=-\left( \lambda /2\right) \left[ 1+O\left( \sqrt{eE}\xi _{2}\right)
^{-1}\right] $ to show that the mean numbers asymptotically, in the
leading-order approximation, are given by the equation%
\begin{equation}
N_{n}^{\mathrm{cr}}\approx e^{-\pi \lambda }\,.  \label{31.5}
\end{equation}%
This local approximation, holds both for Fermions and for Bosons, and
coincides with differential mean numbers of pairs created by a constant
electric field \cite{Nik69-70,Nik79} and locally by slowly varying
time-dependent electric fields ($t$-electric potential steps), such as the $%
T $-constant electric field \cite{GavGit96,AdoGavGit17}, Sauter-type
electric field \cite{GavGit96,AdoGavGit17}, peak electric field \cite%
{AdoGavGit16} and inverse-square electric field \cite{AdoGavGit18}.
Moreover, it also coincides with local approximations by some small gradient
coordinate-dependent electric fields ($x$-electric potential steps), namely
the $L$-constant electric field and Sauter electric field \cite%
{x-case,GavGit16b}, and exponential electric step \cite{GavGitShi17}. For $t$%
-electric potential steps, the local behavior refers to small values of the
longitudinal momentum while for $x$-electric potential steps refers to small
energies, as was seen above. Distribution (\ref{31.5}) is an universal
feature of the differential mean number of particles created from the vacuum
by electric fields, which is uniform (either with respect to the
longitudinal momentum or the energy) only for homogeneous (either in time or
space) electric fields.

The subrange $\left( \mathrm{c}\right) $ corresponds to finite energies $%
\min \left( p_{0}/\sqrt{eE}\right) =\Upsilon U_{\mathrm{R}}/\sqrt{eE}-\sqrt{%
\lambda }$ and values of parameters $\eta _{1}$ and $\eta _{2}$ slightly
distant from the unity $\min \eta _{1}=1+\Upsilon U_{\mathrm{R}}/\left\vert
U_{\mathrm{L}}\right\vert -\pi _{\perp }/\left\vert U_{\mathrm{L}%
}\right\vert $, $\eta _{2}\gtrsim 1-\Upsilon $, resulting in sufficiently
large values to the variables $\left\vert \mathcal{Z}_{j}\right\vert $.
Hence, we can use Eq. (67) from Ref. \cite{AdoGavGit18} for $\Psi \left(
c_{1}-a_{1},c_{1};e^{-i\pi }z_{1}\right) $ and the first line of Eq. (68)
also from \cite{AdoGavGit18} for $\Psi \left( a_{2},c_{2};z_{2}\right) $ to
show that the mean numbers are given by the equation%
\begin{equation}
N_{n}^{\mathrm{cr}}\approx \exp \left( 2\pi \nu _{2}^{+}\right) \,,
\label{31.6}
\end{equation}%
in the leading-order approximation, valid both for Fermions and for Bosons.
In the subrange $\left( \mathrm{b}\right) $, the mean numbers vary between
the asymptotic forms (\ref{31.5}) and (\ref{31.6}). More accurate (but less
simple) representations for the mean numbers in this subrange can be
calculated using the uniform asymptotic approximation given by Eq. (63) from 
\cite{AdoGavGit18}. Despite being a local approximation, the asymptotic form
(\ref{31.6}) tends to uniform approximation (\ref{31.5}) in the limit of
small energies, as it can be seen by expanding $\nu _{2}^{+}$ for small $%
p_{0}/\sqrt{eE}$. Therefore, approximation (\ref{31.6}) can be extended over
all sub-ranges above.

From the symmetry properties of $g$-coefficients, we can generalize the
above results to negative energies and find approximately differential mean
numbers%
\begin{equation}
N_{n}^{\mathrm{cr}}\approx \left\{ 
\begin{array}{ll}
\exp \left( 2\pi \nu _{1}^{-}\right) \,, & \mathrm{if}\ \ U_{\mathrm{L}}+\pi
_{\perp }\leq p_{0}<0\, \\ 
\exp \left( 2\pi \nu _{2}^{+}\right) \,, & \mathrm{if\ \ }0\leq p_{0}\leq U_{%
\mathrm{R}}-\pi _{\perp }\,%
\end{array}%
\right. .  \label{31.7}
\end{equation}

The above representation corresponds to dominant contributions. To calculate
the total number $N$ dominant in the same approximation, one has to perform
integrations over quantum numbers according to Eq. (\ref{3.4}). Then we get:%
\begin{eqnarray}
&&N^{\mathrm{cr}}=\frac{J_{\left( d\right) }V_{\perp }T}{\left( 2\pi \right)
^{d-1}}\int_{\sqrt{\lambda }<K_{\perp }}d\mathbf{p}_{\perp }\left[ I_{%
\mathbf{p}_{\perp }}^{\left( 1\right) }+I_{\mathbf{p}_{\perp }}^{\left(
2\right) }\right] \,,  \notag \\
&&I_{\mathbf{p}_{\perp }}^{\left( 1\right) }=\int_{\pi _{\perp }}^{eE\xi
_{1}}d\pi _{0}\left( \mathrm{L}\right) \exp \left( 2\pi \nu _{1}^{-}\right)
\,,\ \ I_{\mathbf{p}_{\perp }}^{\left( 2\right) }=\int_{\pi _{\perp
}}^{eE\xi _{2}}d\left\vert \pi _{0}\left( \mathrm{R}\right) \right\vert \exp
\left( 2\pi \nu _{2}^{+}\right) \,.  \label{31.8}
\end{eqnarray}%
Performing a change of variables $-\lambda s_{1}=2\nu _{1}^{-}$ in $I_{%
\mathbf{p}_{\perp }}^{\left( 1\right) }$ and $-\lambda s_{2}=2\nu _{2}^{+}$
in $I_{\mathbf{p}_{\perp }}^{\left( 2\right) }$, one can represent both
integrals as follows%
\begin{equation}
I_{\mathbf{p}_{\perp }}^{\left( j\right) }=-\int_{1}^{\infty
}ds_{j}F_{j}\left( s_{j}\right) e^{-\pi \lambda s_{j}}\,,\ \ F_{1}\left(
s_{1}\right) =\frac{d\pi _{0}\left( \mathrm{L}\right) }{ds_{1}}\,,\ \
F_{2}\left( s_{2}\right) =\frac{d\left\vert \pi _{0}\left( \mathrm{R}\right)
\right\vert }{ds_{2}}\,.  \label{31.9}
\end{equation}%
Computing the functions $F_{j}\left( s_{j}\right) $ and restricting
ourselves to the zeroth order term in powers series of $\lambda /eE\xi
_{j}^{2}$, we obtain%
\begin{equation}
I_{\mathbf{p}_{\perp }}^{\left( j\right) }\approx \frac{eE\xi _{j}}{2}%
e^{-\pi \lambda }G\left( \frac{1}{2},\pi \lambda \right) \,,  \label{31.10}
\end{equation}%
where $G\left( \alpha ,z\right) =e^{z}z^{\alpha }\Gamma \left( -\alpha
,z\right) $ and $\Gamma \left( -\alpha ,z\right) $ is the incomplete Gamma
function \cite{NIST}. Neglecting exponentially small contributions, one can
extend the integration domain over the perpendicular momenta $\mathbf{p}%
_{\perp }$ (\ref{31.8}). Thus, one obtains%
\begin{eqnarray}
N^{\mathrm{cr}} &=&V_{\perp }Tn^{\mathrm{cr}}\,,\ \ n^{\mathrm{cr}}=r^{%
\mathrm{cr}}\frac{\mathbb{U}}{2eE}G\left( \frac{d-1}{2},\frac{\pi m^{2}}{eE}%
\right) \,,  \notag \\
r^{\mathrm{cr}} &=&\frac{J_{\left( d\right) }\left( eE\right) ^{d/2}}{\left(
2\pi \right) ^{d-1}}\exp \left( -\frac{\pi m^{2}}{eE}\right) \,,
\label{31.11}
\end{eqnarray}%
where $r^{\mathrm{cr}}$ is the rate of pair creation and $n^{\mathrm{cr}}$
denotes the dominant total density of pairs per unit of time $T$ and volume $%
V_{\perp }$. Then one one can straightforwardly calculate the
vacuum-to-vacuum transition probability (\ref{3.8}),%
\begin{eqnarray}
&&P_{v}=\exp \left( -\mu N^{\mathrm{cr}}\right) \,,\ \ \mu
=\sum_{l=0}^{\infty }\frac{\kappa ^{l}\epsilon _{l+1}}{\left( l+1\right)
^{d/2}}\exp \left( -\frac{\pi m^{2}}{eE}l\right) \,,  \notag \\
&&\epsilon _{l}=G\left( \frac{d-1}{2},l\pi \frac{m^{2}}{eE}\right) G\left( 
\frac{d-1}{2},\pi \frac{m^{2}}{eE}\right) ^{-1}\,.  \label{31.12}
\end{eqnarray}

It is worth noticing that Eqs. (\ref{31.11}) and (\ref{31.12}) are in
agreement with universal forms for total dominant densities of particles
created from the vacuum and vacuum-to-vacuum transition probabilities by
weakly inhomogeneous $x$-electric potential steps, recently formulated in 
\cite{GavGitShi19b}. It can be readily shown that Eqs. (\ref{31.11}) and (%
\ref{31.12}) can be reproduced using universal forms for the above
quantities and the explicit form of the external field (\ref{2.2}).

According to Eq. (\ref{31.11}), the dominant total density of pairs is
proportional to the total work done on a charged particle by the electric
field, $\pi _{0}\left( \mathrm{L}\right) -\pi _{0}\left( \mathrm{R}\right) =%
\mathbb{U}$. This is a common feature to electric fields in the
small-gradient regime and therefore allow us to compare the present results
with total quantities obtained in another examples, in particular in the $L$%
-constant electric field \cite{GavGit16b}, $E\left( x\right) =E\,,\ x\in %
\left[ -L/2,L/2\right] $, where $L$ is the length of the applied constant
field.

Recalling its dominant total density of pairs created by the small-gradient
regime$\ n^{\mathrm{cr}}=Lr^{\mathrm{cr}}$, one can establish relations
between both fields by which they are equivalent in the pair production. For
example, considering the electric field $E$ and assuming that the total
density of pairs created by the $L$-constant electric field and by the
inverse-square electric field are the same, we conclude that both fields are
equivalent in pair production provided the total length of the applied
inverse-square electric over the $x$-direction is given by%
\begin{equation}
L_{\mathrm{eff}}=\frac{\xi _{1}+\xi _{2}}{2}G\left( \frac{d-1}{2},\frac{\pi
m^{2}}{eE}\right) \,.  \label{31.12.3}
\end{equation}%
By definition, $L=L_{\mathrm{eff}}$ for the constant field. Thus, one can
say that both fields are equivalent in pair production provided that they
have the same effective length $L_{\mathrm{eff}}$ over the $x$-axis.

So far, we have discussed the electric field in an almost symmetrical
configuration, characterized by simultaneously large length scales $\xi _{j}$
but slightly close to one another (fixed ratio $\xi _{2}/\xi _{1}$).
However, there may be situations where the field is essentially asymmetrical
by physical conditions, for instance, growing \textquotedblleft
smoothly\textquotedblright\ from infinitely remote negative distances but
decreasing \textquotedblleft abruptly\textquotedblright\ to infinitely
remote positive distances. Situations like that correspond to cases in which
one characteristic length $\xi $ is much larger than another one, in the
situation illustrated above, $\xi _{1}/\xi _{2}\gg 1$. More precisely,
electric field (\ref{2.2}) can be \textquotedblleft
concentrated\textquotedblright\ in a \textquotedblleft
narrow\textquotedblright\ or \textquotedblleft wide\textquotedblright\
region over the $x$-axis depending on the length scales $\xi _{1}$ and $\xi
_{2}$. The larger the characteristic lengths $\xi _{j}$, the
\textquotedblleft smoother\textquotedblright\ the electric field grows from
or decreases to asymptotic regions $x=\mp \infty $, respectively. In this
way, we qualitatively say that the electric field is concentrated in a
\textquotedblleft wide\textquotedblright\ region for $x<0$ if $\xi _{1}$ is
sufficiently large or concentrated in a \textquotedblleft
narrow\textquotedblright\ vicinity of $x=0$ $\left( x<0\right) $ if $\xi
_{1} $ is sufficiently small. For example, for a very asymmetrical
configuration specified by very \textquotedblleft large\textquotedblright\ $%
\xi _{1}$ and very \textquotedblleft small\textquotedblright\ $\xi _{2}$
provided that the parameters $eE\xi _{j}^{2}$ satisfy the relations%
\begin{equation}
eE\xi _{1}^{2}\gg \max \left( 1,\frac{m^{2}}{eE}\right) \,,\ eE\xi
_{2}^{2}\ll \min \left( 1,\frac{m^{2}}{eE}\right) \,,\ \sqrt{eE}\xi _{1}%
\sqrt{eE}\xi _{2}\ll 1\,,  \label{31.13}
\end{equation}%
the results concerning particle creation can be formally extracted from Eqs.
(\ref{31.7}) - (\ref{31.12}) considering the limit $\sqrt{eE}\xi
_{2}\rightarrow 0$. The last inequality implies that the parameter $eE\xi
_{2}^{2}$ is so small that the contribution from the second interval $x\in 
\mathrm{II}$ is negligible for particle creation. To see that, it is enough
to compare the $g\left( _{+}|^{-}\right) $ coefficient calculated for the
asymmetrical electric field%
\begin{equation}
E\left( t\right) =E\left\{ 
\begin{array}{ll}
\left( 1-x/\xi _{1}\right) ^{-2}\,, & x\in \mathrm{I}\, \\ 
0\,, & x\in \mathrm{II}\,%
\end{array}%
\right. ,  \label{31.14}
\end{equation}%
with the one given by Eq. (\ref{2.30}) in the limit $\sqrt{eE}\xi
_{2}\rightarrow 0$ to conclude that both coefficients coincides in the
leading-order approximation.

As a result, dominant contributions to differential and total quantities can
be straightforwardly derived from the aforementioned expressions. This can
be proved following the same approximations and considerations done for the
inverse-square time-dependent electric field \cite{AdoGavGit18}, due to the
close analogy between electric fields (\ref{2.2}), (\ref{31.14}) and their
time-dependent equivalents.

At last but not least, some clarifying remarks concerning the local
properties of differential quantities near the boundaries of $\Omega _{3}$
are in order. As we discussed before, the coefficient $g\left(
_{+}|^{-}\right) $ (or $g\left( _{-}|^{+}\right) $) diverges near the
boundaries, which means that the numbers $N_{n}^{\mathrm{cr}}$ vanish at
these regions. For small-gradient fields, this is clearly seen from the
asymptotic forms given by Eq. (\ref{31.7}). In a vicinity of $\Omega _{2}$,
where $p_{0}/\sqrt{eE}=U_{\mathrm{R}}/\sqrt{eE}-\sqrt{\lambda }-\varepsilon $%
, or in a vicinity of $\Omega _{4}$, where $p_{0}/\sqrt{eE}=U_{\mathrm{L}}/%
\sqrt{eE}+\sqrt{\lambda }+\varepsilon $, with $\varepsilon $ being an
infinitesimally small positive number, the parameters%
\begin{equation}
\nu _{2}^{+}=eE\xi _{2}^{2}\left( -\frac{\lambda ^{1/4}}{\sqrt{2\varepsilon }%
}+O\left( 1\right) \right) \,,\ \ \nu _{1}^{-}=eE\xi _{1}^{2}\left( -\frac{%
\lambda ^{1/4}}{\sqrt{2\varepsilon }}+O\left( 1\right) \right) \,,
\label{31.15}
\end{equation}%
diverge as $\varepsilon \rightarrow 0^{+}$, resulting in exponentially small
contributions to the mean numbers according to Eq. (\ref{31.7}). This result
is in agreement with the general theory \cite{x-case}, in which no particle
production occurs beyond the Klein zone. This property can also be seen in
asymmetrical configurations, corresponding to an electric field growing
\textquotedblleft smoothly\textquotedblright\ from $x=-\infty $ but not
decreasing \textquotedblleft smoothly\textquotedblright\ nor
\textquotedblleft abruptly\textquotedblright\ to $x=+\infty $, so that the
parameter $eE\xi _{1}^{2}$ is sufficiently large while $eE\xi _{2}^{2}$ is
considered finite. Namely,%
\begin{equation}
eE\xi _{1}^{2}\gg \max \left( eE\xi _{2}^{2},\frac{m^{2}}{eE}\right) \,,\ \
eE\xi _{2}^{2}=O\left( \lambda \right) \,.  \label{31.16}
\end{equation}%
In a vicinity of $\Omega _{4}$, the mean numbers are exponentially small on
account of the behavior of $\nu _{1}^{-}$ given by Eq. (\ref{31.15}). As for
the vicinity of $\Omega _{2}$, one must take into account that $a_{2}$ is
large while $z_{2}$ and $c_{2}$ are finite to use asymptotic approximations (%
\ref{app1.1}) from Appendix \ref{App1} to find%
\begin{equation}
\left\vert g\left( _{+}|^{-}\right) \right\vert ^{2}\approx \frac{2\lambda }{%
\pi }e^{\pi \left\vert \nu _{2}^{+}\right\vert }\sinh \left( \pi \left\vert
\nu _{2}^{+}\right\vert \right) \left\vert \sqrt{\frac{2\Delta U_{2}}{\pi
_{\perp }}}K_{c_{2}-1}\left( 2\sqrt{a_{2}z_{2}}\right) +iK_{c_{2}}\left( 2%
\sqrt{a_{2}z_{2}}\right) \right\vert ^{2}\,,  \label{31.17}
\end{equation}%
for Fermions and%
\begin{equation}
\left\vert g\left( _{+}|^{-}\right) \right\vert ^{2}\approx \frac{2\lambda }{%
\pi }\left( 2\Delta U_{2}\right) e^{\pi \left\vert \nu _{2}^{+}\right\vert
}\cosh \left( \pi \left\vert \nu _{2}^{+}\right\vert \right) \left\vert 
\sqrt{\frac{2\Delta U_{2}}{\pi _{\perp }}}K_{c_{2}-1}\left( 2\sqrt{a_{2}z_{2}%
}\right) -iK_{c_{2}}\left( 2\sqrt{a_{2}z_{2}}\right) \right\vert ^{2}\,,
\label{31.18}
\end{equation}%
for Bosons, in leading-order approximation. Here, $K_{\nu }\left( z\right) $
are modified Bessel functions of the second kind \cite{NIST}. According to
Eq. (\ref{31.15}), both coefficients diverge exponentially in a vicinity of $%
\Omega _{2}$, which means that $N_{n}^{\mathrm{cr}}\rightarrow 0$ in this
region. Note that the combination of the Bessel functions are finite, once $%
\sqrt{a_{2}z_{2}}=O\left( \lambda ^{7/4}\right) $ and $c_{2}$ is fixed.
Using the symmetries discussed in the end of Subsection \ref{Sec2.1}, we can
generalize these results to a configuration opposite to the one under
consideration (\ref{31.16}), corresponding to an electric field growing not
too \textquotedblleft smoothly\textquotedblright\ nor too \textquotedblleft
abruptly\textquotedblright\ from $x=-\infty $ but decreasing
\textquotedblleft smoothly\textquotedblright\ to $x=+\infty $, such that $%
eE\xi _{2}^{2}\gg \max \left( eE\xi _{1}^{2},m^{2}/eE\right) \,$and$\ eE\xi
_{1}^{2}=O\left( \lambda \right) $. In particular, the behavior near the
range $\Omega _{4}$ can be formally derived from Eqs. (\ref{31.17}) and (\ref%
{31.18}) substituting $\xi _{2}\rightarrow \xi _{1}$.

\subsection{Sharp-gradient configuration\label{Sec3.2}}

In contrast to the previous configurations, where one or both length scales
are considered large $\xi _{j}$, here we consider the opposite case,
characterized by sufficiently small length scales such that the parameters $%
\left\vert U_{\mathrm{L}}\right\vert \xi _{1}$ and $U_{\mathrm{R}}\xi _{2}$
obey the following conditions%
\begin{equation}
\max \left( \left\vert U_{\mathrm{L}}\right\vert \xi _{1},U_{\mathrm{R}}\xi
_{2}\right) \ll 1\,,\ \ \frac{\xi _{2}}{\xi _{1}}=\mathrm{fixed\,.}
\label{32.1}
\end{equation}%
This configuration corresponds to a very sharp electric field, highly
concentrated about the origin $x=0$, described by a very \textquotedblleft
steep\textquotedblright\ potential step. The potential energy of an electron
in this field is illustrated by the solid blue line in Fig. \ref{Fig1}. This
configuration has a special interest because it corresponds to a
two-parameter regularization of the Klein step and may be useful in a
discussion of the Klein paradox.

From condition (\ref{32.1}) and the fact that energies in the Klein zone are
bounded, see (\ref{3.0}), parameters involving kinetic energies%
\begin{equation}
\max \left( \pi _{0}\left( \mathrm{L}\right) \xi _{1},\left\vert \pi
_{0}\left( \mathrm{R}\right) \right\vert \xi _{2}\right) \ll 1\,,
\label{32.2}
\end{equation}%
as well as the asymptotic momenta $\left\vert p^{\mathrm{L}}\right\vert \xi
_{1}$ and $\left\vert p^{\mathrm{R}}\right\vert \xi _{2}$ are also small in
this case, since $\left\vert p^{\mathrm{L/R}}\right\vert <\left\vert \pi
_{0}\left( \mathrm{L/R}\right) \right\vert $. Thus, to study differential
quantities in the Fermi case it is more convenient to use a representation
of the coefficients $g\left( _{+}|^{-}\right) $ in terms of Whittaker
functions given by Eq. (\ref{app1.0a}) in Appendix \ref{App1}. Thus, we get%
\footnote{%
The differential mean numbers (as well as any physical quantities) are
invariant by the choice of $\chi $. Thus, $\left. N_{n}^{\mathrm{cr}%
}\right\vert _{\chi =+1}=\left. N_{n}^{\mathrm{cr}}\right\vert _{\chi =-1}$.}%
:%
\begin{equation}
N_{n}^{\mathrm{cr}}\approx \frac{4\left\vert p^{\mathrm{L}}\right\vert
\left\vert p^{\mathrm{R}}\right\vert }{\left( \left\vert p^{\mathrm{L}%
}\right\vert -\left\vert p^{\mathrm{R}}\right\vert +\chi \mathbb{U}\right)
^{2}}\frac{\left\vert \pi _{0}\left( \mathrm{R}\right) +\chi \left\vert p^{%
\mathrm{R}}\right\vert \right\vert }{\left\vert \pi _{0}\left( \mathrm{L}%
\right) -\chi \left\vert p^{\mathrm{L}}\right\vert \right\vert }\,,
\label{32.3}
\end{equation}%
in the leading-order approximation. Distribution (\ref{32.3}) has a maximum
at $\left\vert p^{\mathrm{L}}\right\vert -\left\vert p^{\mathrm{R}%
}\right\vert =0$, i.e., at $p_{0}=\left( U_{\mathrm{R}}^{2}-U_{\mathrm{L}%
}^{2}\right) /2\mathbb{U}$,%
\begin{equation}
\max N_{n}^{\mathrm{cr}}\approx 1-\left( \frac{2\pi _{\perp }}{\mathbb{U}}%
\right) ^{2}<1\,,  \label{32.4}
\end{equation}%
which is less than the unity due to the Fermi statistics. Similar results
were obtained for other exactly-solvable backgrounds, such as for the Sauter
field \cite{x-case} and the Peak electric field \cite{GavGitShi17}.

For scalar particles, one can use a representation of $g\left(
_{+}|^{-}\right) $ in terms of Whittaker functions (\ref{app1.0b}) and
limiting form (\ref{app1.4}) to show%
\begin{eqnarray}
&&N_{n}^{\mathrm{cr}}\approx 4\pi ^{2}\Delta U_{2}\Delta U_{1}\left\{ \left[ 
\mathbb{U}\left( Y_{1}Y_{2}+\pi ^{2}/4\right) -2\left( \Delta
U_{1}Y_{1}+\Delta U_{2}Y_{2}\right) \right] ^{2}\right.  \notag \\
&&+\left. \pi ^{2}\left[ \frac{\mathbb{U}}{2}\left( Y_{2}-Y_{1}\right)
+\left( \Delta U_{2}-\Delta U_{1}\right) \right] ^{2}\right\} ^{-1}\,,
\label{32.5}
\end{eqnarray}%
in the leading-order approximation, where $Y_{2}=\psi \left( 1\right) +\ln
4-\ln \left( 2\left\vert p^{\mathrm{R}}\right\vert \xi _{2}\right) $, $%
Y_{1}=\psi \left( 1\right) +\ln 4-\ln \left( 2\left\vert p^{\mathrm{L}%
}\right\vert \xi _{1}\right) $ and $-\psi \left( 1\right) \approx 0.577$ is
the Euler constant. Despite the possibility of a large number of scalar
particles be created from the vacuum due to the Bose statistics, the
differential mean numbers can be less than the unity due to logarithmic
contributions in the denominator. This is more clearly seen considering the
symmetric case, $\xi _{1}=\xi _{2}\equiv \xi $ (then $\Delta U_{1}=\Delta
U_{2}=\mathbb{U}/2$), whose maximum at $p_{0}=\left( U_{\mathrm{R}}^{2}-U_{%
\mathrm{L}}^{2}\right) /2\mathbb{U}$%
\begin{equation}
\max N_{n}^{\mathrm{cr}}\approx \left( \frac{\pi }{\mathbb{Y}^{2}+2\mathbb{Y}%
+\pi ^{2}/4}\right) ^{2}\,,\ \ \mathbb{Y}=-\psi \left( 1\right) -\ln 2+\ln
\left( \xi \sqrt{\frac{\mathbb{U}^{2}}{4}-\pi _{\perp }^{2}}\right) \,,
\label{32.6}
\end{equation}%
can be less than the unity depending on the magnitude of the step $\mathbb{U}
$ and on the length scale $\xi $. This feature, particular to the
inverse-square electric field, is not seen in the Sauter electric field \cite%
{x-case} nor in the Peak electric field \cite{GavGitShi17}, which may create
a large number of Bosons in the small-gradient field regime. In section \ref%
{Sec5}, we compare approximations (\ref{32.3}) and (\ref{32.5}) with
numerical calculations and explore further details concerning particle
creation.

With the aid of the above results, we may calculate total quantities
corresponding to sharp-gradient fields and, in particular, compare them with
results obtained through worldline methods. More precisely, it has been
recently discovered that in the deeply critical regime, defined by%
\begin{equation}
1-\gamma ^{2}\ll \min \left\{ \left( eE/m^{2}\right) ^{2},\left(
eE/m^{2}\right) ^{-2},1\right\} \,,\ \ \gamma =\frac{2m}{\mathbb{U}}\,,
\label{32.6.1}
\end{equation}%
the imaginary part of the one-loop QED effective action exhibits universal
properties similar to those of continuous phase transitions \cite%
{cr-regime1,cr-regime2}. According to our terminology, this occurs when the
Klein zone is sufficiently small, as in the case of strong fields in the
sharp-gradient regime. Indeed, assuming $\xi _{1}=\xi _{2}$ for simplicity
and rewriting the condition (\ref{32.1}) in terms of the Keldysh parameter $%
\gamma $ we obtain a condition compatible with (\ref{32.6.1})%
\begin{equation}
1-\gamma ^{2}\ll 1\,,  \label{32.6.2}
\end{equation}%
\ provided the field amplitude is strong enough, $E>m^{2}/e$. Thus, to
compute total quantities, we simplify calculations by setting $U_{\mathrm{L}%
}=0$, $U_{\mathrm{R}}=\mathbb{U}=2m/\gamma $, and introduce new variables%
\begin{equation}
\frac{p_{0}}{m}=1+\left( 1-\gamma ^{2}\right) \frac{v}{2}\,,\ \ \frac{%
\mathbf{p}_{\perp }^{2}}{m^{2}}=\left( 1-\gamma ^{2}\right) r\,,
\label{32.6.3}
\end{equation}%
so that $\left( p^{\mathrm{L}}\right) ^{2}\approx \left( 1-\gamma
^{2}\right) \left( v-r\right) $ and $\left( p^{\mathrm{R}}\right)
^{2}\approx \left( 1-\gamma ^{2}\right) \left( 2-v-r\right) $ in
leading-order, on account of (\ref{32.6.2}). Moreover, using the condition (%
\ref{32.6.2}) one can expand all quantities in ascending powers of $1-\gamma
^{2}$ to show that the mean number of Fermions created (\ref{32.3}) reads%
\begin{equation}
N_{n}^{\mathrm{cr}}=\left( 1-\gamma ^{2}\right) \sqrt{\left( r-1\right)
^{2}-\left( v-1\right) ^{2}}+O\left( \left( 1-\gamma ^{2}\right)
^{3/2}\right) \,.  \label{32.6.4}
\end{equation}%
Substituting the leading-order term of (\ref{32.6.4}) into (\ref{3.4}) and
performing the change of variables (\ref{32.6.3}), the total number of
Fermions created can be expressed as follows%
\begin{equation*}
N^{\mathrm{cr}}\approx \frac{J_{\left( d\right) }V_{\perp }T}{\left( 2\pi
\right) ^{d}}\frac{m^{d-1}\pi ^{d/2}}{\Gamma \left( d/2-1\right) }\left(
1-\gamma ^{2}\right) ^{1+d/2}\int_{0}^{r_{\max }}\frac{dr}{r^{2}}%
r^{d/2}\int_{v_{\min }}^{v_{\max }}dv\sqrt{\left( r-1\right) ^{2}-\left(
v-1\right) ^{2}}\,,
\end{equation*}%
in which the integration limits $v_{\min }\approx r$, $v_{\max }\approx 2-r$%
, $r_{\max }\approx 1$ are determined by restrictions of the Klein zone,
given by Eq. (\ref{3.0}) and $2\pi _{\perp }\leq \mathbb{U}$, respectively.
Computing the remaining integrals, we finally obtain%
\begin{equation}
N^{\mathrm{cr}}\approx \frac{J_{\left( d\right) }V_{\perp }T}{\left( 2\pi
\right) ^{d-1}}\frac{m^{d-1}\pi ^{d/2}}{\Gamma \left( d/2-1\right) }\frac{4}{%
d\left( d^{2}-4\right) }\left( 1-\gamma ^{2}\right) ^{1+d/2}\,,\ \ d>2\,.
\label{32.6.5}
\end{equation}

Due to the smallness of the scaling factor $1-\gamma ^{2}$, we observe that
the total numbers are substantially small within the Klein zone, which means
that the vacuum-vacuum transition probability $P_{v}$, given by the general
form (\ref{3.8}), can be approximated by the total number of created
particles as $P_{v}\approx 1-N^{\mathrm{cr}}$. At the same time, this
probability can be represented via the imaginary part of a one-loop
effective action $S_{\mathrm{eff}}$ according to Schwinger formula \cite%
{Sch51},%
\begin{equation}
P_{v}=\exp \left( -2\mathrm{Im}S_{\mathrm{eff}}\right) \,.  \label{32.6.6}
\end{equation}%
Therefore, taking into account that $P_{v}\approx 1-2\mathrm{Im}S_{\mathrm{%
eff}}$, we can use the result (\ref{32.6.5}) to establish the following
expression to the imaginary part of the effective action:%
\begin{equation}
\mathrm{Im}S_{\mathrm{eff}}\approx N^{\mathrm{cr}}/2\ \ \mathrm{if\ \ }N^{%
\mathrm{cr}}\ll 1\,.  \label{32.6.7}
\end{equation}%
The above expression coincides, in particular, with results obtained for QED
in $d=3+1$ dimensions; cf. Eq. (13) in Ref. \cite{cr-regime2}. This is an
independent confirmation of the universal behavior of pair creation when the
Klein zone is sufficiently small (or, near the criticality, according to
Refs. \cite{cr-regime1,cr-regime2}). Moreover, it should be noted that
particle creation ceases when the step approaches the noncritical
configuration $\mathbb{U}\rightarrow 2m$, which means $\gamma \rightarrow 1$
and therefore $P_{v}\rightarrow 1$ ($P_{v}\leq 1$). Following the same
considerations, it can be shown that $\mathrm{Im}S_{\mathrm{eff}}\varpropto
\left( 1-\gamma ^{2}\right) ^{1+d/2}$ for the Klein-Gordon case.

Because the Klein paradox is often discussed in particle scattering problems
by inhomogeneous potential steps, it is worth considering some relative
probabilities listed at the beginning of this section to clarify the absence
of the Klein paradox through the correct interpretation of these quantities
in $\Omega _{3}$. To this aim, we use Eq. (\ref{32.3}) for $\left\vert
g\left( _{+}|^{-}\right) \right\vert ^{-2}$, the representation (\ref%
{app1.0a}) for $g\left( _{+}|^{+}\right) $ and the approximations (\ref%
{app1.5}) to show that the relative probability of a pair creation $%
\left\vert w_{n}\left( +-|0\right) \right\vert ^{2}$ and of electron
scattering $\left\vert w_{n}\left( +|+\right) \right\vert ^{2}$ are
approximately given by%
\begin{eqnarray}
&&\left\vert w_{n}\left( +-|0\right) \right\vert ^{2}\approx \frac{%
4\left\vert p^{\mathrm{R}}\right\vert \left\vert p^{\mathrm{L}}\right\vert }{%
\left( \left\vert p^{\mathrm{L}}\right\vert +\left\vert p^{\mathrm{R}%
}\right\vert +\chi \mathbb{U}\right) ^{2}}\frac{\left\vert \pi _{0}\left( 
\mathrm{R}\right) -\chi \left\vert p^{\mathrm{R}}\right\vert \right\vert }{%
\left\vert \pi _{0}\left( \mathrm{L}\right) -\chi \left\vert p^{\mathrm{L}%
}\right\vert \right\vert }\,,  \notag \\
&&\left\vert w_{n}\left( +|+\right) \right\vert ^{2}\approx \left( \frac{%
\left\vert p^{\mathrm{L}}\right\vert -\left\vert p^{\mathrm{R}}\right\vert
+\chi \mathbb{U}}{\left\vert p^{\mathrm{L}}\right\vert +\left\vert p^{%
\mathrm{R}}\right\vert +\chi \mathbb{U}}\right) ^{2}\frac{\left\vert \pi
_{0}\left( \mathrm{R}\right) -\chi \left\vert p^{\mathrm{R}}\right\vert
\right\vert }{\left\vert \pi _{0}\left( \mathrm{R}\right) +\chi \left\vert
p^{\mathrm{R}}\right\vert \right\vert }\,.  \label{32.7}
\end{eqnarray}%
These probabilities can be larger than the unity in a sufficiently wide
range of energies within $\Omega _{3}$. For example, they reach their maxima
at $p_{0}=\left( U_{\mathrm{R}}^{2}-U_{\mathrm{L}}^{2}\right) /2\mathbb{U}$%
\begin{equation}
\max \left\vert w_{n}\left( +-|0\right) \right\vert ^{2}\approx \left( \frac{%
\mathbb{U}}{2\pi _{\perp }}\right) ^{2}\,,\ \ \max \left\vert w_{n}\left(
+|+\right) \right\vert ^{2}\approx \left( \frac{\mathbb{U}}{2\pi _{\perp }}%
\right) ^{2}-1\,,  \label{32.8}
\end{equation}%
and unveil the possibility of particle transmission and particle reflection
larger than the unity if interpreted as transmission and reflection
coefficients, respectively. Such an interpretation is due to a formal
analogy between the above probabilities and reflection and transmission
coefficients calculated for the ranges $\Omega _{1}$ and $\Omega _{5}$; cf.
Eqs. (\ref{3.4b}) and Eqs. (\ref{4.1}), (\ref{4.2}) in the next section. In
the Klein zone, an in-electron (or an in-positron) is subjected to total
reflection, whose amplitude probability is given by $w_{n}\left( +|+\right) $
($w_{n}\left( -|-\right) $ for positrons), and no transmission occurs in
this range. Moreover, from Eqs. (\ref{3.4b}) and (\ref{3.8}) we observe that 
$\left\vert w_{n}\left( +|+\right) \right\vert ^{-2}=1-N_{n}^{\mathrm{cr}%
}=p_{v}^{n}$ describes the probability that the partial vacuum state, with
given quantum numbers $n$, remains a vacuum while $p_{v}^{n}\left\vert
w_{n}\left( +-|0\right) \right\vert ^{2}$ denotes the probability that a
pair of Fermions, with given quantum numbers $n$, will be created.
Therefore, from the second line of Eq. (\ref{2.27}), we obtain the
probability conservation%
\begin{equation}
p_{v}^{n}+p_{v}^{n}\left\vert w_{n}\left( +-|0\right) \right\vert ^{2}=1\,,
\label{32.9}
\end{equation}%
resulting from Pauli's principle, which states that for given quantum
numbers $n$ there are only two possibilities: either the vacuum state
remains vacuum or a pair of Fermions be created in a cell of the space. This
is the correct interpretation of the coefficients $\left\vert g\left(
_{+}|^{-}\right) \right\vert ^{-2}$, $\left\vert g\left( _{+}|^{+}\right)
\right\vert ^{-2}$ and $\left\vert g\left( _{+}|^{-}\right) \right\vert
^{2}\left\vert g\left( _{+}|^{+}\right) \right\vert ^{-2}$ in the Klein zone.

Similar interpretations of relative probabilities discussed above hold for
scalar case, namely $\left\vert g\left( _{+}|^{-}\right) \right\vert
^{2}\left\vert g\left( _{+}|^{+}\right) \right\vert ^{-2}$ is the relative
probability of a particle scattering, while $\left\vert g\left(
_{+}|^{+}\right) \right\vert ^{-2}$ is the relative probability of a
particle-antiparticle creation. To obtain explicit forms for these
quantities, one can use Eq. (\ref{32.5}) for $\left\vert g\left(
_{+}|^{-}\right) \right\vert ^{-2}$ and definitions given by Eqs. (\ref{3.4b}%
). The essential difference in comparison with the Fermi case is the
identification of $p_{v}^{n}$. From Eqs. (\ref{3.4b}) and (\ref{3.8}), the
probability that the partial vacuum state with a quantum numbers $n$ remains
the vacuum reads: $p_{v}^{n}=\left( 1+N_{n}^{\mathrm{cr}}\right)
^{-1}=\left\vert w_{n}\left( +|+\right) \right\vert ^{2}$. From the second
line of Eq. (\ref{2.27}), we obtain the identity%
\begin{equation}
p_{v}^{n}\left[ 1-\left\vert w_{n}\left( +-|0\right) \right\vert ^{2}\right]
^{-1}=1\,,  \label{32.10}
\end{equation}%
that can be interpreted as follows: The conditional probability of a pair
creation with quantum numbers $n$ is the sum of probabilities of creation
for any number $l$ of pairs%
\begin{equation}
P\left( \mathrm{pairs}|0\right) _{n}=p_{v}^{n}\left[ \sum_{l=1}^{\infty
}\left\vert w_{n}\left( +-|0\right) \right\vert ^{2l}\right] \,,
\label{32.11}
\end{equation}%
under the condition that all other partial vacua, labelled by quantum
numbers $m\neq n$, remain vacua. Hence, the conservation of probability (\ref%
{32.10}) is the sum of probabilities of all possible events in a cell of the
space with quantum numbers $n$, namely%
\begin{equation}
P\left( \mathrm{pairs}|0\right) _{n}+p_{v}^{n}=1\,.  \label{32.12}
\end{equation}

\section{Processes beyond the Klein zone\label{Sec4}}

The most elementary quantum processes beyond the Klein zone are particle
scattering in the form of reflection from the potential step and particle
transmission through the potential step, both occurring in the ranges $%
\Omega _{1}$ and $\Omega _{5}$. To study these processes, it is enough to
introduce the relative $R_{+,n}$ and absolute $\tilde{R}_{+,n}=c_{v}R_{+,n}$
reflection amplitudes of right antiparticles%
\begin{equation}
R_{+,n_{5}}=\left\langle 0\left\vert \ ^{-}b_{n_{5}}\left( \mathrm{out}%
\right) \ ^{+}b_{n_{5}}^{\dagger }\left( \mathrm{in}\right) \right\vert
0\right\rangle =g\left( _{+}|^{+}\right) ^{-1}g\left( _{+}|^{-}\right) \,,
\label{4.1}
\end{equation}%
and the relative $T_{+,n}$ and absolute $\tilde{T}_{+,n}=c_{v}T_{+,n}$
transmission amplitudes of right antiparticles%
\begin{equation}
T_{+,n_{5}}=\left\langle 0\left\vert \ _{+}b_{n_{5}}\left( \mathrm{out}%
\right) \ ^{+}b_{n_{5}}^{\dagger }\left( \mathrm{in}\right) \right\vert
0\right\rangle =\eta _{\mathrm{R}}g\left( _{+}|^{+}\right) ^{-1}\,,
\label{4.2}
\end{equation}%
since all remaining probabilities, corresponding reflection $\left\vert
R_{-,n_{5}}\right\vert ^{2}$ and transmission $\left\vert
T_{-,n_{5}}\right\vert ^{2}$ of left antiparticles in $\Omega _{5}$ or else
particle reflection $\left\vert R_{\zeta ,n_{1}}\right\vert ^{2}$ and
particle transmission $\left\vert T_{\zeta ,n_{1}}\right\vert ^{2}$
probabilities in $\Omega _{1}$ can be obtained with the aid of the identities%
\begin{eqnarray}
&&\left\vert R_{+,n}\right\vert ^{2}=\left\vert R_{-,n}\right\vert
^{2}=\left\vert g\left( _{+}|^{-}\right) \right\vert ^{2}\left\vert g\left(
_{+}|^{+}\right) \right\vert ^{-2}\,,\ \ \left\vert T_{+,n}\right\vert
^{2}=\left\vert T_{-,n}\right\vert ^{2}=\left\vert g\left( _{+}|^{+}\right)
\right\vert ^{-2}\,,  \notag \\
&&\left\vert R_{\zeta ,n}\right\vert ^{2}+\left\vert T_{\zeta ,n}\right\vert
^{2}=1\,,\ \ n\in \Omega _{1}\cup \Omega _{5}\,,  \label{4.3}
\end{eqnarray}%
that follows from Eqs. (\ref{2.26b}) and (\ref{2.27}) specialized to $\Omega
_{1}\cup \Omega _{5}$. The representations, in terms of the $g$%
-coefficients, in Eqs. (\ref{4.1}) and (\ref{4.2}) are determined by linear
canonical transformations that can be extracted from Eqs. (4.33) in \cite%
{x-case} after the formal substitutions$\ _{-}^{+}a_{n}\left( \mathrm{out}%
\right) \rightarrow \ _{-}^{+}b_{n}^{\dagger }\left( \mathrm{in}\right) $ and%
$\ _{+}^{-}a_{n}\left( \mathrm{in}\right) \rightarrow \
_{+}^{-}b_{n}^{\dagger }\left( \mathrm{out}\right) $. Moreover, reflection
and transmission amplitudes for particles in $\Omega _{1}$ are given by Eqs.
(5.3) and (5.5) in \cite{x-case}. The expressions for Bosons coincide with
the above equations setting $\eta _{\mathrm{R}}=\eta _{\mathrm{L}}=+1$.

The above probabilities can be studied for any configurations of the
electric field, in particular cases where the field is concentrated over a
finite region along the $x$-axis. Such configurations are characterized by
\textquotedblleft small\textquotedblright\ or even \textquotedblleft
finite\textquotedblright\ length scales $\xi _{j}$ so that the parameters $%
\left\vert U_{\mathrm{L}}\right\vert \xi _{1}$, $U_{\mathrm{R}}\xi _{2}$ are
fixed. First, considering energies and length scales $\xi _{j}$ small enough
so that the parameters $\left\vert p^{\mathrm{L}}\right\vert \xi _{1}$, $%
\left\vert p^{\mathrm{R}}\right\vert \xi _{2}$ are sufficiently small, the
coefficient $\left\vert g\left( _{+}|^{-}\right) \right\vert ^{-2}$ formally
coincides with Eq. (\ref{32.3}) for Fermions and Eq. (\ref{32.5}) for
Bosons. Therefore, the reflection and transmission probabilities acquire the
same form as the relative probabilities in $\Omega _{3}$%
\begin{eqnarray}
\left\vert R_{\zeta ,n}\right\vert ^{2} &\approx &\left( \frac{\left\vert p^{%
\mathrm{L}}\right\vert -\left\vert p^{\mathrm{R}}\right\vert +\chi \mathbb{U}%
}{\left\vert p^{\mathrm{L}}\right\vert +\left\vert p^{\mathrm{R}}\right\vert
+\chi \mathbb{U}}\right) ^{2}\frac{\left\vert \pi _{0}\left( \mathrm{R}%
\right) -\chi \left\vert p^{\mathrm{R}}\right\vert \right\vert }{\left\vert
\pi _{0}\left( \mathrm{R}\right) +\chi \left\vert p^{\mathrm{R}}\right\vert
\right\vert }\,,  \notag \\
\left\vert T_{\zeta ,n}\right\vert ^{2} &\approx &\frac{4\left\vert p^{%
\mathrm{R}}\right\vert \left\vert p^{\mathrm{L}}\right\vert }{\left(
\left\vert p^{\mathrm{L}}\right\vert +\left\vert p^{\mathrm{R}}\right\vert
+\chi \mathbb{U}\right) ^{2}}\frac{\left\vert \pi _{0}\left( \mathrm{R}%
\right) -\chi \left\vert p^{\mathrm{R}}\right\vert \right\vert }{\left\vert
\pi _{0}\left( \mathrm{L}\right) -\chi \left\vert p^{\mathrm{L}}\right\vert
\right\vert }\,,  \label{4.5}
\end{eqnarray}%
for Fermions and coincides, in particular, with the reflection and
transmission coefficients calculated for the Peak electric field \cite%
{GavGitShi17} in the sharp-gradient regime. For Bosons, the transmission
coefficient can be conveniently calculated using Eq. (\ref{32.5}) and the
quadratic relations (\ref{2.27}) specialized to this case, namely $%
\left\vert T_{\zeta ,n}\right\vert ^{2}=\left[ 1+\left\vert g\left(
_{+}|^{-}\right) \right\vert ^{2}\right] ^{-1}$. Once the transmission
coefficient is obtained, the reflection probability coefficient can be
calculated using the conservation of probabilities, $\left\vert R_{\zeta
,n}\right\vert ^{2}=1-\left\vert T_{\zeta ,n}\right\vert ^{2}$.

Next, considering energies and length scales $\xi _{j}$ large enough so that
the parameters $\left\vert p^{\mathrm{L}}\right\vert \xi _{1}$, $\left\vert
p^{\mathrm{R}}\right\vert \xi _{2}$ are sufficiently large, one can use the
asymptotic approximations for the Whittaker functions with large argument
given by Eq. (13.19.3) in \cite{NIST} to show that the reflection and
transmission probabilities coincides with Eqs. (\ref{4.5}) for Fermions.
Notice that despite the formal coincidence between Eqs. (\ref{4.5}) and (\ref%
{32.7}), the reflection and transmission coefficients (\ref{4.5}) are less
than the unity beyond the Klein zone. For energies in the Klein zone, these
coefficients can be larger than unity as we discussed in the previous
section. Hence, if interpreted as reflection and transmission coefficients,
this suggests that more Fermions are reflected from the potential step than
coming in and also more Fermions are transmitted by the potential step than
coming in. This is the Klein paradox, which is removed by the correct
interpretation of the rhs. of Eqs. (\ref{4.5}). As for Bosons, the
probabilities acquire substantially different approximations than the
previous case, namely%
\begin{equation}
\left\vert R_{\zeta ,n}\right\vert ^{2}\approx \left( \frac{\left\vert p^{%
\mathrm{L}}\right\vert -\left\vert p^{\mathrm{R}}\right\vert }{\left\vert p^{%
\mathrm{L}}\right\vert +\left\vert p^{\mathrm{R}}\right\vert }\right)
^{2}\,,\ \ \left\vert T_{\zeta ,n}\right\vert ^{2}\approx \frac{4\left\vert
p^{\mathrm{L}}\right\vert \left\vert p^{\mathrm{R}}\right\vert }{\left(
\left\vert p^{\mathrm{L}}\right\vert +\left\vert p^{\mathrm{R}}\right\vert
\right) ^{2}}\,,  \label{4.6}
\end{equation}%
in leading-order approximation. Eqs. (\ref{4.6}) coincides with reflection
and transmission coefficients calculated for the Sauter electric field \cite%
{x-case} and the Peak electric field \cite{GavGitShi17} in the
sharp-gradient regime. It should be noted that for energies well above or
far below the asymptotic potential energies $\left\vert U_{\mathrm{L}%
}\right\vert $, $U_{\mathrm{R}}$, the reflection $\left\vert R_{\zeta
,n}\right\vert ^{2}$ and transmission $\left\vert T_{\zeta ,n}\right\vert
^{2}$ probabilities tend to zero and one, respectively. Indeed, if $p_{0}\gg
U_{\mathrm{R}}$ or $p_{0}\ll -\left\vert U_{\mathrm{L}}\right\vert $, one
finds $\left\vert R_{\zeta ,n}\right\vert ^{2}=O\left( \pi _{\perp }^{2}%
\mathbb{U}^{2}/p_{0}^{4}\right) $ for Fermions and $\left\vert R_{\zeta
,n}\right\vert ^{2}=O\left( \mathbb{U}^{2}/p_{0}^{2}\right) $ for Bosons,
while $\left\vert T_{\zeta ,n}\right\vert ^{2}=1+O\left( \mathbb{U}%
^{2}/p_{0}^{2}\right) $ for both types of particles.

At last, but not least, it is worth comparing the above results with results
that can be obtained in the context of nonrelativistic Quantum Mechanics,
more precisely, in the study of one-dimensional particle scattering by
inverse-square electric fields. To this end, we set $\pi _{\perp }=m$, $%
p_{0}=m+E$, and choose $\pi _{0}\left( \mathrm{L}\right) =m+E$, $\pi
_{0}\left( \mathrm{R}\right) =m+E-\mathbb{U}$. To consider the
nonrelativistic limit for Bosons, it is enough to study the so-called
kinematic factor $k_{b}=\left\vert p^{\mathrm{R}}\right\vert /\left\vert p^{%
\mathrm{L}}\right\vert $, because the reflection and transmission
coefficients (\ref{4.6}) are expressed in terms of the latter as $\left\vert
R_{\zeta ,n}\right\vert ^{2}=\left( 1-k_{b}\right) ^{2}/\left(
1+k_{b}\right) ^{2}$ and $\left\vert T_{\zeta ,n}\right\vert
^{2}=4k_{b}/\left( 1+k_{b}\right) ^{2}$. In the nonrelativistic limit $E\ll
m $, $k_{b}$ is approximately given by \cite{x-case,Landauvol3}%
\begin{equation}
k_{b}\approx k^{\mathrm{NR}}\left( 1-\frac{\mathbb{U}}{4m}\right) \,,\ \ k^{%
\mathrm{NR}}=\sqrt{\frac{E-\mathbb{U}}{E}}\,.  \label{4.7}
\end{equation}%
It is noteworthy that Eqs. (\ref{4.6}) formally coincides with reflection
and transmission coefficients calculated for the rectangular potential step
in the context of the nonrelativistic Quantum Mechanics; see e.g. the
textbook \cite{Landauvol3}, Sec. 25.

In contrast to the Klein-Gordon case, the reflection and transmission
coefficients for Fermions (\ref{4.5}) does not admit simple representations
in terms of a kinematic factor because of their more complex
representations. Nevertheless, for sufficiently small steps $\mathbb{U}\ll
E+m$, the ratio%
\begin{equation}
\frac{\left\vert \pi _{0}\left( \mathrm{R}\right) -\left\vert p^{\mathrm{R}%
}\right\vert \right\vert }{\left\vert \pi _{0}\left( \mathrm{L}\right)
-\left\vert p^{\mathrm{L}}\right\vert \right\vert }=1+\frac{\mathbb{U}}{%
\left\vert p^{\mathrm{L}}\right\vert }+O\left( \mathbb{U}^{2}/\left\vert p^{%
\mathrm{L}}\right\vert ^{2}\right) \,,  \label{4.8}
\end{equation}%
allow us to simplify the transmission coefficient (\ref{4.5}) as follows%
\begin{equation}
\left\vert T_{\zeta ,n}\right\vert ^{2}\approx \frac{4k_{b}\left( 1+\mathbb{U%
}/\left\vert p^{\mathrm{L}}\right\vert \right) }{\left( 1+k_{b}+\mathbb{U}%
/\left\vert p^{\mathrm{L}}\right\vert \right) ^{2}}\,.  \label{4.9}
\end{equation}%
In Eqs. (\ref{4.8}), (\ref{4.9}), we selected $\chi =+1$ for simplicity. In
the nonrelativistic limit, $\mathbb{U}/\left\vert p^{\mathrm{L}}\right\vert
\approx \mathbb{U}/\sqrt{2mE}$ in leading-order. Combining the latter
approximation and Eq. (\ref{4.7}) with Eq. (\ref{4.9}) one can easily obtain
a nonrelativistic expression for the transmission coefficient. The
reflection coefficient for Fermions can be obtained through the probability
conservation (\ref{4.3}).

\section{Comparing asymptotic estimates with exact results\label{Sec5}}

In this section we supplement the study with comparisons between
differential quantities, calculated numerically, and asymptotic
approximations discussed throughout the text. To this aim, we simplify
computations by setting $\mathbf{p}_{\perp }=0$ and work with the system of
units where $\hslash =c=m=1$. In all plots below, the length scales $\xi
_{j} $, energies $p_{0}$ and electric field amplitudes $E$ are relative to
electron's mass $m$ and Schwinger's critical field $E_{c}=m^{2}/e$,
respectively.

\begin{figure}[th]
\begin{center}
\includegraphics[scale=0.46]{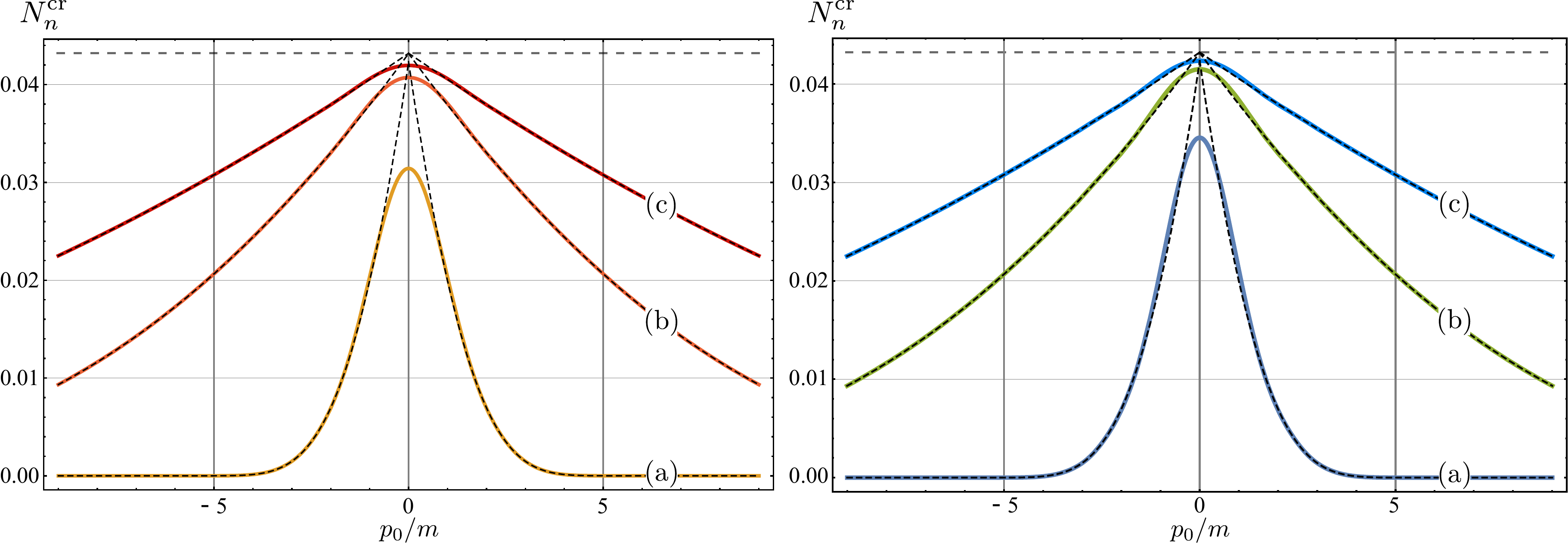}
\end{center}
\caption{Differential mean numbers $N_{n}^{\mathrm{cr}}$ of Fermions (left
panel) and Bosons (right panel) created from the vacuum by symmetrical
inverse-square fields (\protect\ref{2.2}). The solid lines refer to
numerical calculation of the exact expressions (\protect\ref{2.31}), (%
\protect\ref{2.32c}) while the dashed curves the asymptotic approximations (%
\protect\ref{31.7}). In $\left( \mathrm{a}\right) $, $\left( \mathrm{b}%
\right) $ and $\left( \mathrm{c}\right) $, $m\protect\xi _{1}=m\protect\xi %
_{2}=10$, $50$ and $100$, respectively, and $E=E_{c}$ in all cases. The
horizontal dashed lines denotes the uniform distribution $e^{-\protect\pi 
\protect\lambda }$.}
\label{Fig2}
\end{figure}

\begin{figure}[th]
\begin{center}
\includegraphics[scale=0.45]{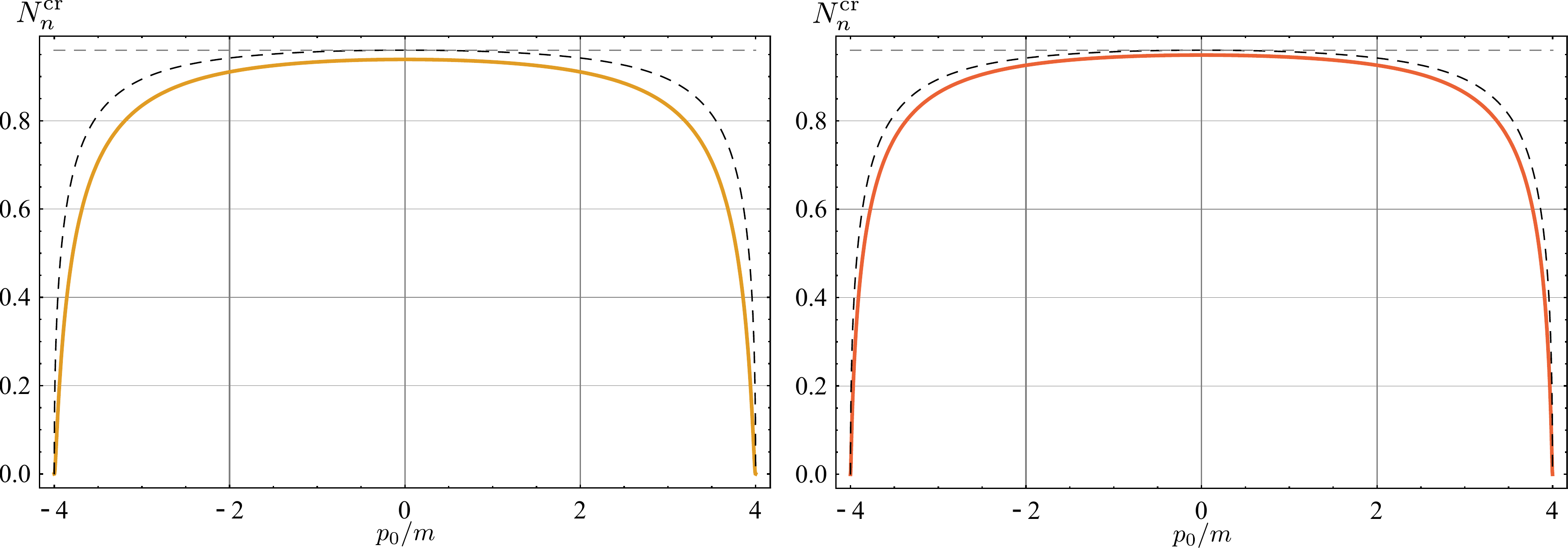}
\end{center}
\caption{Differential mean numbers of Fermions created from the vacuum $%
N_{n}^{\mathrm{cr}}$ by symmetrical inverse-square electric fields. The
exact results (solid lines) are given by the absolute square value of Eq. (%
\protect\ref{2.31}) and the asymptotic ones (dashed curves), by Eq. (\protect
\ref{32.3}). In the left panel, $E=250E_{c}$ and $m\protect\xi_{1}=m\protect%
\xi_{2}=1/50$ while in the right panel, $E=500E_{c}$ and $m\protect\xi_{1}=m%
\protect\xi_{2}=1/100$. The magnitude of the potential energy step is
constant for both plots $\mathbb{U}/m=5=\mathrm{const.}$ and so does the
extension of the Klein zone, which is $|p_{0}|/m\leq 4$. The horizontal
dashed lines denotes the maximum value (\protect\ref{32.4}).}
\label{Fig3}
\end{figure}

\begin{figure}[th]
\begin{center}
\includegraphics[scale=0.45]{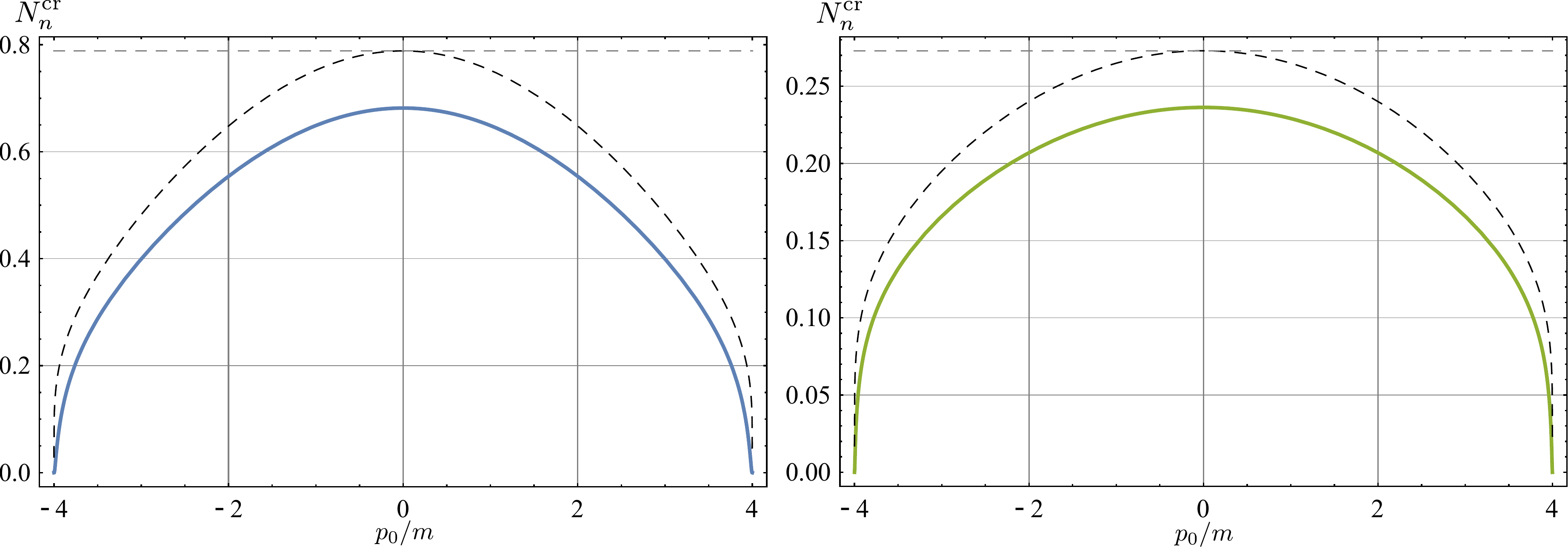}
\end{center}
\caption{Differential mean numbers of Bosons created from the vacuum $N_{n}^{%
\mathrm{cr}}$ by symmetrical inverse-square electric fields. The exact
results (solid lines) are given by the absolute square value of Eq. (\protect
\ref{2.32c}) and the asymptotic ones (dashed curves), by Eq. (\protect\ref%
{32.5}). In the left panel, $E=250E_{c}$ and $m\protect\xi _{1}=m\protect\xi %
_{2}=1/50$ while in the right panel, $E=500E_{c}$ and $m\protect\xi _{1}=m%
\protect\xi _{2}=1/100$. The magnitude of the potential energy step is
constant for both plots $\mathbb{U}/m=5=\mathrm{const.}$ and so does the
extension of the Klein zone, which is $\left\vert p_{0}\right\vert /m\leq 4$%
. The horizontal dashed lines denotes the maximum value (\protect\ref{32.6}%
). }
\label{Fig4}
\end{figure}

Starting with quantities defined in the Klein zone, we present in Figs. \ref%
{Fig2}, \ref{Fig3}, and \ref{Fig4} plots of differential mean numbers $%
N_{n}^{\mathrm{cr}}$ given by exact expressions (\ref{2.31}) and (\ref{2.32c}%
) (solid lines) and by specific asymptotic approximations (dashed curves) as
functions of the energy $p_{0}$, for some values of the length scales $\xi
_{j}$ and field amplitudes $E$. On the Fig. \ref{Fig2}, one can compare
exact results with asymptotic approximations obtained in small-gradient
regime (\ref{31.7}) while on the Figs. \ref{Fig3} and \ref{Fig4}, we compare
exact results with asymptotic approximations obtained for electric fields in
sharp-gradient regime (\ref{32.3}) and (\ref{32.5}).

The numerical results represent differential quantities obtained for
electric fields near the small-gradient or the sharp-gradient regimes. For
example, in Fig. \ref{Fig2}, the larger the length scales $\xi _{j}$ the
more accurate asymptotic forms (\ref{31.7}), which means that electric
fields in Fig. \ref{Fig2} are closer to the small-gradient regime. Moreover,
we see that the differential means tend to the uniform distribution $e^{-\pi
\lambda }$ as $\xi _{j}$ increases. This is expected since inverse-square
electric field (\ref{2.2}) tends to the $L$-constant field \cite{GavGit16b}
as the length scales $\xi _{j}$ increase, whose differential means are given
by $e^{-\pi \lambda }$ provided it acts on the vacuum over a sufficiently
wide region in the space. Still in Fig. \ref{Fig2}, we observe that the
means $N_{n}^{\mathrm{cr}}$ approach to the uniform distribution $e^{-\pi
\lambda }$ for small energies, while they approach to asymptotic forms (\ref%
{31.7}) as the energy increases, irrespective the value of the length scales 
$\xi _{j}$. However, increasing the length scales $\xi _{j}$ and the field
amplitudes $E$ the parameters $\left\vert U_{\mathrm{L}}\right\vert \xi _{1}$%
, and $U_{\mathrm{R}}\xi _{2}$ increase as well, which significantly improve
the accuracy of asymptotic approximations (\ref{31.7}) as was discussed in
Sec. \ref{Sec3.1}. For the values considered on the Fig. \ref{Fig2}, the
lines $\left( \mathrm{a}\right) $, $\left( \mathrm{b}\right) $ and $\left( 
\mathrm{c}\right) $ correspond to $\mathbb{U}\xi /2=100$,\ $2500$ and $%
10^{4} $, respectively.

In Figs. \ref{Fig3} and \ref{Fig4} we represent mean numbers in the case of
an inverse-square electric field near the sharp-gradient regime. In these
plots, we chose sufficiently large electric amplitudes $E$ and sufficiently
small length scales $\xi _{j}$, simulating very strong, sharp and critical
electric fields. For all plots on the Figs. \ref{Fig3} and \ref{Fig4}, the
magnitude of the potential energy step is fixed, namely $\mathbb{U}/m=5$.
According to the above results, we see that the accuracy of asymptotic forms
(\ref{32.3}) and (\ref{32.5}) increases as the length scales $\xi _{j}$
decreases. As discussed in Sec. \ref{Sec3.2}, this results from the fact
that the parameters $\left\vert U_{\mathrm{L}}\right\vert \xi _{1}$ and $U_{%
\mathrm{R}}\xi _{2}$ decrease as $\xi _{j}$ decrease and the smaller their
values the more accurate the approximations. This explains why the dashed
lines are closer to the solid lines on the right panels than on the left
panels on the both figures.

Although the asymptotic forms are less accurate in the scalar case, the
accuracy of the approximations can be improved incorporating next-to-leading
order terms into Eq. (\ref{32.5}). Furthermore, it should be noted that for
the values of $E/E_{c}$ and $m\xi $ considered on the Fig. \ref{Fig4}, the
differential mean numbers are less than the unity. However, for
inverse-square electric fields in intermediate regimes\footnote{%
Electric fields that are not in the small-gradient regime nor in the
sharp-gradient regime.}, this may not be the case: a very large number of
Bosons can be created from the vacuum. In these cases, the notion of
inverse-square electric fields as an external one is limited.

\begin{figure}[th]
\begin{center}
\includegraphics[scale=0.4]{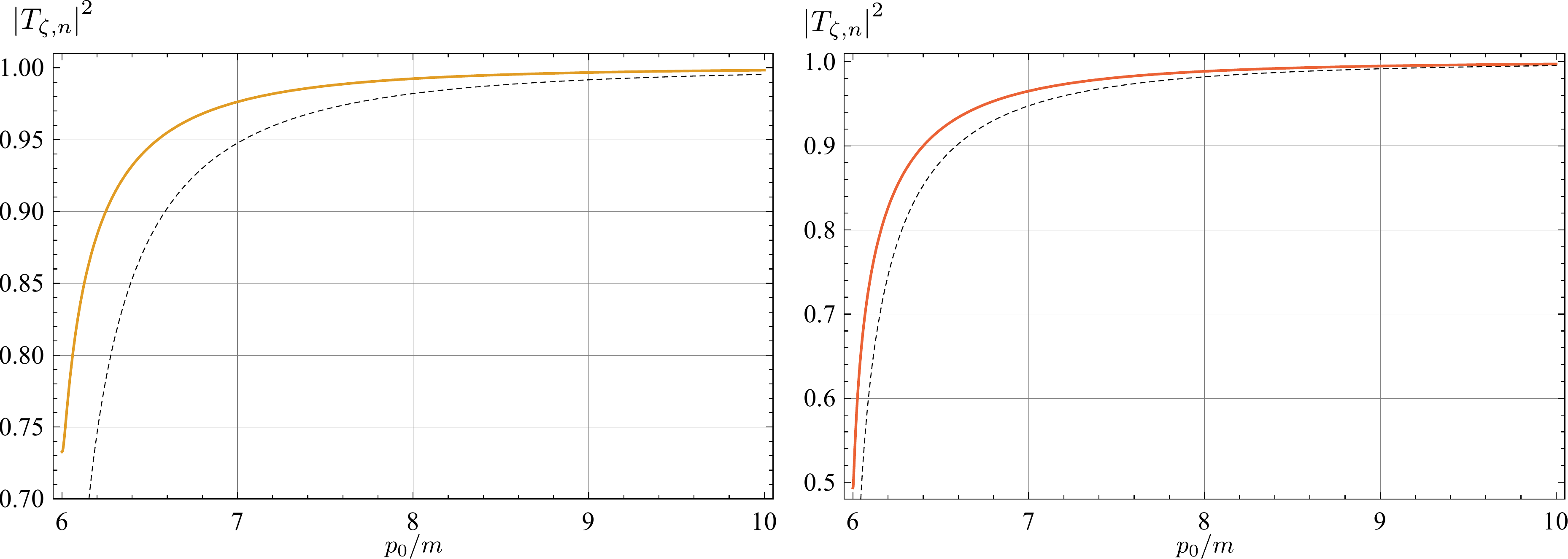}
\end{center}
\caption{Probabilities of Fermions transmission through symmetrical inverse
potential steps in $\Omega _{1}$. Exact results, given by Eqs. (\protect\ref%
{4.3}) and (\protect\ref{2.32a}), are represented by solid lines while
asymptotic ones, given by Eq. (\protect\ref{4.5}), are represented by dashed
curves. In the left panel, $E=250E_{c}$ and $m\protect\xi _{1}=m\protect\xi %
_{2}=1/50$ while in the right panel, $E=500E_{c}$ and $m\protect\xi _{1}=m%
\protect\xi _{2}=1/100$. The lower bound of $\Omega _{1}$ is the same for
both plots which, in this system of units, is $p_{0}/m=6$.}
\label{Fig5}
\end{figure}

\begin{figure}[th]
\begin{center}
\includegraphics[scale=0.4]{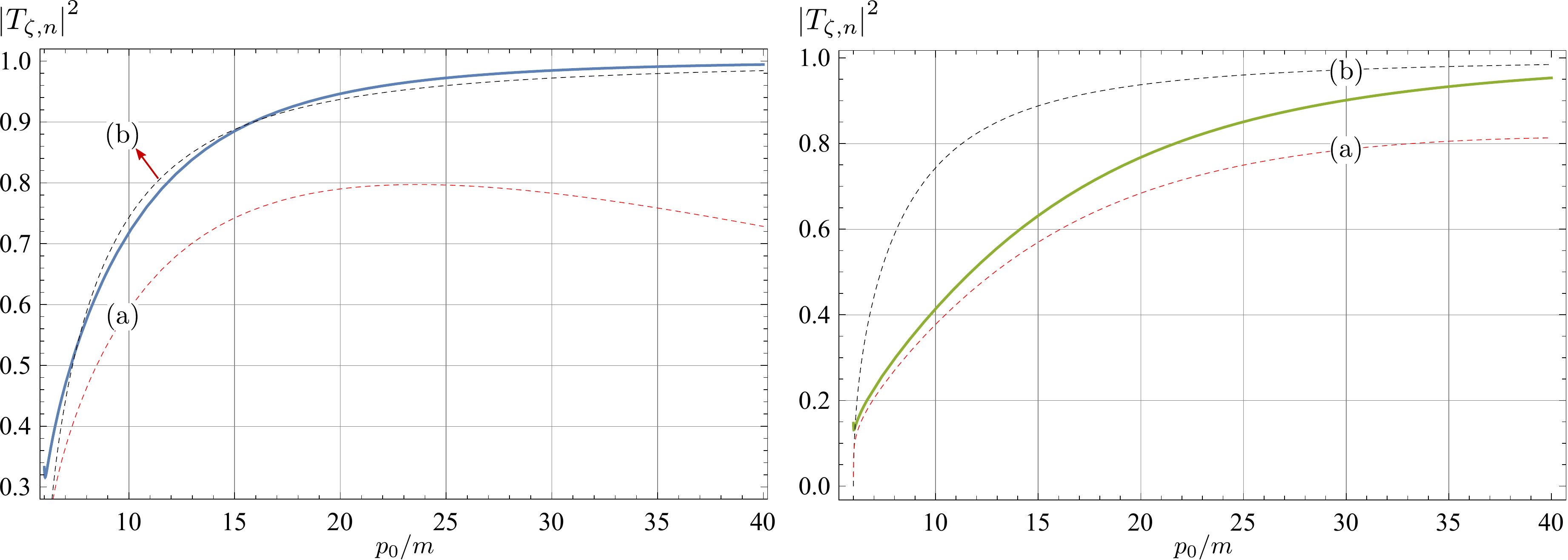}
\end{center}
\caption{Probabilities of Bosons transmission through symmetrical inverse
potential steps in $\Omega _{1}$. Exact results, given by Eqs. (\protect\ref%
{4.3}) and (\protect\ref{2.32a}), are represented by solid lines while the
asymptotic ones are represented by dashed curves. $\left( \mathrm{a}\right) $%
: approximations given by the coefficient $\left\vert g\left(
_{+}|^{-}\right) \right\vert ^{-2}$ in Eq. (\protect\ref{32.5}) and the
identity $\left\vert T_{\protect\zeta ,n}\right\vert ^{2}=\left[
1+\left\vert g\left( _{+}|^{-}\right) \right\vert ^{2}\right] ^{-1}$. $%
\left( \mathrm{b}\right) $: approximations given by Eq. (\protect\ref{4.6}).
In the left panel, $E=250E_{c}$ and $m\protect\xi _{1}=m\protect\xi %
_{2}=1/50 $ while in the right panel, $E=500E_{c}$ and $m\protect\xi _{1}=m%
\protect\xi _{2}=1/100$. The lower bound of $\Omega _{1}$ is the same for
both plots which, in this system of units, is $p_{0}/m=6$.}
\label{Fig6}
\end{figure}

Besides the differential mean numbers, there are other differential
quantities worth of consideration, such as transmission probabilities
defined beyond the Klein zone, as was mentioned in the previous section.
Thus, on the Figs. \ref{Fig5} and \ref{Fig6} we present transmission
probabilities both for scalar and Fermi cases given analytically by exact
expressions (\ref{4.3}), (\ref{2.32a}) and by appropriate asymptotic
representations discussed in Sec. \ref{Sec4}.

Thus, one can see that the asymptotic approximations agree with exact
results within a certain level of accuracy. In a sufficient wide energy
interval, on the Fig. \ref{Fig5} we see that approximations (\ref{4.5}) for
Fermions are more accurate on the right panel than on the left panel. This
is a consequence of the fact that the results on the right panel refer to an
electric field sharper than the one of the left panel, which means that the
parameter $\mathbb{U}\xi /2$ is smaller in the former case than in the
latter. According to the discussions presented in the Sec. \ref{Sec4}, the
smaller the parameter $\mathbb{U}\xi /2$ the more accurate are transmission
probabilities (\ref{4.5}). For the values considered on the Fig \ref{Fig5},
we see that $\mathbb{U}\xi /2=0.1$ on the left panel while $\mathbb{U}\xi
/2=0.05$ on the right panel. This explains why the dashed lines on the right
panel are closer to the solid lines than on the left panel. Nevertheless, it
should be noted that as the energy $p_{0}/m$ grows the parameters $%
\left\vert p^{\mathrm{L}}\right\vert \xi $ and $\left\vert p^{\mathrm{R}%
}\right\vert \xi $ increase as well, which means that approximation (\ref%
{4.5}) becomes more accurate irrespective the value of $\mathbb{U}\xi /2$,
as long as it is finite. Technically, this follows from the fact that the
asymptotic approximations for the Whittaker functions with small or large
arguments and fixed parameters notably reproduce the same approximation (\ref%
{4.5}) for Fermions.

The results for Bosons are slightly different than the ones for Fermions,
since there are two approximations: one calculated using the coefficients $%
\left\vert g\left( _{+}|^{-}\right) \right\vert ^{-2}$ given by Eq. (\ref%
{32.5}) and the identify $\left\vert T_{\zeta ,n}\right\vert ^{2}=\left[
1+\left\vert g\left( _{+}|^{-}\right) \right\vert ^{2}\right] ^{-1}$ and the
second one, given by Eq. (\ref{4.6}). That is why both panels on the Fig. %
\ref{Fig6} display two dashed lines. The dashed lines labelled by $\left( 
\mathrm{a}\right) $ refer to the approximation calculated with the help of
Eq. (\ref{32.5}) while the ones labelled with $\left( \mathrm{b}\right) $,
to Eq. (\ref{4.6}). According to the results represented on the left panel,
we see that approximations (\ref{4.6}) are much more accurate than the ones
calculated with the help of Eq. (\ref{32.5}). This is quite expected since
the electric field parameterized by values considered on that plot\ is not
\textquotedblleft sharp\textquotedblright\ enough. In other words, the
parameter $\mathbb{U}\xi /2$ is not small enough in comparison to the unity
for making results of electric fields in the sharp-gradient regime agree
with exact results. Thus, the approximation for electric fields beyond
sharp-gradient regime (\ref{4.6}) is more accurate in this case. One ought
to say that this is a peculiarity of Bosons since for Fermions the
approximations obtained for an electric field in the sharp-gradient regime (%
\ref{4.5}) are closer to exact results. The results displayed on the right
panel of Fig. \ref{Fig6} are even more interesting since the exact results
\textquotedblleft interpolate\textquotedblright\ between both
approximations. For sufficiently small energies, the approximations obtained
from Eq. (\ref{32.5}) are more accurate than the ones calculated with the
help of Eq. (\ref{4.6}) as was expected, since the parameters $\left\vert p^{%
\mathrm{L}}\right\vert \xi $, $\left\vert p^{\mathrm{R}}\right\vert \xi $
and $\mathbb{U}\xi /2$ are small enough. Thus, we may say that an electric
field parameterized by the values considered on the right panel are in the
sharp-gradient regime. However, as the energy grows, the parameters $%
\left\vert p^{\mathrm{L}}\right\vert \xi $ and $\left\vert p^{\mathrm{R}%
}\right\vert \xi $ grow so that approximation (\ref{4.6}) eventually will be
more accurate than the one obtained from Eq. (\ref{32.5}). This explains why
the exact results tend to the dashed lines $\left( \mathrm{b}\right) $ as
the energy grows. Finally, based on the numerical values on the Figs. \ref%
{Fig5} and \ref{Fig6}, we conclude that for electric fields characterized by
amplitudes $E\lesssim 250E_{c}$ and by length scales $m\xi \gtrsim 1/50$,
approximations for electric fields beyond the sharp-gradient regime (\ref%
{4.6}) agree sufficiently well with exact results for Bosons while for
Fermions accuracy approximations (\ref{4.5}) is good enough only for large
energies. As for electric fields parameterized by $E\gtrsim 500E_{c}$ and $%
m\xi \lesssim 1/100$, the field is sharp enough so that approximations (\ref%
{4.5}) agree sufficiently well over a wide range of energies for Fermions.
For Bosons, approximations of sharp-gradient regime work sufficiently well
only for small energies. For large energies, approximations (\ref{4.6}) must
be considered instead. Note that similar results can be obtained for
negative energies in $\Omega _{5}$. Lastly, one may arrive at the same
conclusions studying reflection probabilities instead of transmission
probabilities, either in $\Omega _{1}$ or in $\Omega _{5}$.

\section{Role of \textquotedblleft growing\textquotedblright\ \&
\textquotedblleft decaying\textquotedblright\ processes in the vacuum
instability\label{Sec7}}

As an application of the above results, one can analyze contributions from
\textquotedblleft growing\textquotedblright\ and \textquotedblleft
decaying\textquotedblright\ areas that accompany the arising uniform in
space constant electric field to the vacuum instability. To this end, we
consider here a field configuration composed of three independent regions,
growing inversely squared in the first region $x\in \mathrm{I}=\left(
-\infty ,x_{\mathrm{L}}\right) $, remaining constant in the second region $%
x\in \mathrm{Int}=\left[ x_{\mathrm{L}},x_{\mathrm{R}}\right] ,$\ and
decreasing inversely squared in the last region $x\in \mathrm{II}=\left( x_{%
\mathrm{R}},+\infty \right) $. Such a field--hereafter referred by composite
electric field--corresponds to a generalization of the inverse-square
electric field (\ref{2.2}) by having, instead of a peak at $x=0$, an
intermediate region. Afterward, we compare the vacuum instability caused by
the composite electric field with the one which is caused by the electric
field that \textquotedblleft suddenly grows\textquotedblright\ and
\textquotedblleft suddenly decays\textquotedblright\ at precise positions on
the $x$-axis, say at $x=x_{\mathrm{L}}$\ and $x=x_{\mathrm{R}}$,
respectively.

Let us consider the $L$-constant electric field \cite{GavGit16b}%
\begin{equation}
E\left( x\right) =E\left\{ 
\begin{array}{ll}
0\,, & x\in \mathrm{I}\, \\ 
1\,, & x\in \mathrm{Int}\, \\ 
0\,, & x\in \mathrm{II}\,%
\end{array}%
\right. ,  \label{7.1}
\end{equation}%
which, by the definition, \textquotedblleft grows\textquotedblright\ and
\textquotedblleft decays\textquotedblright\ sharply\ at $x=x_{\mathrm{L}}$
and $x=x_{\mathrm{R}}$, respectively, and the composite electric field%
\begin{equation}
E\left( x\right) =E\left\{ 
\begin{array}{ll}
\left[ 1-\left( x-x_{\mathrm{L}}\right) /\xi _{1}\right] ^{-2}\,, & x\in 
\mathrm{I}\, \\ 
1\,, & x\in \mathrm{Int}\, \\ 
\left[ 1+\left( x-x_{\mathrm{R}}\right) /\xi _{2}\right] ^{-2}\,, & x\in 
\mathrm{II}\,%
\end{array}%
\right. ,  \label{7.2}
\end{equation}%
whose dependence on $x$, within\textrm{\ }$\mathrm{I}$\textrm{\ }and\textrm{%
\ }$\mathrm{II}$\textrm{, }simulate \textquotedblleft
growing\textquotedblright\ and \textquotedblleft decaying\textquotedblright\
processes. For both cases, the constants $x_{\mathrm{L}}=-L/2<0$ and $x_{%
\mathrm{R}}=L/2>0$ sets the spatial region where the field is constant and
shall be the same for both fields. Potential energies of an electron in each
field are:%
\begin{equation}
U\left( x\right) =eE\left\{ 
\begin{array}{ll}
x_{\mathrm{L}}\,, & x\in \mathrm{I}\, \\ 
x\,, & x\in \mathrm{Int}\, \\ 
x_{\mathrm{R}}\,, & x\in \mathrm{II}\,%
\end{array}%
\right. ,  \label{7.3}
\end{equation}%
and%
\begin{equation}
U\left( x\right) =eE\left\{ 
\begin{array}{ll}
-\left( \xi _{1}-x_{\mathrm{L}}\right) +\xi _{1}\left[ 1-\left( x-x_{\mathrm{%
L}}\right) /\xi _{1}\right] ^{-1}\,, & x\in \mathrm{I}\, \\ 
x\,, & x\in \mathrm{Int}\, \\ 
\xi _{2}+x_{\mathrm{R}}-\xi _{2}\left[ 1+\left( x-x_{\mathrm{R}}\right) /\xi
_{2}\right] ^{-1}\,, & x\in \mathrm{II}\,%
\end{array}%
\right. ,  \label{7.4}
\end{equation}%
respectively.

We are interested in differential quantities characterizing vacuum
instability, in particular, mean numbers $N_{n}^{\mathrm{cr}}$. To this end,
it is enough to analyze relevant $g$-coefficients within the Klein zone for
both examples. For the $L$-constant field (\ref{7.1}), necessary
coefficients have been calculated before; see Eqs. (2.27) in Ref. \cite%
{GavGit16b}. As for the composite field (\ref{7.2}), it should noted that
the existence of the intermediate region $\mathrm{Int}=\left[ x_{\mathrm{L}%
},x_{\mathrm{R}}\right] $ increase the magnitude of the step%
\begin{equation}
\mathbb{U}=eE\left( \xi _{1}+\xi _{2}+L\right) \,,  \label{7.5}
\end{equation}%
which, in turn, modifies asymptotic kinetic energies $\pi _{0}\left( \mathrm{%
L/R}\right) $ (\ref{2.12}) and variables $z_{j}\left( x\right) $ (\ref{2.11}%
) only by additive constants%
\begin{eqnarray}
\pi _{0}\left( \mathrm{L}\right) &=&p_{0}+eE\left( \xi _{1}-x_{\mathrm{L}%
}\right) \,,\ \ z_{1}\left( x\right) =2i\left\vert p^{\mathrm{L}}\right\vert %
\left[ \xi _{1}-\left( x-x_{\mathrm{L}}\right) \right] \,,\ \ x\in \mathrm{I}%
\,,  \notag \\
\pi _{0}\left( \mathrm{R}\right) &=&p_{0}-eE\left( \xi _{2}+x_{\mathrm{R}%
}\right) \,,\ \ z_{2}\left( x\right) =2i\left\vert p^{\mathrm{R}}\right\vert %
\left[ \xi _{2}+\left( x-x_{\mathrm{R}}\right) \right] \,,\ \ x\in \mathrm{II%
}\,,  \label{7.6}
\end{eqnarray}%
while the asymptotic momenta $\left\vert p^{\mathrm{L/R}}\right\vert $ and
parameters $\kappa _{j}$, $\mu _{j}$ are defined in the same way as in Eqs. (%
\ref{2.12}), (\ref{2.15})%
\begin{eqnarray}
&&\left\vert p^{\mathrm{L/R}}\right\vert =\sqrt{\pi _{0}\left( \mathrm{L/R}%
\right) ^{2}-\pi _{\perp }^{2}}\,,\ \ \mu _{j}=\left( -1\right) ^{j}\left(
ieE\xi _{j}^{2}-\chi /2\right) \,,  \notag \\
&&\kappa _{1}=ieE\xi _{1}^{2}\frac{\pi _{0}\left( \mathrm{L}\right) }{%
\left\vert p^{\mathrm{L}}\right\vert }\,,\ \ \kappa _{2}=-ieE\xi _{2}^{2}%
\frac{\pi _{0}\left( \mathrm{R}\right) }{\left\vert p^{\mathrm{R}%
}\right\vert }\,,  \label{7.7}
\end{eqnarray}%
but with $\pi _{0}\left( \mathrm{L/R}\right) $ given by Eqs. (\ref{7.6}).
The above modifications does not interfere on asymptotic properties of the
solutions in the intervals $\mathrm{I}$, \textrm{II} and therefore does not
change the classification of solutions with special left and right
asymptotics. Hence, the exact solutions of wave equations for the intervals $%
\mathrm{I}$, $\mathrm{II}$ are Whittaker functions, classified according to
Eqs. (\ref{2.21}) and whose arguments and parameters are given by Eqs. (\ref%
{7.6}), (\ref{7.7}).

As for the intermediate region $x\in \mathrm{Int}$, Dirac spinors (or KG
wave functions) are proportional to Weber Parabolic Cylinder functions
(WPCF) \cite{Erdelyi} once general solutions of the second-order
differential equation (\ref{2.10}) are expressed in terms of these functions 
\cite{GavGit16b}%
\begin{equation}
\varphi _{n}\left( x\right) =\alpha _{+}u_{+}\left( \mathfrak{Z}\left(
x\right) \right) +\alpha _{-}u_{-}\left( \mathfrak{Z}\left( x\right) \right)
\,,\ \ x\in \mathrm{Int}\,.  \label{7.8}
\end{equation}%
Here $u_{+}\left( \mathfrak{Z}\right) =D_{\rho }\left( \mathfrak{Z}\right) $
and $u_{-}\left( \mathfrak{Z}\right) =D_{-\rho -1}\left( i\mathfrak{Z}%
\right) $ are WPCF while $\alpha _{\pm }$ are arbitrary constants. The
argument $\mathfrak{Z}$ and parameter $\rho $ are defined as%
\begin{equation}
\mathfrak{Z}\left( x\right) =\left( 1-i\right) \left( \sqrt{eE}x-\frac{p_{0}%
}{\sqrt{eE}}\right) \,,\ \ \rho =-\nu -\frac{\chi +1}{2}\,,\ \ \nu =\frac{%
i\lambda }{2}\,.  \label{7.9}
\end{equation}%
Thus, with the aid of (\ref{7.8}) and the solutions for the intervals 
\textrm{I}, \textrm{II} (\ref{2.21})\ (with the substitutions described
above), one may demand continuity of the wave functions and its derivatives
at $x=x_{\mathrm{L}}$ and $x=x_{\mathrm{R}}$ (similarly to the derivation of
Eqs. (\ref{2.30}), (\ref{2.31})) to obtain the following form for the
coefficient $g\left( _{+}|^{-}\right) $:%
\begin{eqnarray}
g\left( _{+}|^{-}\right) &=&\eta _{\mathrm{L}}\sqrt{\frac{\left\vert \pi
_{0}\left( \mathrm{L}\right) -\chi \left\vert p^{\mathrm{L}}\right\vert
\right\vert }{8eE\left\vert p^{\mathrm{L}}\right\vert \left\vert \pi
_{0}\left( \mathrm{R}\right) +\chi \left\vert p^{\mathrm{R}}\right\vert
\right\vert \left\vert p^{\mathrm{R}}\right\vert }}\exp \left[ -\frac{i\pi }{%
2}\left( \kappa _{1}+\kappa _{2}+\nu +\frac{\chi }{2}\right) \right]  \notag
\\
&\times &\left[ \mathcal{F}_{2}^{-}\left( x_{2}\right) \mathcal{G}%
_{1}^{+}\left( x_{1}\right) -\mathcal{F}_{2}^{+}\left( x_{2}\right) \mathcal{%
G}_{1}^{-}\left( x_{1}\right) \right] \,,  \label{7.10}
\end{eqnarray}%
in which%
\begin{eqnarray}
&&\mathcal{G}_{j}^{\pm }\left( x\right) =u_{\pm }\left( \mathfrak{Z}\right) 
\frac{d}{dx}W_{-\kappa _{j},\mu _{j}}\left( e^{-i\pi }z_{j}\right)
-W_{-\kappa _{j},\mu _{j}}\left( e^{-i\pi }z_{j}\right) \frac{d}{dx}u_{\pm
}\left( \mathfrak{Z}\right) \,,  \notag \\
&&\mathcal{F}_{j}^{\pm }\left( x\right) =u_{\pm }\left( \mathfrak{Z}\right) 
\frac{d}{dx}W_{\kappa _{j},\mu _{j}}\left( z_{j}\right) -W_{\kappa _{j},\mu
_{j}}\left( z_{j}\right) \frac{d}{dx}u_{\pm }\left( \mathfrak{Z}\right) \,.
\label{7.11}
\end{eqnarray}

\begin{figure}[th]
\begin{center}
\includegraphics[scale=0.45]{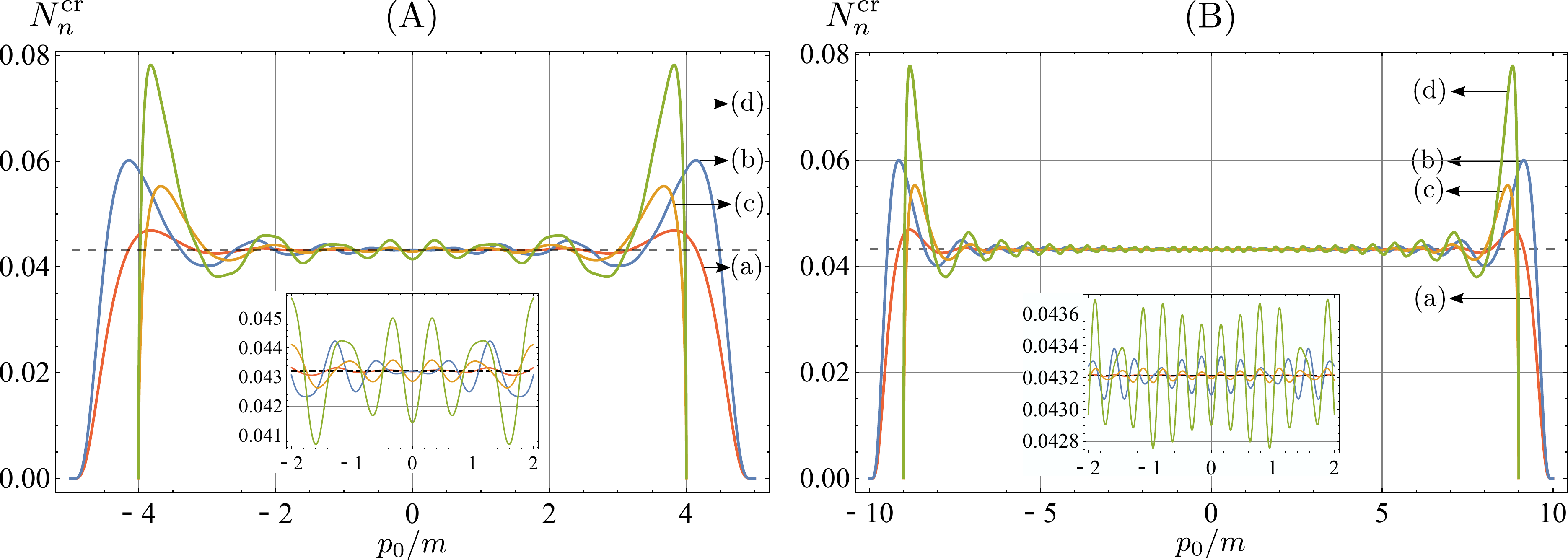}
\end{center}
\caption{Differential mean numbers $N_{n}^{\mathrm{cr}} $ of particles
created from the vacuum by the\ composite electric field (\protect\ref{7.2})
(solid lines (a), (b)) and by the $L$-constant electric field (\protect\ref%
{7.1}) (solid lines (c), (d)), with field amplitudes $E=E_{c}$ and length
scales $m\protect\xi _{1}=m\protect\xi _{2}=1$. In the left panel (A), $%
mL=10 $, while in the right panel (B), $mL=20$. The distributions have
different energy ranges because the extent of the Klein zone depends on the
external field under consideration: for the $L$-constant field, $\left\vert
p_{0}\right\vert /m\leq 4$ in (A) and $\left\vert p_{0}\right\vert /m\leq 9$
in (B) while for the composite field, $\left\vert p_{0}\right\vert /m\leq 5$
in (A) and $\left\vert p_{0}\right\vert /m\leq 10$ in (B). The horizontal
dashed lines denotes the uniform distribution $e^{-\protect\pi }$.}
\label{Fig7.1}
\end{figure}

\begin{figure}[th]
\begin{center}
\includegraphics[scale=0.45]{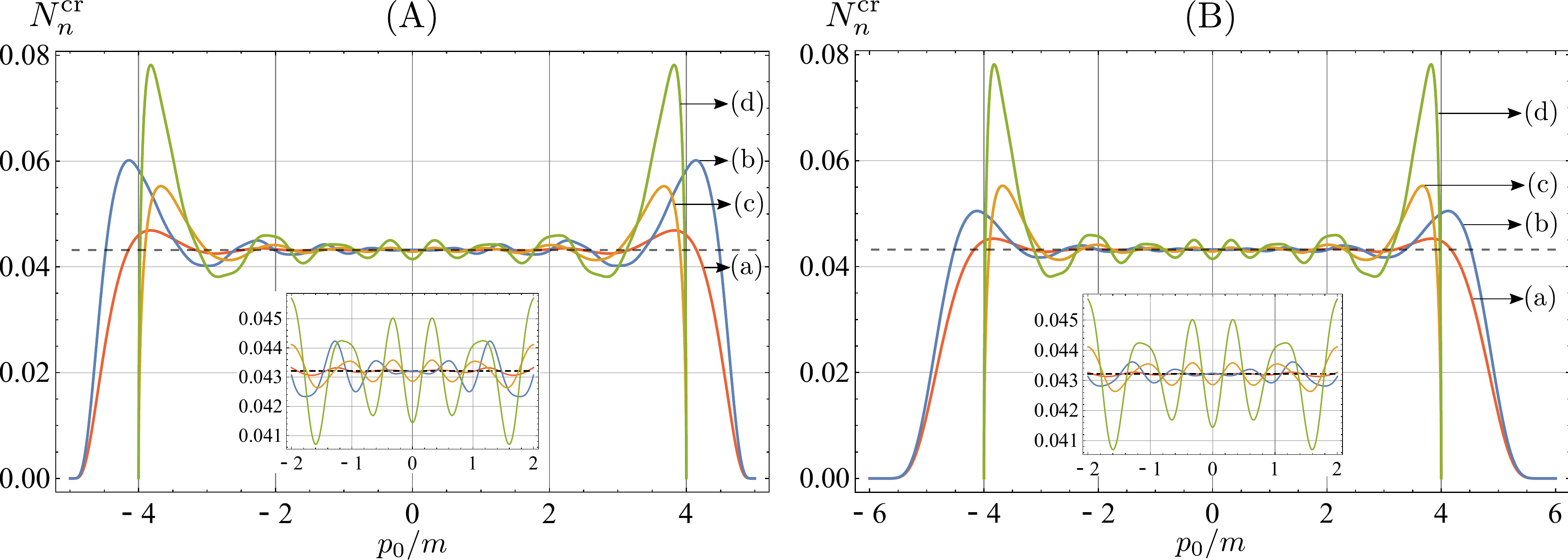}
\end{center}
\caption{Differential mean numbers $N_{n}^{\mathrm{cr}}$ of particles
created from the vacuum by the\ composite electric field (\protect\ref{7.2})
(solid lines (a), (b)) and by the $L$-constant electric field (\protect\ref%
{7.1}) (solid lines (c), (d)), with field amplitudes $E=E_{c}$ and fixed
length scale $mL=10$. In the left panel (A), $m\protect\xi _{1}=m\protect\xi %
_{2}=1$, while in the right panel (B), $m\protect\xi _{1}=m\protect\xi %
_{2}=2 $. The distributions have different energy ranges because the extent
of the Klein zone depends on the external field under consideration: for the 
$L$-constant field, $\left\vert p_{0}\right\vert /m\leq 4$ in both panels
while for the composite field, $\left\vert p_{0}\right\vert /m\leq 5$ in (A)
and $\left\vert p_{0}\right\vert /m\leq 6$ in (B). The horizontal dashed
lines denotes the uniform distribution $e^{-\protect\pi }$.}
\label{Fig7.2}
\end{figure}

\begin{figure}[th]
\begin{center}
\includegraphics[scale=0.45]{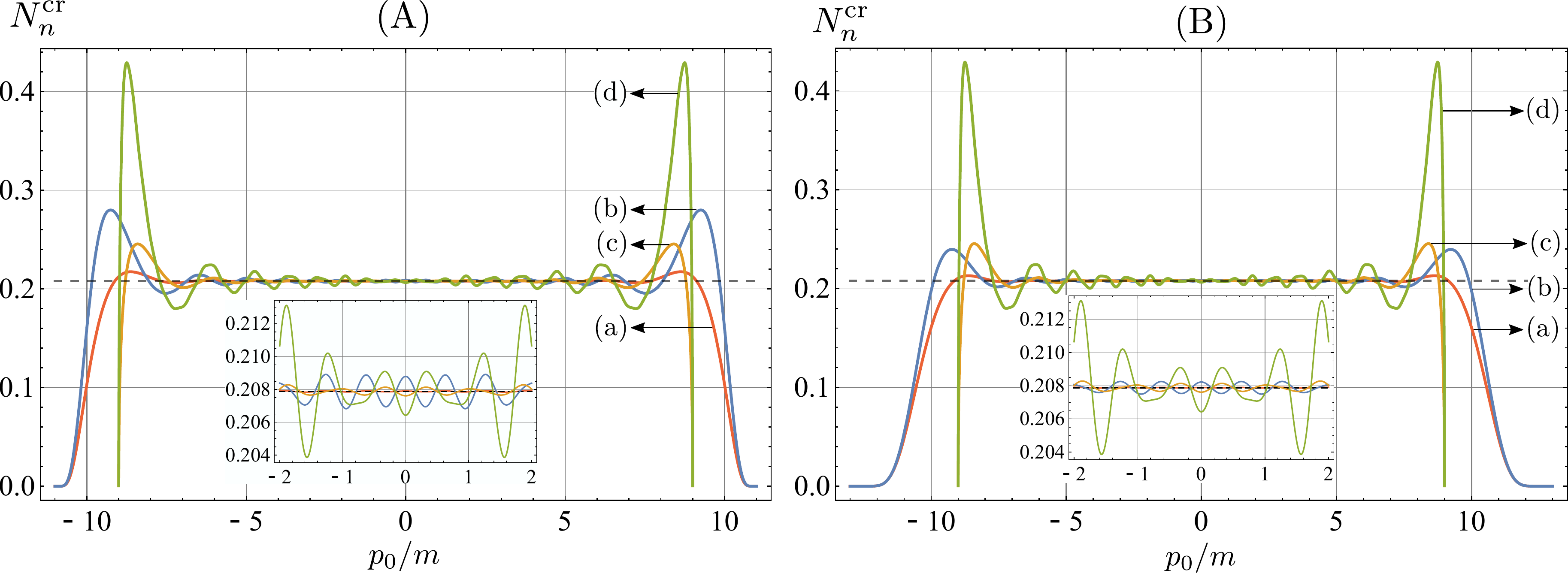}
\end{center}
\caption{Differential mean numbers $N_{n}^{\mathrm{cr}}$ of particles
created from the vacuum by the\ composite electric field (\protect\ref{7.2})
(solid lines (a), (b)) and by the $L$-constant electric field (\protect\ref%
{7.1}) (solid lines (c), (d)) with fixed field amplitudes $E=2E_{c}$ and
length $mL=10$. In the left panel (A), $m\protect\xi _{1}=m\protect\xi %
_{2}=1 $ while in the right panel (B), $m\protect\xi _{1}=m\protect\xi %
_{2}=2 $. The distributions have different energy ranges because the Klein
zone extent depends on the external field under consideration: for the $L$%
-constant field, $\left\vert p_{0}\right\vert /m\leq 9$ in both panels while
for the composite field, $\left\vert p_{0}\right\vert /m\leq 11$ in (A) and $%
\left\vert p_{0}\right\vert /m\leq 13$ in (B). The horizontal dashed lines
denotes the uniform distribution, $e^{-\protect\pi /2}$.}
\label{Fig7.3}
\end{figure}

Henceforward, we shall compare mean numbers of particles created from the
vacuum $N_{n}^{\mathrm{cr}}$ by the $L$-constant electric field (\ref{7.1})
and by the composite electric field (\ref{7.2}) in situations whose
intermediate length $L$ and the field amplitude $E$ are larger than the
stabilization characteristic number $\max \left( 1,m^{2}/eE\right) $, namely 
$\sqrt{eE}L>\max \left( 1,m^{2}/eE\right) $. The remaining parameters,
related to the length scales $\xi _{j}$, are finite. These configurations
allow us to compare and analyze how the mean numbers approach or deviate
from the uniform distribution $e^{-\pi \lambda }$, as the field is supplied
by growing and decaying processes. Figures \ref{Fig7.1}, \ref{Fig7.2} and %
\ref{Fig7.3} display exact mean numbers $N_{n}^{\mathrm{cr}}$, as a function
of the energy $p_{0}$, corresponding to the $L$-constant field (solid lines,
(c) and (d)) and to the composite electric field (solid lines, (a) and (b))
for some values of the parameters $\sqrt{eE}L$, $\sqrt{eE}\xi _{j}$, $%
E/E_{c} $. For the computation of the mean numbers, we use the coefficient
given by Eq. (2.27) in Ref. \cite{GavGit16b} for the $L$-constant field
while Eq. (\ref{7.10}) for the composite field. Energies, length scales and
field amplitudes $E$ are expressed in units of the electron mass $m$ and
Schwinger's critical field $E_{c}$, respectively. Moreover, we keep the same
conventions employed in Sec. \ref{Sec5}, namely we set $\mathbf{p}_{\perp
}=0 $ and work with the system of units where $\hslash =c=m=1$.

Within the range of values chosen for parameters associated with length
scales $L,\xi _{j}$, and field amplitude $E$, we observe that the mean
numbers of particles created $N_{n}^{\mathrm{cr}}$ oscillate around the
uniform distribution $e^{-\pi \lambda }$ in all cases. This is a direct
consequence of the parameter $\sqrt{eE}L$ being larger than the
stabilization characteristic number $\max \left( 1,m^{2}/eE\right) $, which
is reduced to the unity here because $E\geq E_{c}$. According to general
results obtained for the $L$-constant field \cite{GavGit16b}, the mean
number stabilizes to the uniform distribution $e^{-\pi \lambda }$ provided $%
\sqrt{eE}L$ is sufficiently large; the larger its value, the closer from the
uniform result. This is particularly clear in the plots of Fig. \ref{Fig7.1}%
, in which the parameters $m\xi _{1}=m\xi _{2}$ and the field amplitude $E$
are fixed but the length $L$ of the intermediate interval is smaller in (A)
than in (B).

To understand the role of growing and decaying length scales $\xi _{j}$ on
the mean numbers $N_{n}^{\mathrm{cr}}$, we compare results between external
fields in two additional configurations, both having the same length of the
intermediate region $L$ but field amplitudes $E$ and length scales $\xi _{j}$
assuming the following values: $E=E_{c}$, $m\xi _{1}=m\xi _{2}=1$ and $m\xi
_{1}=m\xi _{2}=2$ in Fig. \ref{Fig7.2} while $E=2E_{c}$, $m\xi _{1}=m\xi
_{2}=1$ and $m\xi _{1}=m\xi _{2}=2$ in Fig. \ref{Fig7.3}. According to the
results in Fig. \ref{Fig7.2}, we observe that increasing the length scales $%
\xi _{j}$ lead to results closer to the uniform distribution, as it can be
seen comparing the amplitude of oscillations of the lines (a), (b) in the
left panel (A) with those in the right panel (B), both in Fig. \ref{Fig7.2}.
This feature does not depend on the amplitude of the electric field, as it
occurs for different values of the field amplitudes; cf. lines (a) and (b)
in the left panel with the ones in the right panel in Fig. \ref{Fig7.3}.
Moreover, comparing results from the composite field (lines (a) and (b))
with the ones from the $L$-constant field (lines (c) and (d)) we see that
the former are closer to the uniform distribution $e^{-\pi \lambda }$ than
the latter, irrespective the field amplitude $E$ or length scales $\xi _{j}$%
. In other words, results from composite fields present smaller--in
amplitude--oscillations around $e^{-\pi \lambda }$ as compared to results
from the $L$-constant field. Such a feature can be seen in all cases
displayed in Figs. \ref{Fig7.1}, \ref{Fig7.2}, \ref{Fig7.3} and, besides,
does not depend on particle's statistics, since it occurs both for Fermions
as for Bosons. Furthermore, we also observe that increasing the amplitude of
the electric field $E$ leads to results closer to the uniform distribution,
as it can be seen comparing plots in the left panel of Fig. \ref{Fig7.2}
with those in the left panel of Fig. \ref{Fig7.3}. The same can be concluded
comparing right panels. This feature occurs for both external fields.

These results allow us to conclude that growing and decaying processes plays
a significant role in the stabilization process of differential quantities,
once mean numbers resulting from external fields supplied by growing and
decaying processes reach the stabilization distribution $e^{-\pi \lambda }$\
more accurately and, as a matter of fact, in a wider range of energies than
fields deprived of such processes. The explanation for these results stems
from stabilization conditions associated with the composite field (\ref{7.2}%
) and with the $L$-constant field (\ref{7.1}) being different for the same
length $L$ of the intermediate region. Recalling that the stabilization
condition for the $L$-constant field is $\sqrt{eE}L\gg \max \left(
1,m^{2}/eE\right) $ \cite{GavGit16b} and rephrasing it as%
\begin{equation}
\sqrt{\mathbb{U}L}\gg \max \left( 1,\frac{m^{2}}{eE}\right) \,,\ \ \mathbb{U}%
=eEL\,,  \label{7.12}
\end{equation}%
we realize that when applied to symmetric composite fields $\xi _{1}=\xi
_{2} $, it leads to a number $\sqrt{\mathbb{U}\tilde{L}}$ larger than the
lhs. of (\ref{7.2}) because the length $\tilde{L}\equiv 2\xi +L$ and the
magnitude of the step $\mathbb{U}$, given by Eq. (\ref{7.5}), in this case
are larger than simply $\sqrt{eE}L$. Therefore, it is not unexpected that
the mean numbers of pairs created by the composite field $N_{n}^{\mathrm{cr}}
$ are closer to uniform distribution than the ones created by the $L$%
-constant field, provided $L$ is the same for both fields. At last, but not
least, it is worth pointing out that features similar to the ones above
discussed also occur for time-dependent composite electric fields, as
reported by us previously in \cite{AFGG19,AdoGavGit18}.

\section{Concluding remarks\label{Sec6}}

Using nonperturbative approach developed in QED with $x$-electric potential
steps \cite{x-case}, we have calculated elementary zero-order processes
(with respect to radiative interaction) in inverse-square critical electric
fields. Quantities characterizing the vacuum instability and particle
scattering are studied in every detail, in particular, in situations where
the external field fits in small-gradient and sharp-gradient configurations.
The calculations were done in the Klein zone and beyond, mostly in the
ranges $\Omega _{1}$ and $\Omega _{5}$. The processes considered in the
Klein zone includes not only differential mean numbers, total mean numbers
and vacuum-to-vacuum transition probabilities, but also relative amplitudes
of particle scattering, pair creation and pair annihilation. Results
obtained for configurations in the small-gradient regime are consistent with
universal expressions for total quantities in arbitrary weakly inhomogeneous 
$x$-electric potential steps. For configurations in the sharp-gradient
regime, differential quantities are consistent with ones obtained for
another exactly-solvable examples in the same regime, such as the Sauter
electric field and the Peak electric field. Moreover, computing total
quantities within the Klein zone allow us to extract the imaginary part of
the QED effective action and confirm, in particular, the universal behavior
of pair creation near the criticality, obtained previously in the
literature. For sharp-gradient fields, we still studied the nonrelativistic
limit of reflection and transmission coefficients beyond the Klein zone
whose expressions can be compared, once calculated, with results obtained in
scattering problems by inverse-square electric fields in nonrelativistic
Quantum Mechanics. We also commented on the absence of the Klein paradox for
inverse-square electric fields in sharp-gradient configurations.

Comparing exact results with asymptotic approximations allow us to study
parameters characterizing fields in the small-gradient or sharp-gradient
regime, as well as the accuracy of the approximations. For fields in the
small-gradient regime, the asymptotic approximations agree sufficiently well
with exact results over a wide range of energies in the Klein zone. There
are no significant differences between results for Fermions and Bosons in
this case. This is not true for fields in the sharp-gradient regime, since
approximations in the Fermi case are more accurate than in the scalar case
in general. However, the accuracy of all approximations increase as the
field amplitudes $E$ and the length scales $\xi _{j}$ decrease. Studying
transmission probabilities beyond the Klein zone, we find that
approximations for fields in the sharp-gradient regime agree sufficiently
well with exact results provided the amplitudes $E$ are large enough and the
length scales $\xi _{j}$ small enough. For the scalar case, approximations
for the sharp-gradient regime represent better results only for small
energies while for large energies approximations beyond the sharp-gradient
regime leads to more accurate results.

We study the role of growing and decaying processes in the vacuum
instability considering various electric field configuration, composed by
inverse-square fields and by an $x$-independent electric field between them.
Using exact expressions for differential quantities, we compute mean numbers
for cases whose length scales $L$ and field amplitudes $E$ are sufficiently
large. Comparing results obtained for the composite field and for the $L$%
-constant field with the uniform distribution $e^{-\pi \lambda }$, we
conclude that external fields supplied by growing and decaying processes
leads to mean numbers closer to the stabilization distribution and in a
wider range of energies than fields deprived of such processes. These
results are supported by the fact that parameters characterizing the
stabilization condition are larger for composite fields than for $L$%
-constant fields.

We conclude this work emphasizing that inverse-square electric field is an
additional example of external background where all characteristics
underlying vacuum instability and particle scattering can be performed
exactly. We believe that exact results presented here may be useful in
studies of particle creation by electric fields of more complex spatial
distributions, which may not be exactly solvable but decreasing spatially as
the inverse-square electric field.

\section{Acknowledgements}

We acknowledge the support from the Russian Science Foundation, project
number 19-12-00042. We are grateful to Greger Torgrimsson who drew our
attention to the work \cite{cr-regime2}.

\appendix%

\section{Some $g$-coefficients and their asymptotic representations\label%
{App1}}

In this Appendix, we list exact expressions for the coefficients $g\left(
_{+}|^{-}\right) $ and $g\left( _{+}|^{+}\right) $ in terms of Whittaker
functions as well as some asymptotic representations of CHF that that may be
useful in the study of differential quantities, complementary to the ones
discussed in our previous publication \cite{AdoGavGit18}.

Substituting the relations (\ref{2.14}) in Eqs. (\ref{2.30}) and (\ref{2.32a}%
), the coefficients $g\left( _{+}|^{-}\right) $, $g\left( _{+}|^{+}\right) $
for Fermions acquires the following representation%
\begin{eqnarray}
g\left( _{+}|^{-}\right) &=&\eta _{\mathrm{L}}\sqrt{\frac{\left\vert \pi
_{0}\left( \mathrm{L}\right) -\chi \left\vert p^{\mathrm{L}}\right\vert
\right\vert }{\left\vert p^{\mathrm{R}}\right\vert \left\vert \pi _{0}\left( 
\mathrm{R}\right) +\chi \left\vert p^{\mathrm{R}}\right\vert \right\vert
\left\vert p^{\mathrm{L}}\right\vert }}e^{-i\pi \left( \kappa _{1}+\kappa
_{2}\right) /2}\tilde{\Delta}\left( _{+}|^{-}\right) \left( 0\right) \,, 
\notag \\
g\left( _{+}|^{+}\right) &=&\eta _{\mathrm{L}}\sqrt{\frac{\left\vert \pi
_{0}\left( \mathrm{L}\right) -\chi \left\vert p^{\mathrm{L}}\right\vert
\right\vert }{\left\vert p^{\mathrm{R}}\right\vert \left\vert \pi _{0}\left( 
\mathrm{R}\right) -\chi \left\vert p^{\mathrm{R}}\right\vert \right\vert
\left\vert p^{\mathrm{L}}\right\vert }}e^{-i\pi \left( \kappa _{1}+\kappa
_{2}\right) /2}\tilde{\Delta}\left( _{+}|^{+}\right) \left( 0\right) \,,
\label{app1.0a}
\end{eqnarray}%
where%
\begin{eqnarray*}
\tilde{\Delta}\left( _{+}|^{-}\right) \left( x\right) &=&\left\vert p^{%
\mathrm{L}}\right\vert W_{\kappa _{2},\mu _{2}}\left( z_{2}\right) \frac{d}{%
dz_{1}}W_{-\kappa _{1},\mu _{1}}\left( e^{-i\pi }z_{1}\right) \\
&+&\left\vert p^{\mathrm{R}}\right\vert W_{-\kappa _{1},\mu _{1}}\left(
e^{-i\pi }z_{1}\right) \frac{d}{dz_{2}}W_{\kappa _{2},\mu _{2}}\left(
z_{2}\right) \,, \\
\tilde{\Delta}\left( _{+}|^{+}\right) \left( x\right) &=&\left\vert p^{%
\mathrm{L}}\right\vert W_{-\kappa _{2},\mu _{2}}\left( e^{-i\pi
}z_{2}\right) \frac{d}{dz_{1}}W_{-\kappa _{1},\mu _{1}}\left( e^{-i\pi
}z_{1}\right) \\
&+&\left\vert p^{\mathrm{R}}\right\vert W_{-\kappa _{1},\mu _{1}}\left(
e^{-i\pi }z_{1}\right) \frac{d}{dz_{2}}W_{-\kappa _{2},\mu _{2}}\left(
e^{-i\pi }z_{2}\right) \,.
\end{eqnarray*}%
The corresponding expressions for Bosons read%
\begin{equation}
g\left( _{+}|^{\zeta }\right) =\frac{e^{-i\pi \left( \kappa _{1}+\kappa
_{2}\right) /2}}{\sqrt{\left\vert p^{\mathrm{R}}\right\vert \left\vert p^{%
\mathrm{L}}\right\vert }}\left. \tilde{\Delta}\left( _{+}|^{\zeta }\right)
\left( 0\right) \right\vert _{\chi =0}\,.  \label{app1.0b}
\end{equation}

The above representations are particularly useful for obtaining approximate
expressions for differential quantities when the argument of the Whittaker
functions are small. For example, using the connection formulae \cite%
{Buchholz}%
\begin{eqnarray}
W_{\kappa ,\mu }\left( z\right) &=&\frac{\pi }{\sin 2\pi \mu }\left[ -\frac{%
M_{\kappa ,\mu }\left( z\right) }{\Gamma \left( \frac{1}{2}-\mu -\kappa
\right) \Gamma \left( 1+2\mu \right) }+\frac{M_{\kappa ,-\mu }\left(
z\right) }{\Gamma \left( \frac{1}{2}+\mu -\kappa \right) \Gamma \left(
1-2\mu \right) }\right] \,,  \notag \\
W_{-\kappa ,\mu }\left( e^{\pm i\pi }z\right) &=&\frac{\pi }{\sin 2\pi \mu }%
\left[ -\frac{\exp \left[ \pm i\pi \left( \mu +1/2\right) \right] M_{\kappa
,\mu }\left( z\right) }{\Gamma \left( \frac{1}{2}-\mu +\kappa \right) \Gamma
\left( 1+2\mu \right) }+\frac{\exp \left[ \pm i\pi \left( -\mu +1/2\right) %
\right] M_{\kappa ,-\mu }\left( z\right) }{\Gamma \left( \frac{1}{2}+\mu
+\kappa \right) \Gamma \left( 1-2\mu \right) }\right] \,,  \label{app1.0c}
\end{eqnarray}%
and the power-series expansion of the Whittaker functions regular at the
origin \newline
$M_{\kappa ,\mu }\left( z\right) =z^{\mu +1/2}\left[ 1-z\kappa /\left(
1+2\mu \right) +O\left( z^{2}\right) \right] $, one can obtain approximate
expressions for the Whittaker functions $W_{\kappa ,\mu }\left( z\right)
,W_{-\kappa ,\mu }\left( e^{\pm i\pi }z\right) $ near the origin, specially
when $\kappa $ and $\mu $ are fixed\footnote{%
A number of limiting forms for $\left. W_{\kappa ,\mu }\left( z\right)
\right\vert _{z\rightarrow 0}$ derived from Eqs. (\ref{app1.0c}) can be
found in \cite{NIST}.}. Selecting $\chi =+1$ and considering $\Delta
U_{j}\xi _{j}\ll 1$, one can expand the Gamma functions and exponents to
show that%
\begin{equation}
W_{\kappa ,\mu }\left( z\right) \approx 1\,,\ \ W_{-\kappa ,\mu }\left(
e^{-i\pi }z\right) \approx 1\,,  \label{app1.5}
\end{equation}%
in leading-order approximation. These approximations are useful for
Fermions. Under the same conditions, choosing $\chi =0$ one finds%
\begin{equation}
W_{\kappa ,\mu }\left( z\right) \approx \sqrt{\frac{z}{\pi }}\left( -\ln
z+\psi \left( 1\right) +\ln 4\right) \,,  \label{app1.4}
\end{equation}%
in leading-order approximation. These approximations are useful for Bosons.
Here $-\psi \left( 1\right) \approx 0.577$ is Euler's constant.

For large $a$, $\left\vert \arg a\right\vert \leq \pi -0^{+}$, and fixed $c$%
, $z$, the uniform asymptotic representations for the CHF $\Psi \left(
a,c;z\right) $ are given in terms of modified Bessel functions of the second
kind $K_{\nu }\left( z\right) $, as given by Eq. (13.8.11) in \cite{NIST}.
Using these approximations for $z$ small and $az$ fixed, the CHF and its
derivative are approximately given by%
\begin{eqnarray}
&&\Psi \left( a,c;z\right) \sim 2\left( \frac{z}{a}\right) ^{\left(
1-c\right) /2}\frac{e^{z/2}}{\Gamma \left( a\right) }K_{c-1}\left( 2\sqrt{az}%
\right) \,,  \notag \\
&&\frac{d}{dz}\Psi \left( a,c;z\right) \sim -2\left( \frac{z}{a}\right)
^{-c/2}\frac{e^{z/2}}{\Gamma \left( a\right) }K_{c}\left( 2\sqrt{az}\right)
\,,  \label{app1.1}
\end{eqnarray}%
in leading-order approximation.

For $z$ small and $a,c$ fixed, one can use Kummer connection formula \cite%
{NIST}%
\begin{equation}
\Psi \left( a,c;z\right) =\frac{\Gamma \left( 1-c\right) }{\Gamma \left(
a-c+1\right) }\Phi \left( a,c;z\right) +\frac{\Gamma \left( c-1\right) }{%
\Gamma \left( a\right) }z^{1-c}\Phi \left( a-c+1,2-c;z\right) \,,
\label{app1.2}
\end{equation}%
and the power series expansion of regular CHF at the origin $\Phi \left(
a,c;z\right) =1+\left( a/c\right) z+O\left( z^{2}\right) $ to obtain an
approximate expression of $\Psi \left( a,c;z\right) $. If $a$ and $c$ are
also small, one may choose a value to $\chi $ ($\chi =+1$ for example) and
expand the Gamma functions to obtain%
\begin{eqnarray}
&&\Psi \left( a_{2},c_{2};z_{2}\right) \approx 1\,,\ \ \frac{d}{dz_{2}}\Psi
\left( a_{2},c_{2};z_{2}\right) \approx i\nu _{2}^{+}\,,  \notag \\
&&\Psi \left( c_{1}-a_{1},c_{1};e^{-i\pi }z_{1}\right) \approx -e^{i\pi
c_{1}}z_{1}^{1-c_{1}}\,,  \notag \\
&&\frac{d}{dz_{1}}\Psi \left( c_{1}-a_{1},c_{1};e^{-i\pi }z_{1}\right)
\approx e^{i\pi c_{1}}z_{1}^{1-c_{1}}\left( \frac{c_{1}-1}{z_{1}}\right) \,,
\label{app1.3}
\end{eqnarray}%
for Fermions, in leading-order approximation.

\section{Unitary operator connecting in- and out-vacua in Klein zone\label%
{App2}}

A fundamental property of linear canonical transformations between sets of
creation and annihilation operators is the existence of an unitary operator $%
V$ \cite{Berezin} that connects both sets in the form $\alpha \left( \mathrm{%
out}\right) =V^{\dagger }\tilde{\alpha}\left( \mathrm{in}\right) V$, where $%
\alpha \left( \mathrm{out}\right) $ denotes any out-operator and $\tilde{%
\alpha}\left( \mathrm{in}\right) $ its corresponding in-operator. The
general method for calculating its explicit form has been given in Refs. 
\cite{BagGiS75,Gitman,GavGitTom08}. Here we employ this method for
calculating the corresponding unitary operator in the Klein zone $V_{\Omega
_{3}}$ in terms of in-operators, as a supplement to the representation
calculated in terms of out-operators; cf. Eq. (7.20) in Ref. \cite{x-case}.
Starting with the representation%
\begin{eqnarray}
V_{\Omega _{3}} &=&\exp \left[ \ ^{-}a_{n}^{\dagger }\left( \mathrm{in}%
\right) B\ _{-}b_{n}^{\dagger }\left( \mathrm{in}\right) \right] \exp \left[
\ ^{-}a_{n}\left( \mathrm{in}\right) A\ ^{-}a_{n}^{\dagger }\left( \mathrm{in%
}\right) \right] \,,  \notag \\
&\times &\exp \left[ \ _{-}b_{n}^{\dagger }\left( \mathrm{in}\right) D\
_{-}b_{n}\left( \mathrm{in}\right) \right] \exp \left[ \ _{-}b_{n}\left( 
\mathrm{in}\right) C\ ^{-}a_{n}\left( \mathrm{in}\right) \right] \,,
\label{app2.13}
\end{eqnarray}%
for Fermions and%
\begin{eqnarray}
V_{\Omega _{3}} &=&\exp \left[ \ ^{+}a_{n}^{\dagger }\left( \mathrm{in}%
\right) B\ _{+}b_{n}^{\dagger }\left( \mathrm{in}\right) \right] \exp \left[
\ ^{+}a_{n}\left( \mathrm{in}\right) A\ ^{+}a_{n}^{\dagger }\left( \mathrm{in%
}\right) \right] \,,  \notag \\
&\times &\exp \left[ \ _{+}b_{n}^{\dagger }\left( \mathrm{in}\right) D\
_{+}b_{n}\left( \mathrm{in}\right) \right] \exp \left[ \ _{+}b_{n}\left( 
\mathrm{in}\right) C\ ^{+}a_{n}\left( \mathrm{in}\right) \right] \,,
\label{app2.14}
\end{eqnarray}%
for Bosons, where $A$, $B$, $C$ and $D$ are constants, we use the identities%
\begin{eqnarray}
\exp \left( \pm a_{n}Aa_{n}^{\dagger }\right) \left( 
\begin{array}{c}
a_{i} \\ 
a_{i}^{\dagger }%
\end{array}%
\right) \exp \left( \mp a_{n}Aa_{n}^{\dagger }\right) &=&\left( 
\begin{array}{c}
e^{\pm \kappa A}a_{i} \\ 
a_{i}^{\dagger }e^{\mp \kappa A}%
\end{array}%
\right) \,,  \notag \\
\exp \left( \pm b_{n}^{\dagger }Db_{n}\right) \left( 
\begin{array}{c}
b_{i} \\ 
b_{i}^{\dagger }%
\end{array}%
\right) \exp \left( \mp b_{n}^{\dagger }Db_{n}\right) &=&\left( 
\begin{array}{c}
e^{\mp D}b_{i} \\ 
b_{i}^{\dagger }e^{\pm D}%
\end{array}%
\right) \,,  \notag \\
\exp \left( \pm a_{n}^{\dagger }Bb_{n}^{\dagger }\right) \left( 
\begin{array}{c}
a_{i} \\ 
b_{i}%
\end{array}%
\right) \exp \left( \mp a_{n}^{\dagger }Bb_{n}^{\dagger }\right) &=&\left( 
\begin{array}{c}
a_{i}\mp Bb_{i}^{\dagger } \\ 
b_{i}\pm \kappa a_{i}^{\dagger }B%
\end{array}%
\right) \,,  \notag \\
\exp \left( \pm b_{n}Ca_{n}\right) \left( 
\begin{array}{c}
a_{i}^{\dagger } \\ 
b_{i}^{\dagger }%
\end{array}%
\right) \exp \left( \mp b_{n}Ca_{n}\right) &=&\left( 
\begin{array}{c}
a_{i}^{\dagger }\pm b_{i}C \\ 
b_{i}^{\dagger }\mp \kappa Ca_{i}%
\end{array}%
\right) \,,  \label{app2.15}
\end{eqnarray}%
and the canonical transformations given by Eqs. (7.4), (A3) in Ref. \cite%
{x-case}, to show that%
\begin{eqnarray}
A &=&\left\{ 
\begin{array}{l}
-\ln \left[ g\left( _{+}|^{-}\right) g\left( _{+}|^{+}\right) ^{-1}\right]
=-\ln \left[ g\left( ^{+}|_{-}\right) g\left( ^{-}|_{-}\right) ^{-1}\right]
\,,\ \mathrm{Fermi\,,} \\ 
\ln \left[ g\left( _{-}|^{+}\right) g\left( _{-}|^{-}\right) ^{-1}\right]
=\ln \left[ g\left( ^{-}|_{+}\right) g\left( ^{+}|_{+}\right) ^{-1}\right]
\,,\ \mathrm{Bose\,,}%
\end{array}%
\right.  \notag \\
B &=&\left\{ 
\begin{array}{l}
-g\left( _{+}|^{+}\right) ^{-1}\,,\ \mathrm{Fermi\,,} \\ 
g\left( _{-}|^{-}\right) ^{-1}\,,\ \mathrm{Bose\,,}%
\end{array}%
\right. \,,\ \ C=\left\{ 
\begin{array}{l}
g\left( ^{-}|_{-}\right) ^{-1}\,,\ \mathrm{Fermi\,,} \\ 
-g\left( ^{+}|_{+}\right) ^{-1}\,,\ \mathrm{Bose\,,}%
\end{array}%
\right.  \notag \\
D &=&\left\{ 
\begin{array}{l}
\ln \left[ g\left( _{-}|^{+}\right) g\left( _{+}|^{+}\right) ^{-1}\right]
=\ln \left[ g\left( ^{-}|_{+}\right) g\left( ^{-}|_{-}\right) ^{-1}\right]
\,,\ \mathrm{Fermi\,,} \\ 
-\ln \left[ g\left( _{+}|^{-}\right) g\left( _{-}|^{-}\right) ^{-1}\right]
=-\ln \left[ g\left( ^{+}|_{-}\right) g\left( ^{+}|_{+}\right) ^{-1}\right]
\,,\ \mathrm{Bose\,,}%
\end{array}%
\right.  \label{app2.16}
\end{eqnarray}%
In terms of elementary relative amplitudes of particle scattering $%
w_{n}\left( +|+\right) $, antiparticle scattering $w_{n}\left( -|-\right) $,
creation of a pair $w_{n}\left( +-|0\right) $ and annihilation of a pair $%
w_{n}\left( 0|-+\right) $ given by Eqs. (7.17) and (A-9) in \cite{x-case},
the unitary operator for Fermions (\ref{app2.13}) is expressed by Eq. (\ref%
{3.6}) while for Bosons it takes the form%
\begin{eqnarray}
V_{\Omega _{3}} &=&\exp \left[ \ ^{+}a_{n}^{\dagger }\left( \mathrm{in}%
\right) w_{n}\left( +-|0\right) \ _{+}b_{n}^{\dagger }\left( \mathrm{in}%
\right) \right] \exp \left[ \ ^{+}a_{n}\left( \mathrm{in}\right) \ln w\left(
+|+\right) _{n}\ ^{+}a_{n}^{\dagger }\left( \mathrm{in}\right) \right] 
\notag \\
&\times &\exp \left[ -\ _{+}b_{n}^{\dagger }\left( \mathrm{in}\right) \ln
w\left( -|-\right) _{n}\ _{+}b_{n}\left( \mathrm{in}\right) \right] \exp %
\left[ \ _{+}b_{n}\left( \mathrm{in}\right) w\left( 0|-+\right) _{n}\
^{+}a_{n}\left( \mathrm{in}\right) \right] \,.  \label{app2.17}
\end{eqnarray}%
With the help of the representations (\ref{3.6}) and (\ref{app2.17}), the
vacuum-vacuum transition probability $P_{v}$ (\ref{3.5}) acquires the final
form (\ref{3.8}).

\end{document}